\documentclass[prd,aps,preprint,amsmath,nofootinbib,amssymb,eqsecnum,showkeys,tightenlines]{revtex4-1}

\usepackage{amsmath, latexsym, amssymb, hyperref, graphicx, color, multirow, makecell, diagbox}
\usepackage[table]{xcolor}
\usepackage{tabularx}
\usepackage[T1]{fontenc}
\usepackage[capitalize]{cleveref}
\definecolor{nicered}{rgb}{.7,.1,.1}
\definecolor{nicegreen}{rgb}{.1,.5,.1}
\definecolor{darkblue}{rgb}{0,0,.5}
\hypersetup{colorlinks, citecolor=nicegreen,linkcolor=nicered, urlcolor=darkblue}
\usepackage{multirow}
\usepackage{slashed}
\numberwithin{equation}{section}
\usepackage{verbatim}
\usepackage{graphicx}
\usepackage{CJKutf8}
\usepackage{fontawesome}
\graphicspath{{picture/}}

\begin{document}
\preprint{}

\title{Polarization measurement for the dileptonic channel of $W^+ W^-$ scattering using generative adversarial network}

\author{Jinmian Li$^{1}$}
\email{jmli@scu.edu.cn}

\author{Cong Zhang$^{1}$}
\email{zhangcong.phy@gmail.com}

\author{Rao Zhang$^{1}$}
\email{zhangrao@stu.scu.edu.cn}

\affiliation{$^1$ College of Physics, Sichuan University, Chengdu 610065, China}

\begin{abstract}

Measuring the polarization fractions of the $W^+W^-$ scattering reveals the interactions of the Higgs boson as well as new neutral states that are related to the standard model electroweak symmetry breaking. 
The dileptonic channel has a relatively lower background rate, but the kinematics of its final states can not be fully reconstructed due to the presence of two neutrinos. 
We propose neural networks to establish maps between the distributions of measurable quantities and the distributions of the lepton angles in $W$ boson rest frames. 
New physics contributions and collision energy can largely affect the kinematic properties of the $W^+W^-$ scattering beside the lepton angles. To make the network in ignorance of that information, the loss function is modified in two different ways. 
We show that the networks are promising in reproducing the lepton angle distributions, and the precision of the fitted polarization fractions obtained from network predictions is comparable to that obtained with the truth lepton angle. 
Although the best-fit values of polarization fractions do not change much after including the background uncertainty, the precisions is substantially reduced. 
Our trained models are available at GitHub \href{https://github.com/scu-heplab/wlwl-polarization}{\faGithub}.
\end{abstract}


\maketitle

\section{Introduction}\label{sec:intro}

Vector Boson Scattering (VBS)~\cite{Rauch:2016pai,Green:2016trm,Buarque:2021dji,Covarelli:2021gyz} represents sensitive probe to any new physics that is interacting with the electroweak sector of the Standard Model (SM).
If the Higgs sector is extended or the couplings between the Higgs boson and the gauge bosons deviate from the SM predictions, the scattering amplitudes for the longitudinal mode of the VBS will increase with center-of-mass energy and violate unitarity. 

At hadron colliders, the VBS processes result in final states with two gauge bosons and a pair of forward-backward jets. 
VBS channels have been observed at the LHC run-II, including the dileptonic same-sign $W^\pm W^\pm$~\cite{Aaboud:2019nmv,Sirunyan:2017ret}, fully leptonic $ZZ$~\cite{Sirunyan:2017fvv,Aad:2020zbq}, fully leptonic $WZ$~\cite{Aaboud:2018ddq, Sirunyan:2019ksz}, and semileptonic $WV/ZV$ with the $V$ decaying hadronically~\cite{Aad:2019xxo,Sirunyan:2019der}. 
Investigating the polarization modes of the VBS processes is an important step afterward. 
The polarization of the vector bosons can be measured by their decay products. 
The interferences among different polarization channels disappear when the azimuthal angles of the decay products are integrated over. 
Although the selection cuts in analyses induce a certain amount of interferences, it is still possible to extract polarization fractions by fitting the data with simulated templates. 
There have been many studies of polarization measurement for the $W^+ W^-$ channel~\cite{Han:2009em,Ballestrero:2017bxn}, the fully leptonic $W^\pm W^\pm$ channel~\cite{Ballestrero:2020qgv}, the fully leptonic $WZ/ZZ$ channel~\cite{Ballestrero:2019qoy}, as well as the $WW/ZZ$ from the SM Higgs decay~\cite{Maina:2020rgd} and generic processes with boosted hadronically decaying $W$ boson~\cite{De:2020iwq}. 
The CMS Collaboration studied the prospects for measuring the longitudinal modes of $W^\pm W^\pm$ and $WZ$ channels at the future HL-LHC~\cite{CMS-PAS-SMP-14-008,CMS-PAS-FTR-18-005}.   
Some recent studies take advantage of deep learning techniques. 
Taking the final state momenta as input, the network is able to either regress the lepton angle in the gauge boson rest frame~\cite{Searcy:2015apa,Grossi:2020orx} or classify events from different polarizations~\cite{Lee:2018xtt,Lee:2019nhm}. 

In this work, we study the polarization measurement for the dileptonic $W^+ W^-$ channel, as it has a large production cross section at the LHC and is relevant to the neutral scalar bosons. 
Although resolving the polarization of the hadronically decaying $W$ boson is possible~\cite{De:2020iwq}, it suffers from large uncertainties and backgrounds. 
Focusing on the $W W$ scattering with fully leptonic decay in the SM, the fractions of the $W$ boson polarization can be determined from distributions of many kinematic variables~\cite{BuarqueFranzosi:2019boy,Ballestrero:2020qgv}, {\it} e.g. the transverse momenta of leptons, the invariant mass of two leptons and so on. 
Studies in Refs.~\cite{Lee:2018xtt,Lee:2019nhm} use the neural network to discriminate different polarization modes of VBS processes and the output of the network can be used to extract the fraction of each polarization mode. 
However, those methods utilize information that also depends on other properties of the process besides the polarization. Thus they can not be applied directly to VBS processes with significant beyond SM (BSM) contributions. 
The most exclusive variable that characterizes the vector boson polarization is the angle between the charged lepton in the gauge boson rest frame and the gauge boson direction of motion (denoted by $\theta^*_\ell$ hereafter). 
Because of the presence of two neutrinos in the final state, the lepton angles can not be fully reconstructed. There have been attempts to use a neural network to regress two lepton angles in the gauge boson rest frames~\cite{Searcy:2015apa,Grossi:2020orx} for the same sign $W^\pm W^\pm$ scattering. 

We further develop the machine learning methods by adopting the transformer network~\cite{10.5555/3295222.3295349} and the generative adversarial network. 
The Transformer network is known to be quite successful in extracting features of polarizations for VBS processes~\cite{Li:2020fna}. 
The generative adversarial network is used to regress the distribution of lepton angle $\theta^*_\ell$. 
As a result, the network can be used to measure the polarization fraction of the $W^+ W^-$ scattering in a wide class of models.  
For illustration, we apply the network to a simplified model with an effective operator and the two-Higgs-doublet model (2HDM). 
In particular, there is an extra neutral Higgs boson in the 2HDM which induces resonant $W^+ W^-$ production, so the kinematic properties of the $W^+ W^- j j$ final state in the 2HDM are quite different from the SM ones.  
We show that our network works well in regressing the lepton angles for both BSM scenarios.  
The polarization fraction can be extracted from fitting the distribution of the predicted lepton angle to a linear combination of pure longitudinal/transverse templates.
However, to reduce the background events, some preselection cuts need to be applied before constructing the templates. Those cuts will affect the shapes of templates. 
We find that the shapes of templates (according to our preselection) are similar in different models, but exhibit some dependences on collision energy. 
This means different sets of templates should be reconstructed at different collision energy. 

This paper is organized as follows. 
The analysis framework is explained in Sec.~\ref{sec:framework}, including the setups of the network, event preparation, and fitting procedure. In Sec.~\ref{sec:model13} and Sec.~\ref{sec:model100}, we study the performance of the network applied on different models and different collision energies. 
The effects of backgrounds are discussed in Sec.~\ref{sec:bkg}. 
We summarize our work and conclude in Sec~\ref{sec:conclude}.

\section{Analysis framework}\label{sec:framework}

\subsection{Definitions of loss functions and the network} \label{sec::network}

The following issues need to be addressed in network construction: 
\begin{itemize}
\item Because of missing information for two neutrinos in the final state, it is not possible to fully determine the two lepton angles in the rest frames of $W$ bosons for each event.
The lepton decay angles of events with the same values of observables (including momenta of leptons and jets, as well as missing transverse momentum) form a distribution.
Our network is built to establish a map between the distributions of measurable quantities and the distributions of the lepton angles, based on a large number of events.
\item Since we expect that the VBS process is affected by unknown new physics, the network should be able to extract the $W$ boson polarization for processes that have kinematic properties quite different from the SM. This means that the features extracted by the network should be only related to $\theta^*_\ell$ and decorrelated from other process-dependent variables.
\item Extracting the polarization fraction requires fitting to given templates. The shapes of templates are affected by preselection cuts; thus, they will be different at different collision energies of hadron collider. On the other hand, the network used to extract polarization information needs to provide features that do not change with collision energy.
\end{itemize}

Because of the first issue, we can not use the mean square error loss function, which can only reproduce the average value of the lepton angle distribution (for given values of observables) and lead to the deviation between the truth level distribution and the predicted distribution~\cite{Searcy:2015apa,Grossi:2020orx}.
Events with the same measurable momenta of final states while having different $\theta^*_{\ell^\pm}$ are grouped into subsets denoted by $\pmb{e}^i$. 
The measurable momenta for the subset $\pmb{e}^i$ are denoted by $\pmb{p}^i$ (same for all events in $\pmb{e}^i$) and the set of $\theta^*_{\ell^\pm}$ for events in $\pmb{e}^i$ is denoted by $\pmb{t}^i$. 
The goal of the network is to establish a map that maximizes the probability of $P(\pmb{t}^i | \pmb{p}^i)$ while minimizing the probability $P(\pmb{t}^i | \pmb{p}^j)$ for $j\neq i$, where $i$ and $j$ run over all subsets. 
The loss function of the Conditional Generative Adversarial Network (CGAN)~\cite{DBLP:journals/corr/MirzaO14} meets the needs:
\begin{equation}
L_{\text{CGAN}}= \mathop{\min}\limits_{G}\ \mathop{\max}\limits_{D} \left( \mathbb{E} [ \log{D}(\pmb{t}^{\prime i} | \textbf{p}^i) ]+\mathbb{E}_{z \in {\mathcal{N}(0,1)} }[\log(1-D(G(z, \textbf{p}^i)| \textbf{p}^i) )] \right), \label{eq:losscgan}
\end{equation}
where $z$ is sampled from a Gaussian distribution and $\mathbb{E}$ denotes the average over all events. 
The distribution of ($\theta^*_{\ell^+}$, $\theta^*_{\ell^-}$) in subset $\pmb{t}^{i}$ is replaced by a two dimensional Gaussian distribution $\pmb{t}^{\prime i}$ (which centers on the mean of ($\theta^*_{\ell^+}$, $\theta^*_{\ell^-}$) with standard derivative 0.01) for simplicity. The discriminative network ($D$) evaluates the consistency between the $\pmb{p}^i$ and a lepton angle distribution. 
The generative network ($G$) aims to reproduce the $\pmb{t}^{\prime i}$ distribution with the input of $z$ and $\pmb{p}^i$. 
The GAN enables us to obtain the lepton angle by sampling instead of taking the average, and it transforms the random distribution $z$ into meaningful distributions based on the information obtained from training samples.

To address the second problem, we adopt the Mutual Information (MI) variable to measure the nonlinear correlation between features and the target variables. For any two sets of variables $X$ and $Y$, the MI is defined as 
\begin{equation}
I(X;Y) = \sum\limits_{x\in{X}}\sum\limits_{y\in{Y}}\mathbb{P}_{(x, y)}\log(\frac{\mathbb{P}_{(x, y)}}{\mathbb{P}_x\mathbb{P}_y}), 
\end{equation}
where $\mathbb{P}_{(x,y)}$ is the joint probability density function, and $\mathbb{P}_x$ and $\mathbb{P}_y$ are the marginal probability density functions.   
The $I(X;Y)$ is larger if $X$ and $Y$ share similar information, while $I(X;Y)=0$ if $X$ and $Y$ are independent of each other. 
However, MI is difficult to calculate in practice. We use the following approximation to estimate the MI~\cite{DBLP:journals/corr/abs-1801-04062} instead
\begin{equation}
I(X;Y) = \sup\limits_{\omega} [\mathbb{E}_{\mathbb{P}_{XY}}[T_\omega] - \log(\mathbb{E_{\mathbb{P}_{X}\otimes\mathbb{P}_{Y}}}[e^{T_\omega}])], \label{eq:lossmi}
\end{equation}
where $T_\omega$ is an arbitrary function described by a neural network in which the weights $\omega$ are trained to provide the least upper limit for the $I(X;Y)$.
In our study, the loss function is written such that the MI between the two leptons angles $\theta^*_{\ell^\pm}$ and the features (the Transformer output, which has dimension 64, will be discussed later) is maximized, while the MI between the $W$ boson pair momentum (including invariant mass $m(WW)$, energy $E(WW)$, rapidity $y(WW)$, and azimuth $\phi(WW)$) and the features are minimized to reduce the dependence of network performance on $W$ boson pair production mechanism. 
So the $L_{\text{MI}}=I(F;WW)-I(F;\Theta^*)$ is added to the category loss of the Transformer network. The $F$, $WW$ and $\Theta^*$ denote the sets of feature variables, $W$ boson pair momentum, and two leptons angles, respectively. 
The $T_\omega$ networks in $I(F;WW)$ and $I(F;\Theta^*)$ are fully connected neural networks, which consist of four layers with $[272=4\times(64+4), 272, 272, 1]$ and $[264=4\times(64+2), 264, 264, 1]$ numbers of neurons, respectively, and the ReLU function acts on all layers except the last layer of the two fully connected neural networks.

The transverse momenta of particles in the final state are approximately linearly related to the initial collision energy. 
Thus we further add the Pearson Correlation Coefficient (PCC) 
to the loss function to reduce the dependences of the feature variables on collision energy. 
The PCC is defined as 
\begin{equation}
\rho_{XY}=\lvert\frac{\sum\limits_{i}(X_i-\bar{X})(Y_i-\bar{Y})}{\sqrt{\sum\limits_{i}(X_i-\bar{X})^2}\sqrt{\sum\limits_{i}(Y_i-\bar{Y})^2}}\rvert, \label{eq:losspcc}
\end{equation}
where $i$ runs over all events, and $\bar{X}$ and $\bar{Y}$ denote the average of the variables. 
In our case, the features from the Transformer output are taken as $X$ and the variable $Y$ indicates the transverse momenta of the $W$ bosons, leptons, as well as forward-backward jets~\footnote{We also include the pseudorapidities of the $W$ bosons, although they are not linearly correlated with collision energy.}.  
Thus, $64 \times 8$ $\rho_{XY}$'s can be calculated. The average $\bar{\rho}_{XY}$ is added to the loss of the Transformer network. 

Having defined the loss function, we can construct a network based on the Transformer network and the CGAN. 
The Transformer network~\cite{10.5555/3295222.3295349} with a multi-head self-attention mechanism provides a variety of different attentions and improves the learning ability; thus, it can be used to effectively extract the internal connections of the features. In Ref.~\cite{Li:2020fna}, it is found to be efficient in extracting polarization information for the $W^+W^-$ scattering in both semi-leptonic and dileptonic channels. 
We adopt the same Transformer network as used for the dileptonic channel in this work. 
The low-level inputs (momenta of final state particles) are transformed into a 64-dimensional feature variable which is supposed to contain the full polarization information. 
In this study, the loss function of the Transformer network is modified according to the discussions above, using $L_{\text{MI}}$ and $\bar{\rho}_{XY}$. 
The CGAN uses the feature as a condition and reproduces the two-dimensional lepton angle distribution.
The Generator takes the input of the condition (64 dimensions) and a 64-dimensional Gaussian distribution and aims to regenerate the lepton angle distribution. 
The Discriminator takes the input of the condition and $\Theta$(or $\Theta^*$)~\footnote{The predicted lepton angles $\theta^*_{\ell^\pm}$ distribution is denoted by $\Theta$ and the truth lepton angles $\theta^*_{\ell^\pm}$ distribution is denoted by $\Theta^*$. } and determines whether the input $\Theta$(or $\Theta^*$) is consistent with the condition. 

\begin{figure}[thb]
\includegraphics[width=0.9\textwidth]{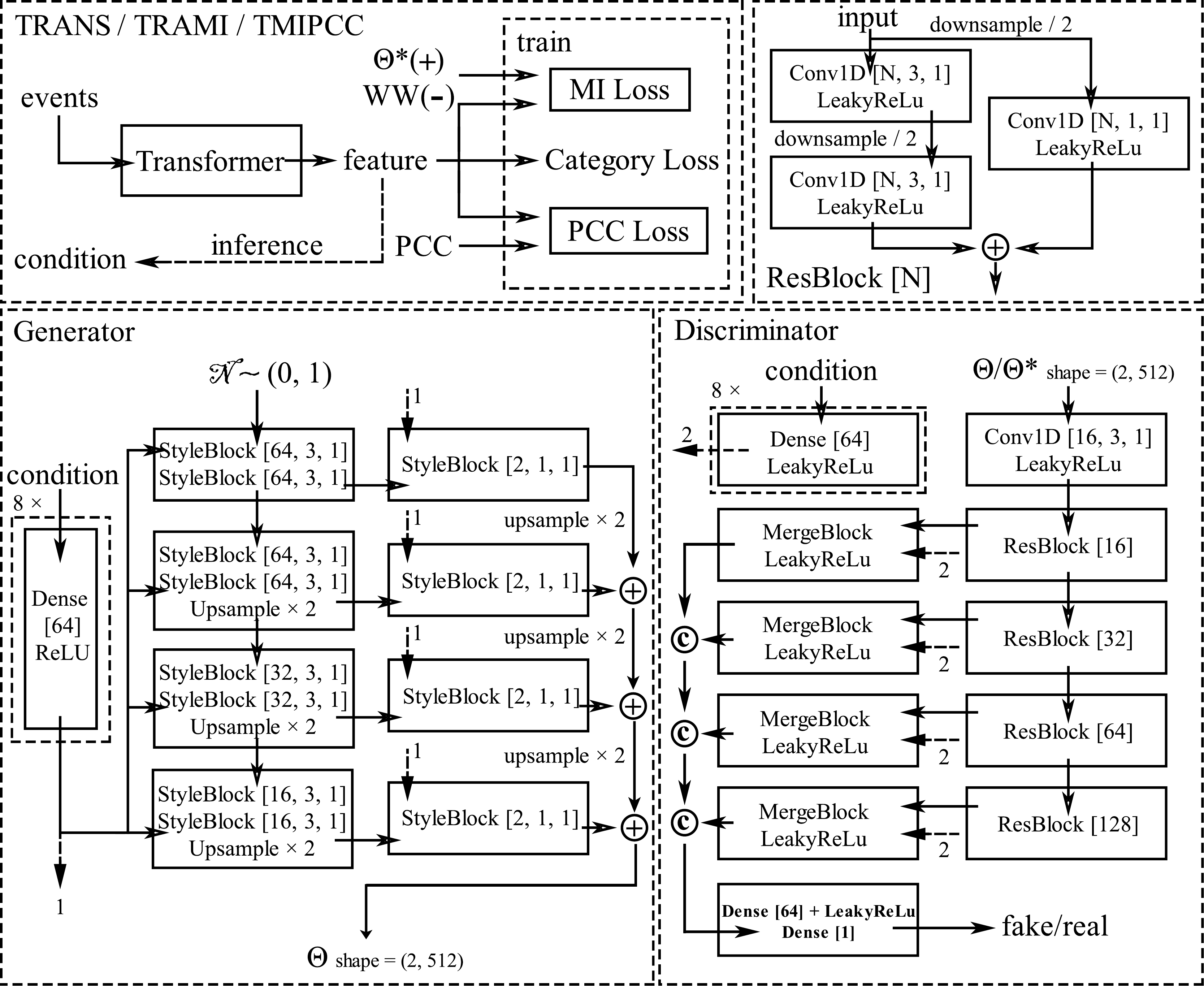}
\caption{ Workflow of the network. \label{fig:network}}
\end{figure}

More details of the data processing and architecture of the network are depicted in Fig.~\ref{fig:network}. 
The upper-left plot illustrates the processing pipeline of the modified Transformer network. 
Note that the variables in sets of $\Theta^*$, $WW$, and PCC, which can only be calculated on the Monte Carlo events are only used for training. During the inference stage, only the inputs of the measurable momenta of final states are required. 
The lower plots show the architectures of the generative network and the discriminative network. 
In both networks, the condition is processed by a dense network with eight layers. The outputs of the dense networks are reused multiple times in the network as indicated by labels 1 and 2. 
The ResBlock is proposed to address the degradation problem~\cite{he2015deep} in training deep networks. We illustrate its decomposition in the upper-right plot. 
The StyleBlock combines the condition with its input by convolution operation (for more detail, see Ref.~\cite{DBLP:journals/corr/abs-1912-04958}). 
The parameters NF, SF, and S in one-dimensional convolution Conv1D[NF, SF, S] and StyleBlock[NF, SF, S] are the number of filters, filter size, and stride, respectively.
The MergeBlock multiplies the condition with lepton angle information in matrix form. 
The LeakyReLu is an activation function, which is consistent with the ReLU for $x\geq 0$ and equals to $0.1  x$ for $x\textless 0$.
The symbol \textcircled{\textbf{+}} means the sum of corresponding elements; \textcircled{\textbf{c}} means concatenating by channel; upsampling$\times$2 (downsampling$/$2) means using linear interpolation to expand (reduce) the dimension of input by twice (half).

\subsection{Event simulation and network training} \label{sec:event}

New physics models are implemented in \texttt{FeynRules}~\cite{Alloul:2013bka} (in our case, we consider the effective field theory~\cite{Alloul:2013naa} and the 2HDM).
Events at the LHC are simulated within the \texttt{MG5\_aMC@NLO} framework~\cite{Alwall:2014hca}, including those with fixed helicities of gauge bosons in the final state~\cite{BuarqueFranzosi:2019boy}~\footnote{The definitions of the polarizations are reference frame-dependent. We take the initial parton center of the mass frame as the reference frame in this work. }. 
The \texttt{MadSpin}~\cite{Artoisenet:2012st} is turned on to preserve the polarization information in the decay products of the gauge bosons. 
The \texttt{Pythia8}~\cite{Sjostrand:2007gs} is used for the parton shower, hadronization, and decays of hadrons. 
The final state jets are reconstructed by \texttt{Fastjet}~\cite{Cacciari:2011ma} using the anti-$k_T$ algorithm with cone size parameter $R = 0.4$. 
The detector effects are simulated by \texttt{Delphes3}~\cite{deFavereau:2013fsa} with the ATLAS configuration card, where $b$-tagging efficiency is set to 70\%, and the mistagging rates for the charm- and light-flavor jets are 0.15 and 0.008, respectively~\cite{ATLAS:2016gsw}. 

The $W^+ W^-$ scattering is simulated at order of $\mathcal{O}(\alpha_{EW}^4)$ \footnote{The $\alpha_{\text{EW}}$ and $\alpha_s$ denote the electroweak coupling constant and strong coupling constant, respectively. } in the SM model.  
There are also $W^+ W^- j j$ productions at $\mathcal{O}(\alpha_{EW}^2 \alpha^2_s)$ with much higher rates, but they do not correspond to VBS. 
They will be treated as the background for the $W^+ W^-$ scattering, 
because the interference contributions at $\mathcal{O}(\alpha_{EW}^3 \alpha_s$) are found to be small~\cite{Biedermann:2017bss,Ballestrero:2017bxn,Campanario:2020xaf}. 
As for simulating the processes in BSM, the new physics coupling ($\alpha_{NP}$) is assumed to be close to the electroweak coupling. The processes at the order of $\mathcal{O}(\alpha_{EW}^a \alpha_{NP}^b)$ with $a+b=4$ are considered. 

In order to separate signal and background events in the dileptonic channel, the following preselections are applied:
\begin{itemize}
\item exactly two opposite sign leptons with $p_T(\ell) >20~\text{GeV}, ~ |\eta(l)|<2.5$; 
\item at least two jets with $ p_T(j) >20~\text{GeV}, ~ |\eta(j)|<4.5$; 
\item the two jets with leading $p_T$ should give large invariant mass ($m_{jj}>500$ GeV) and have large pseudorapidity separation ($|\Delta \eta|_{jj}>3.6$); 
\item no $b$-tagged jet in the final state.
\end{itemize}

The preselected events are used for training and testing the network. 
The network input consists of momenta ($p_x, p_y, p_z, E$) of two leptons,
forward and backward jets, the vectorial sum of all detected particles, and the vectorial sum of jets that are not assigned
as forward-backward jets. 
The transformer network that is used to extract the features of different polarizations is trained on the events of the SM $W^+W^-$ scattering (with given final state polarizations) at the 13 TeV LHC~\footnote{This leads to different performances of a network at 13 and 100 TeV when the features are correlated with the collision energy.}.  
Moreover, as discussed in the previous subsection, the following variables are calculated for each Monte Carlo event (used at the training stage):
\begin{align}
&\text{MI variables:} ~\theta^*_{\ell^+}, ~\theta^*_{\ell^-}, ~m(WW), ~E(WW), ~y(WW),~ \phi(WW),  \nonumber\\
&\text{PCC variables:} ~p_T(W^\pm), ~p_T(\ell^\pm), ~p_T(j_{\text{fb}}), ~\eta(W^\pm).  \nonumber
\end{align}  
To show the performance gain of adding Eq.~\ref{eq:lossmi} and Eq.~\ref{eq:losspcc} to the loss function, three versions of networks are trained: 
\begin{itemize}
\item network with normal Transformer loss function, denoted by TRANS;
\item network with $L_{\text{MI}}$ being added to the loss function, denoted by TRAMI;
\item network with both $L_{\text{MI}}$ and $\bar{\rho}_{XY}$ being added to the loss function, denoted by TMIPCC.
\end{itemize}

The CGAN is trained independently of the Transformer (with modified loss).  It takes the input condition provided by the well-trained TRANS network, TRAMI network, and TMIPCC network, respectively. 
Events of the $W^+W^-$ scattering in both the SM and BSM at several collision energies are used for training the CGAN. 
The BSM scenarios include the EFT and the 2HDM with several choices of benchmark parameters, as will be discussed later.  

\begin{table}[htb]
\begin{center}
\begin{tabular}{|c||c|c|c|c|c|c|} \hline
 &             Accuracy    & MI(F;$\Theta^*$)  & MI(F;$WW$) & $\bar{\rho}_{XY}$ & MI($\Theta;\Theta^*)$ \\ \hline
TRANS &  0.44095   &  0.07139 & 1.26769 & 0.23673 & 0.318141 \\
TRAMI &   0.42746   & 0.84366 & 0.04764 & 0.19595 & 0.548189 \\
TMIPCC & 0.42559  & 0.84437  & 0.0473 & 0.08593 & 0.486755  \\ \hline
\end{tabular}
\caption{\label{tab:train} Network performances after training. Accuracy is defined as the number of correctly predicted events (true positives + true negatives) divided by the total number of events. 
MI(F;$\Theta^*$),  MI(F;$WW$), MI($\Theta;\Theta^*)$ are the mutual information calculated over all training samples. $F$ denotes the 64 dimensional features; $WW$ denotes the variables of $W$ pair momentum $m(WW)$, $E(WW)$, $y(WW)$, $\phi(WW)$. }
\end{center}
\end{table}

In Tab.~\ref{tab:train}, we present the classification accuracies of the Transformer networks and correlation information for the well-trained full networks (Transformer$+$CGAN). 
The classification accuracy can reach 44\% for the TRANS network and decrease a bit in TRAMI and TMIPCC. 
However, the great enhancement (reduction) of MI(F;$\Theta^*$) (MI(F;$WW$)) in TRAMI indicates that the features have been changed dramatically. 
The $\bar{\rho}_{XY}$ is effectively reduced in TMIPCC, although there is also a mild reduction in TRAMI. 
As for the correlation between the truth lepton angles and the predicted ones, we find it is much increased in TRAMI, although adding the $\bar{\rho}_{XY}$ reduces the value by a small amount.
We note that the absolute value of MI is not useful, only the relative size has physical meaning. 

\begin{figure}[htbp]
\includegraphics[width=0.24\textwidth]{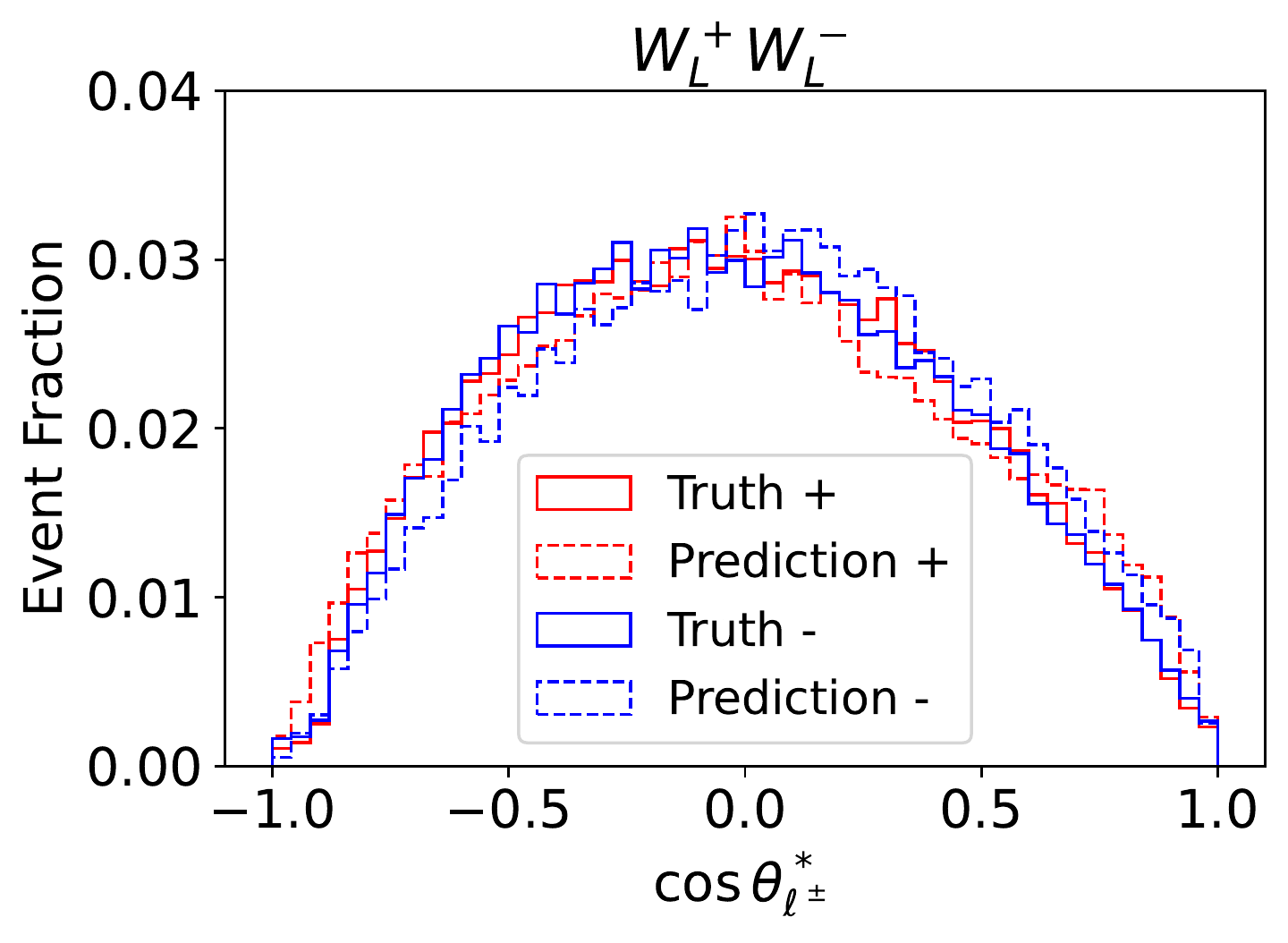}
\includegraphics[width=0.24\textwidth]{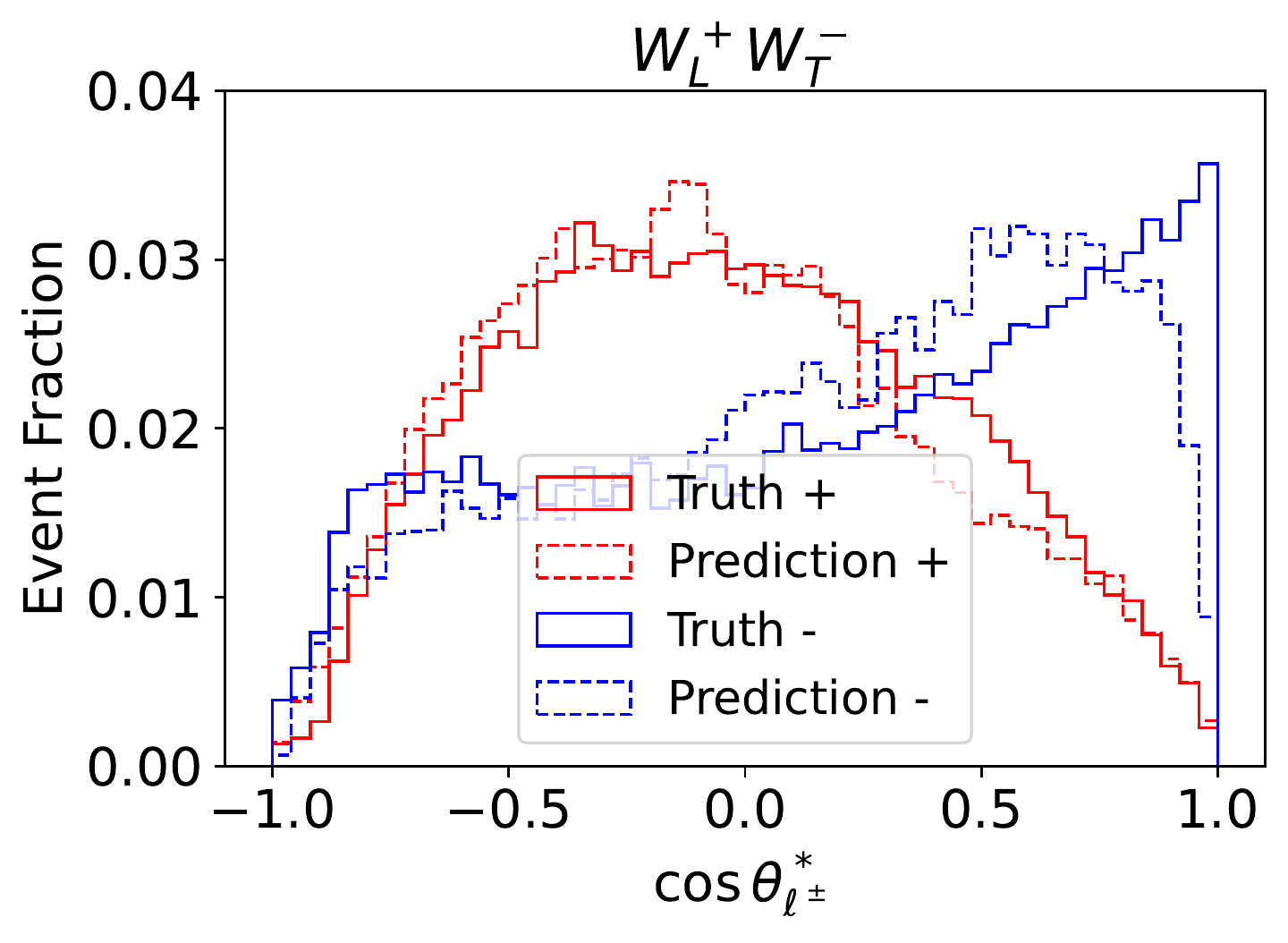}
\includegraphics[width=0.24\textwidth]{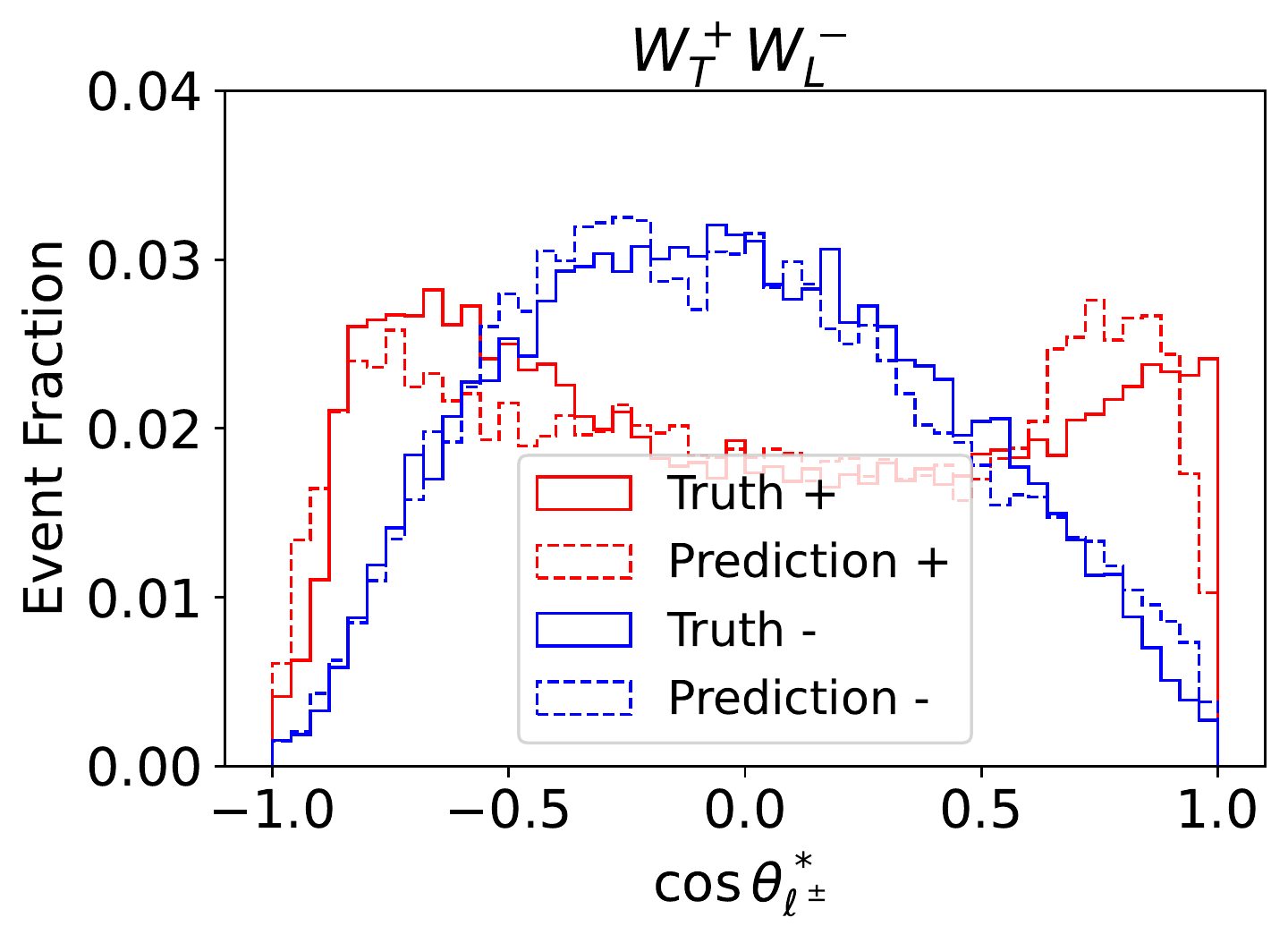}
\includegraphics[width=0.24\textwidth]{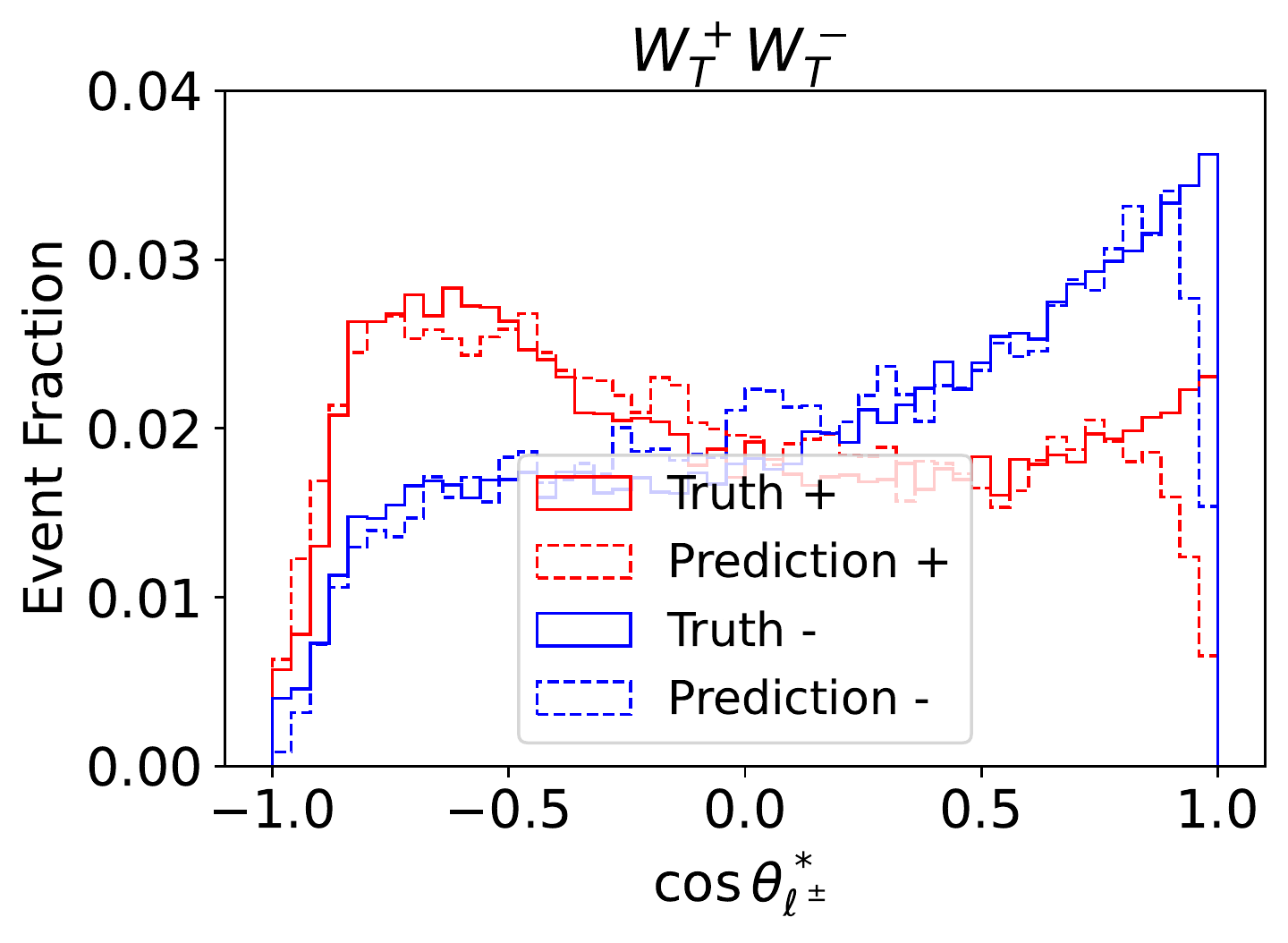}
\caption{Comparison of truth level $\cos \theta^*_{\ell^\pm}$ distributions and the TRAMI network output distributions for different polarization modes of the SM $W^+W^-$ scattering at 13 TeV. \label{fig:1dtemp}}
\end{figure}

For demonstration, we show the distributions of the lepton angles $\cos \theta^*_{\ell^\pm}$ predicted by the TRAMI network for different polarization modes of the SM $W^+W^-$ scattering at 13 TeV in Fig.~\ref{fig:1dtemp}. 
The truth level $\cos \theta^*_{\ell^\pm}$ distributions are also presented for comparison. 
Ideally, the truth level distributions for the transverse and longitudinal polarized $W$ are $(1\pm \cos \theta^*_{\ell})^2$ and $\sin^2 \theta^*_{\ell}$, respectively. In practice, those shapes are distorted by the preselection cuts, especially around $ \cos \theta^*_{\ell} \sim \pm 1$. 
We can conclude that the TRAMI network can reproduce the lepton angle distributions well, although its performance of the $W^+_L W^-_T/ W^+_T W^-_L$ processes is slightly worse than that of the $W^+_L W^-_L/ W^+_T W^-_T$ processes, 
but the situation may change for different networks. In TMIPCC, the performance of the $W^+_L W^-_T/ W^+_T W^-_L$ processes is improved, while that of the $W^+_L W^-_L$ process becomes worse. 
We provide the trained networks in the GitHub repository \href{https://github.com/scu-heplab/wlwl-polarization}{\faGithub}. 

\subsection{Templates and fitting procedure}

\begin{figure}[thb]
\includegraphics[width=0.24\textwidth]{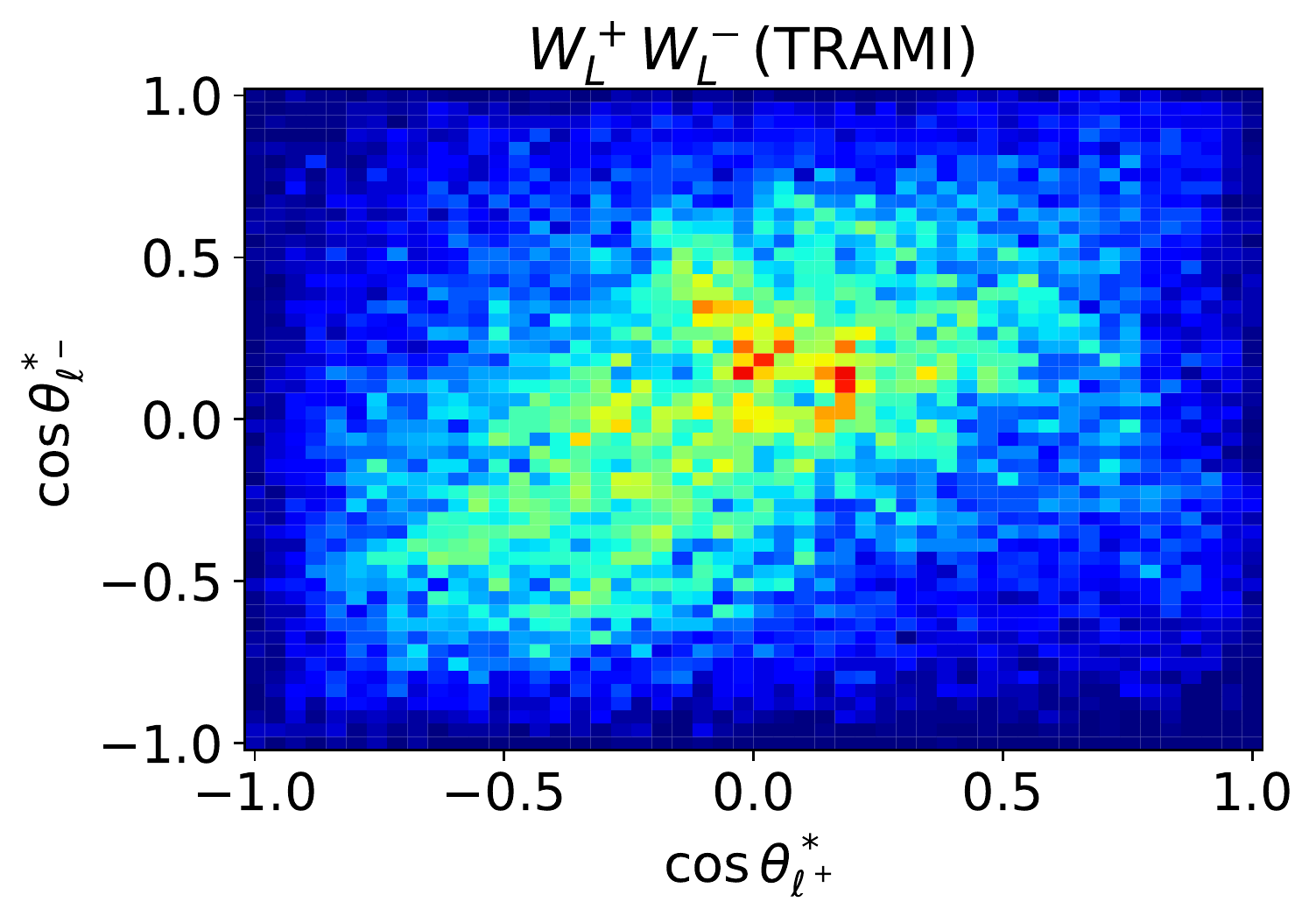}
\includegraphics[width=0.24\textwidth]{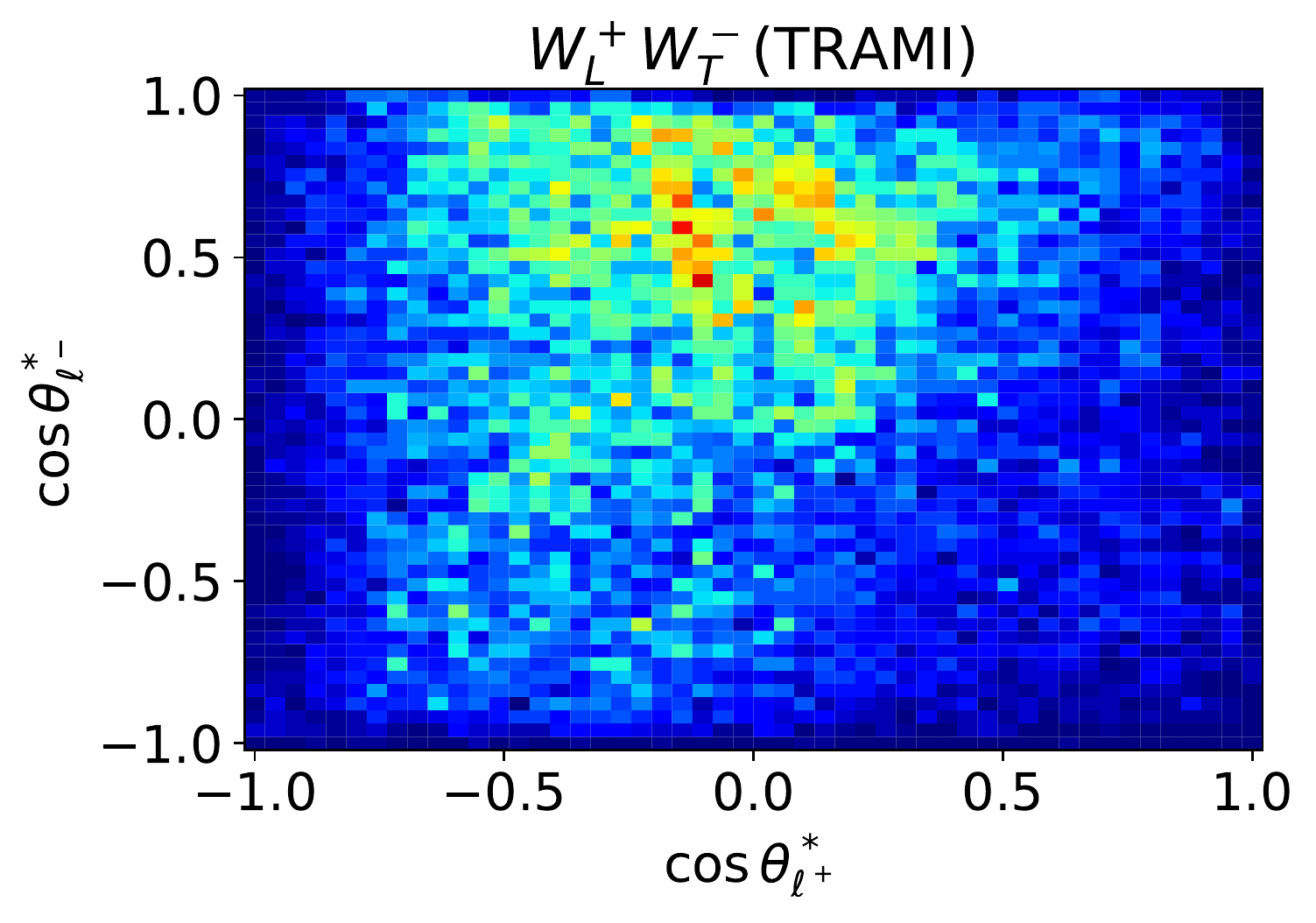}
\includegraphics[width=0.24\textwidth]{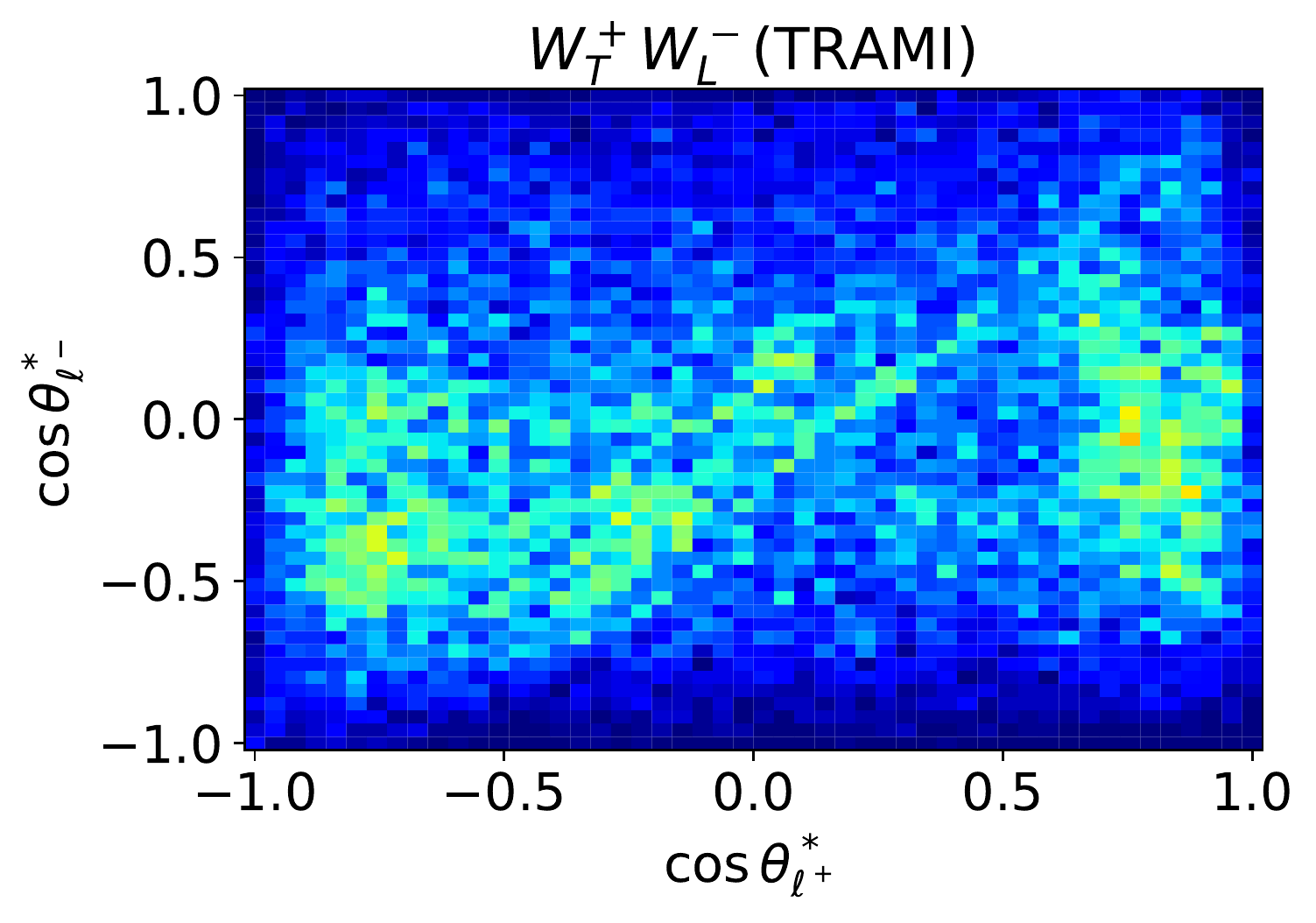}
\includegraphics[width=0.24\textwidth]{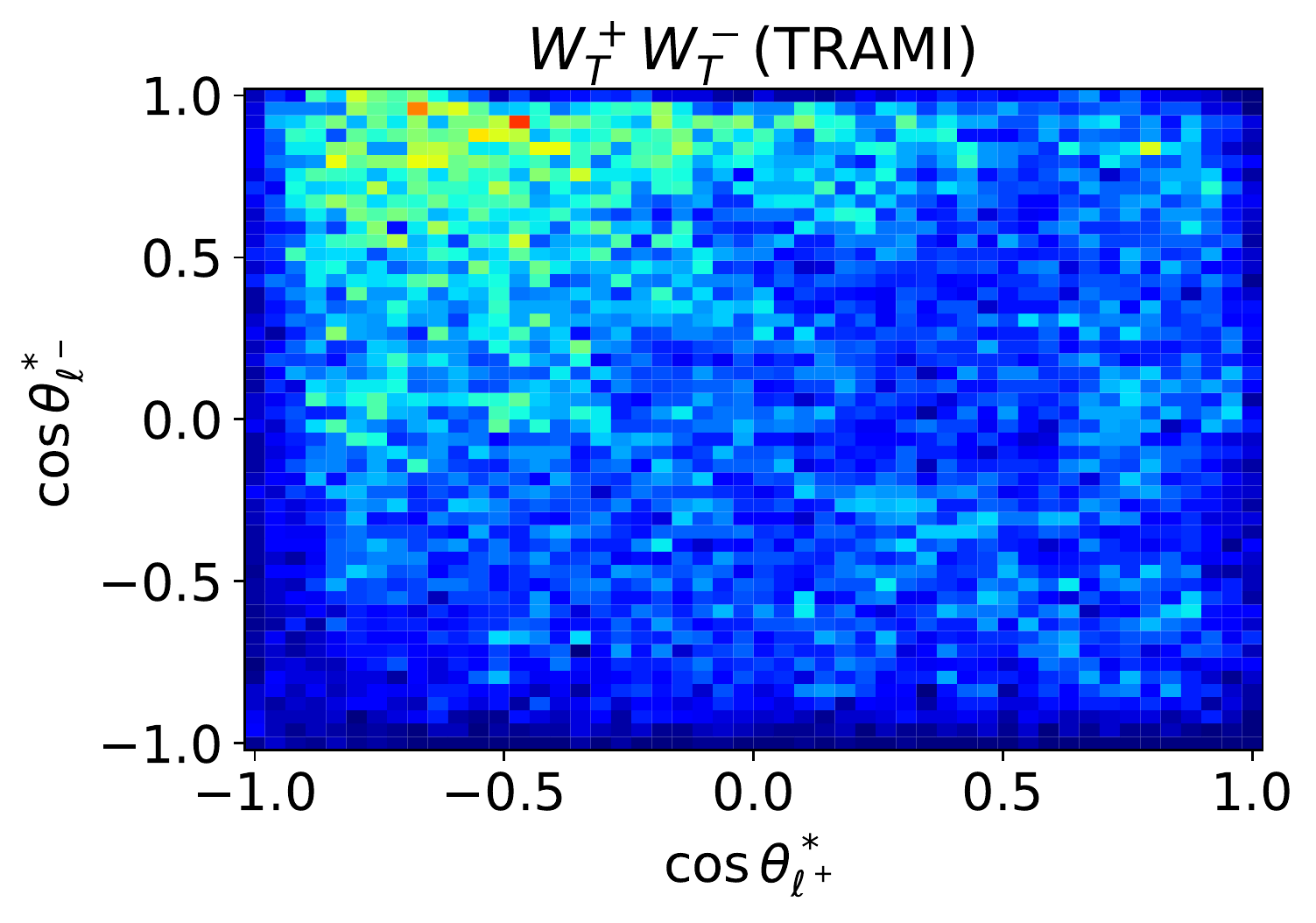}\\
\includegraphics[width=0.24\textwidth]{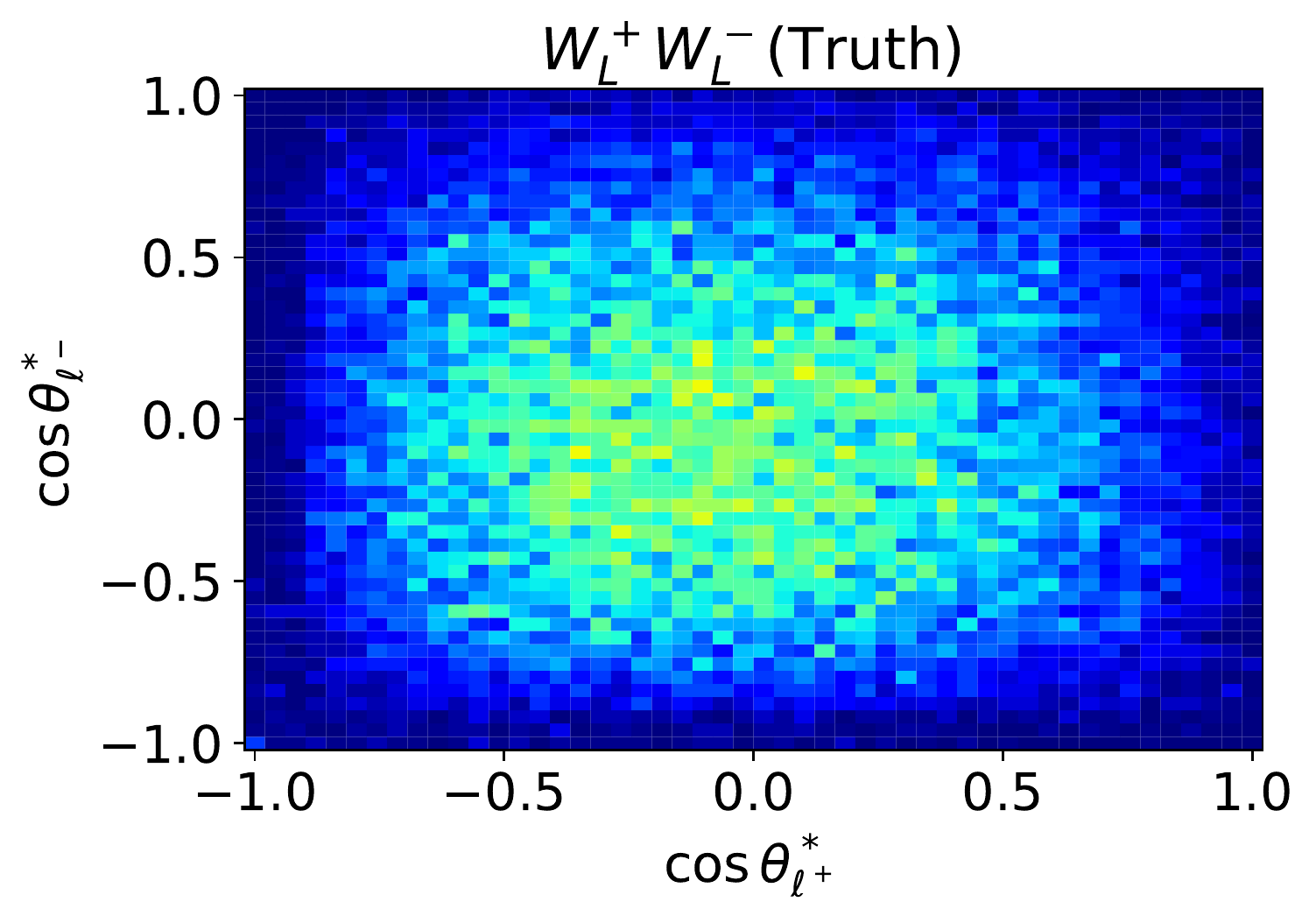}
\includegraphics[width=0.24\textwidth]{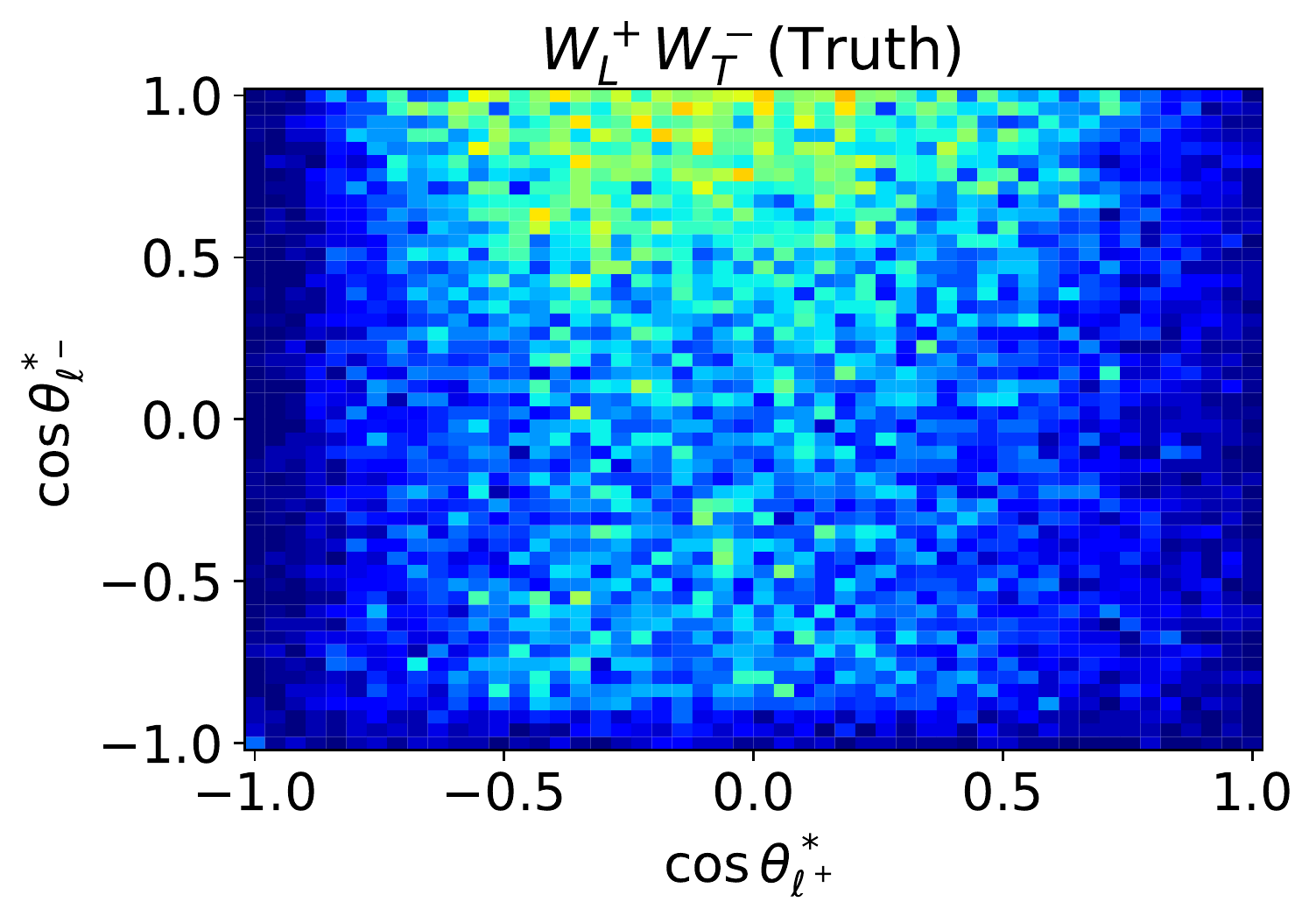}
\includegraphics[width=0.24\textwidth]{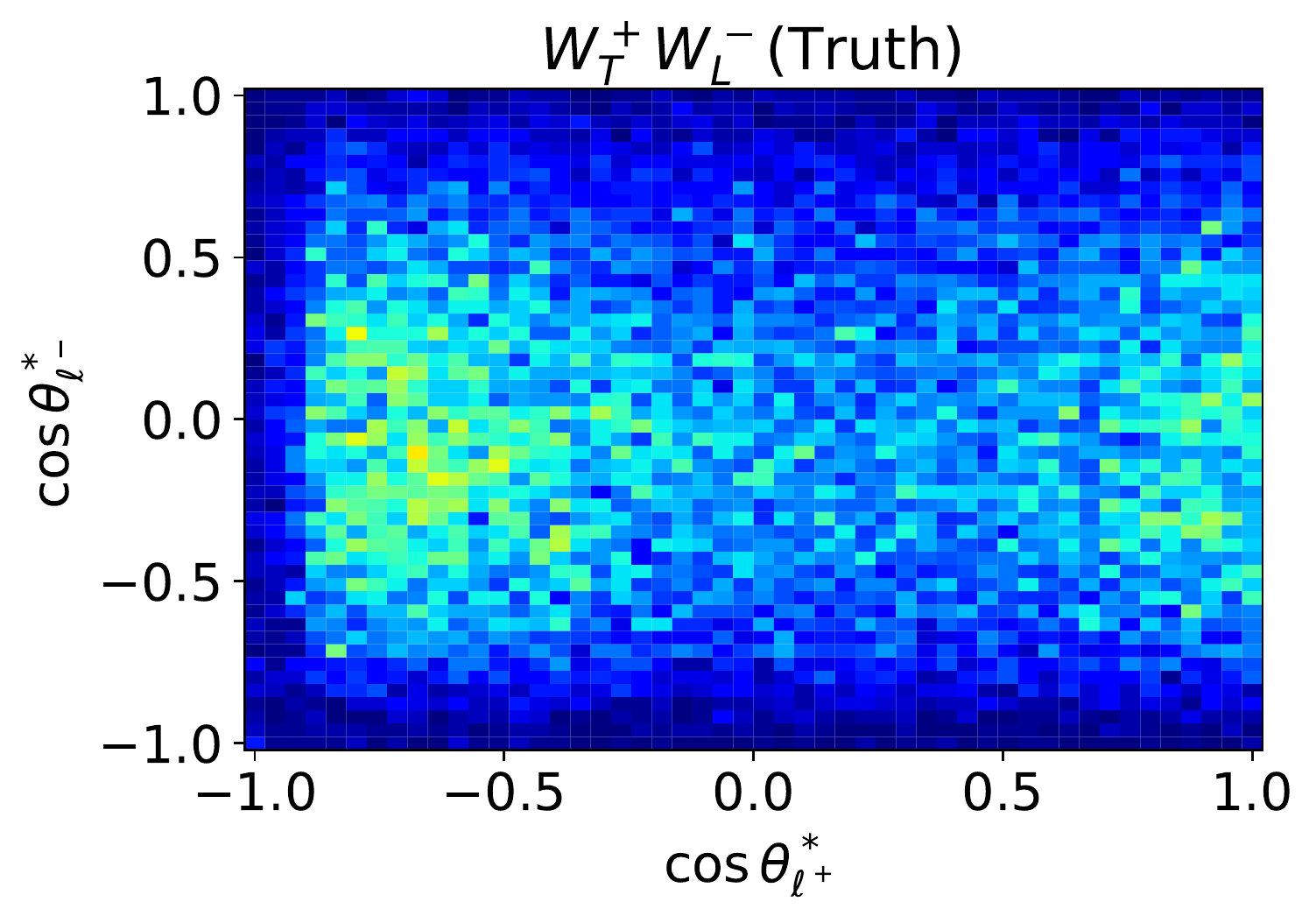}
\includegraphics[width=0.24\textwidth]{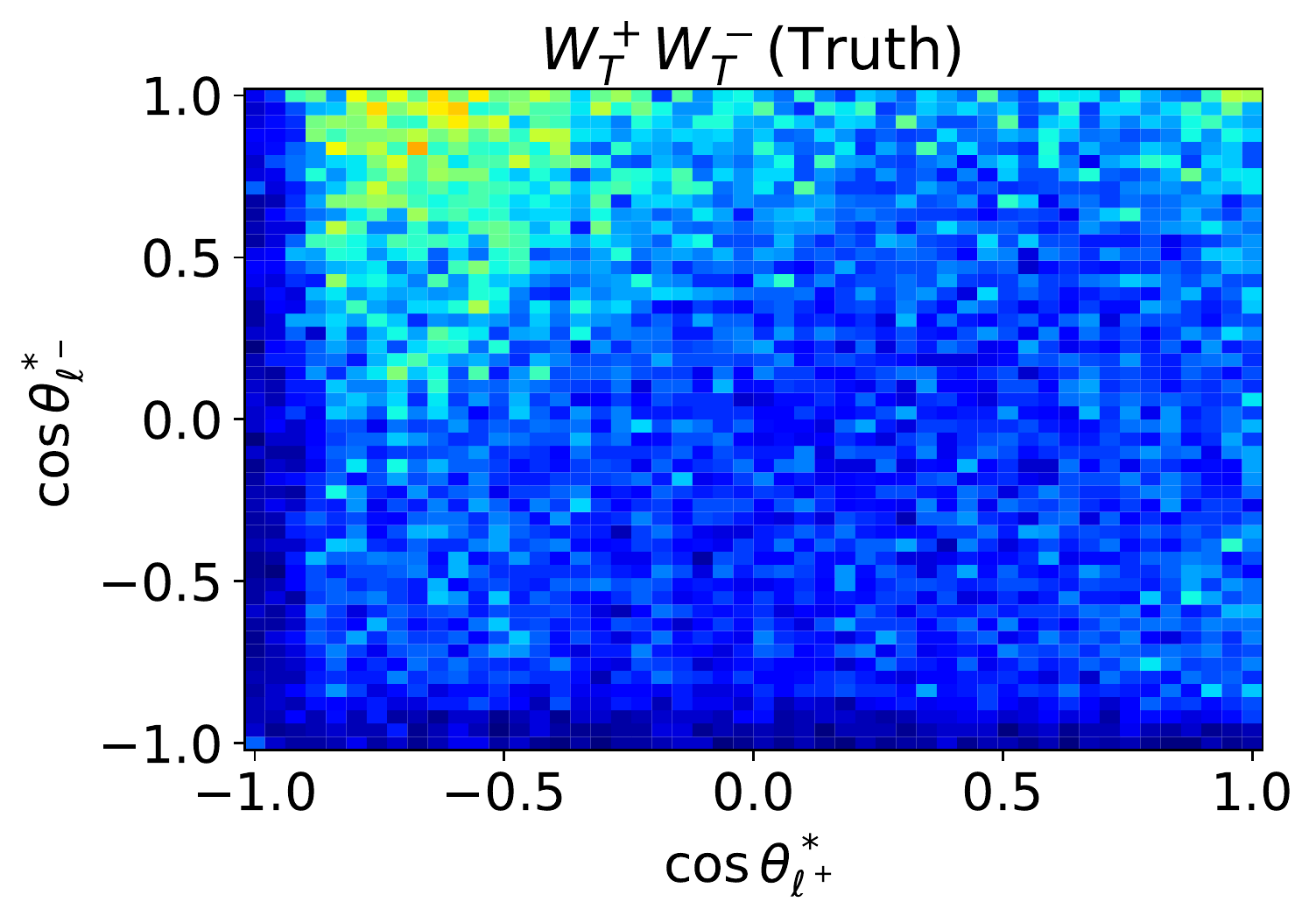}
\caption{The two dimensional distributions of $\cos \theta^*_{\ell^+}$ versus $\cos \theta^*_{\ell^-}$ for different polarization modes of the SM $W^+W^-$ scattering at 13 TeV. Upper panels:  lepton angles $\cos \theta^*_{\ell^\pm}$ calculated by the TRAMI network; Lower panels: truth level lepton angles $\cos \theta^*_{\ell^\pm}$. \label{fig:2dtemp}}
\end{figure}

To obtain the polarization fractions, we need to fit the predicted two-dimensional $\cos \theta^*_{\ell^+} - \cos \theta^*_{\ell^-}$ distribution to the predefined templates for the $W^+_L W^-_L$, $W^+_L W^-_T$, $W^+_T W^-_L$ and  $W^+_T W^-_T$ polarizations respectively. 
The templates are obtained by applying the network to the events of the SM $W^+W^-$ scattering with fixed final state polarization. 
However, due to the presence of preselections, the templates exhibit some dependences on collision energy. We will need to use different sets of templates for different collision energies. 
In Fig.~\ref{fig:2dtemp}, we plot the two-dimensional templates obtained from both the TRAMI network predictions and the truth level lepton angles. 
The network predictions act as nice proxies for the truth lepton angles. 

Having established the template for each polarization state ($T^i$, $i$=LL, TL, LT, TT), given a two-dimensional $\cos \theta^*_{\ell^+} - \cos \theta^*_{\ell^-}$ distribution $O$, we can perform the binned $\chi^2$-fit to estimate the fraction ($f_i$) of each polarization mode.
Because of limited statistics, the two-dimensional lepton angle distribution is divided into $10 \times 10$ bins.
The particle swarm optimization~\cite{2003Particle} is adopted to minimize the $\chi^2$ with constraints: $\sum _i f_i =1$ and $f_i \in [0,1]$.


\section{Test on the different models at 13 TeV} \label{sec:model13}

\subsection{The $W^+ W^-$ polarization in the SM and EFT}

\begin{figure}[thb]
\includegraphics[width=0.3\textwidth]{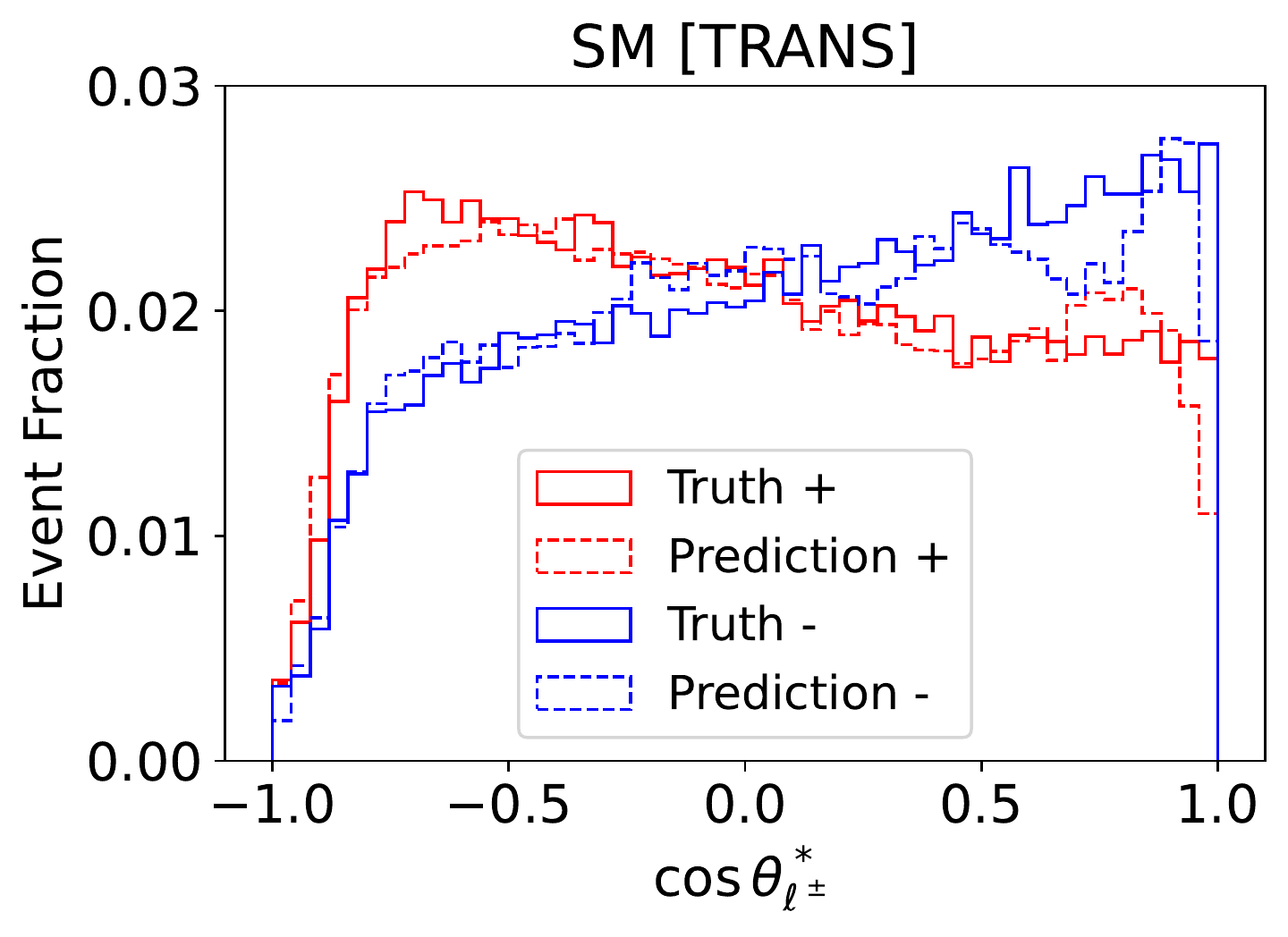}
\includegraphics[width=0.3\textwidth]{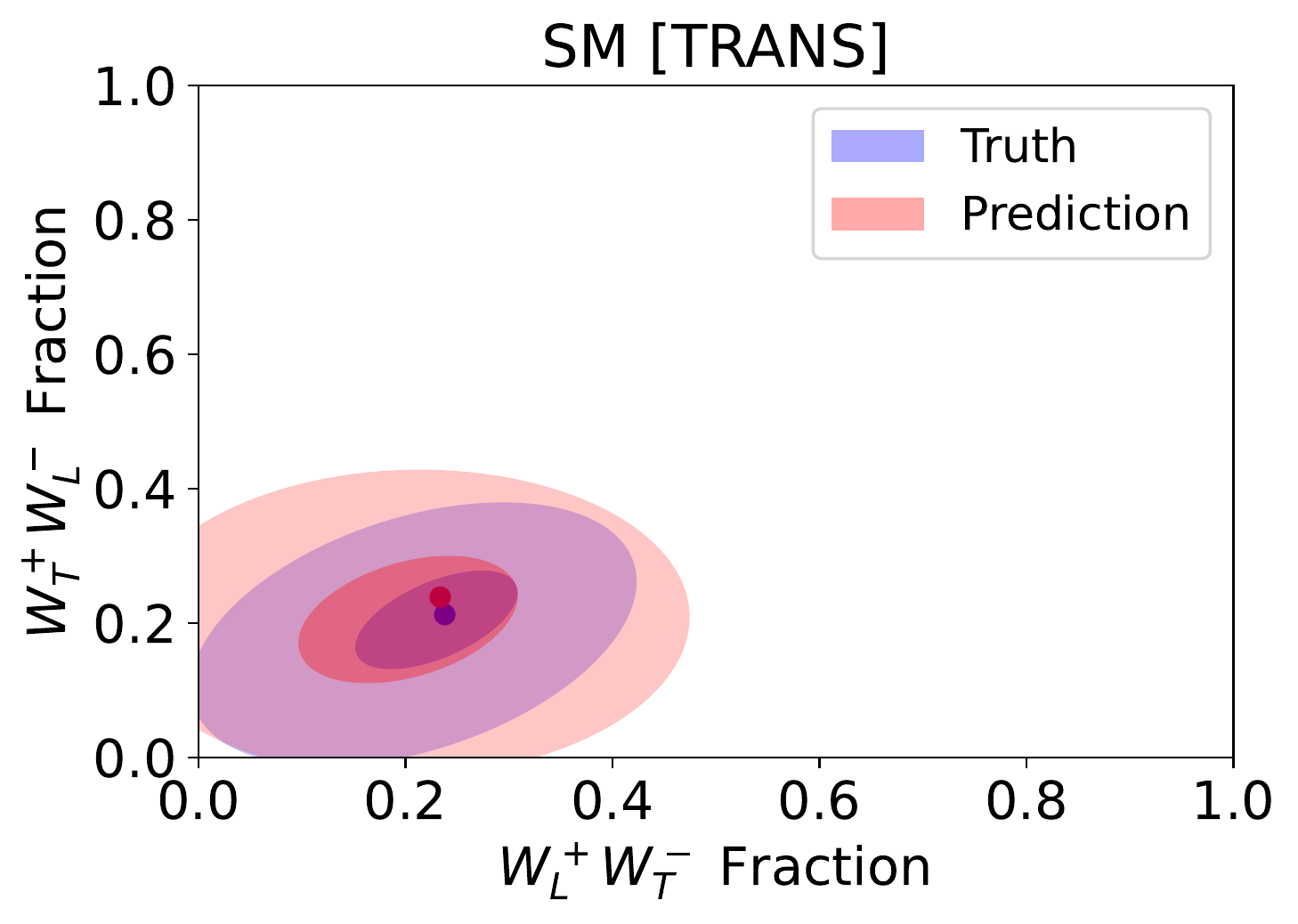}
\includegraphics[width=0.3\textwidth]{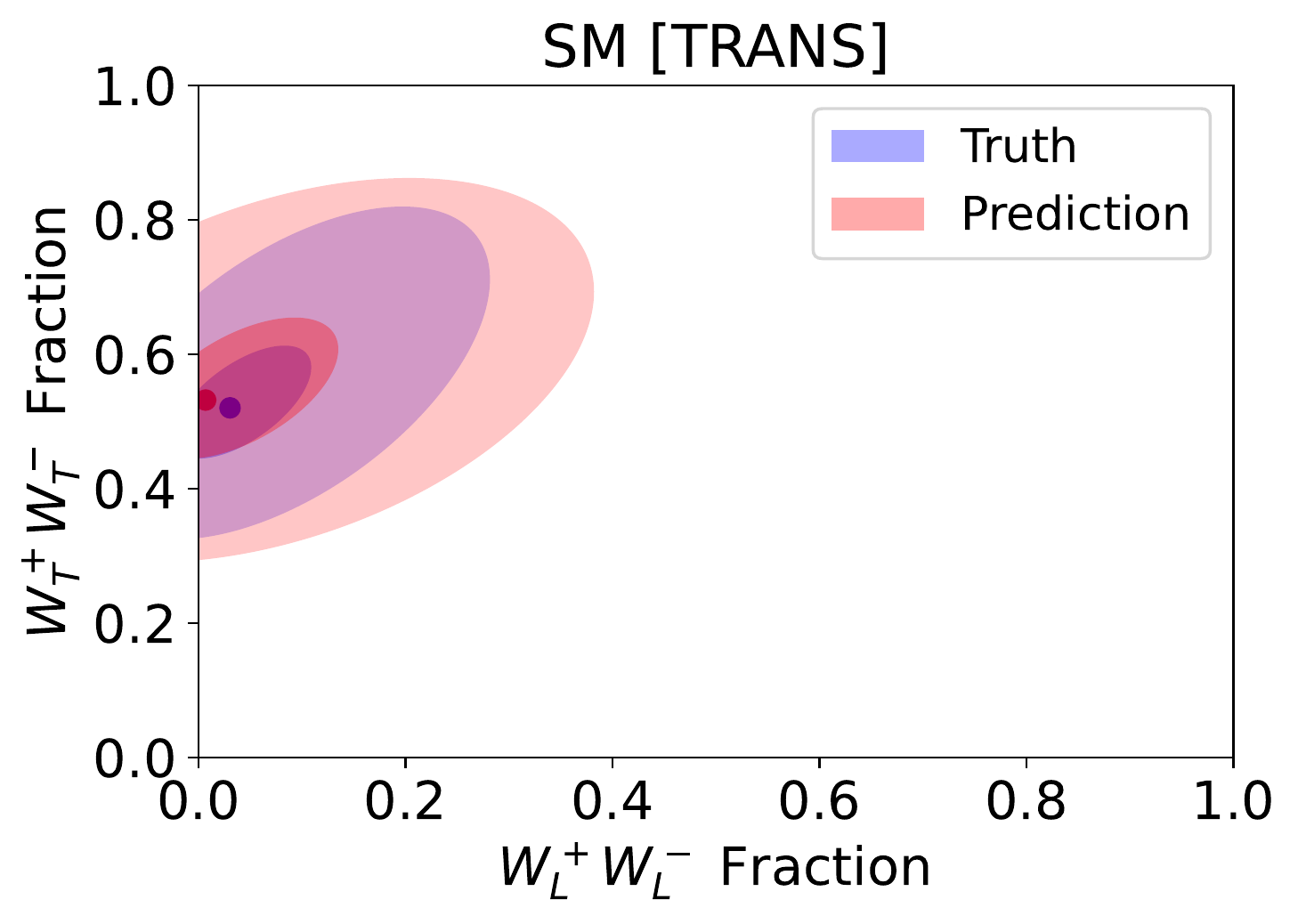}\\
\includegraphics[width=0.3\textwidth]{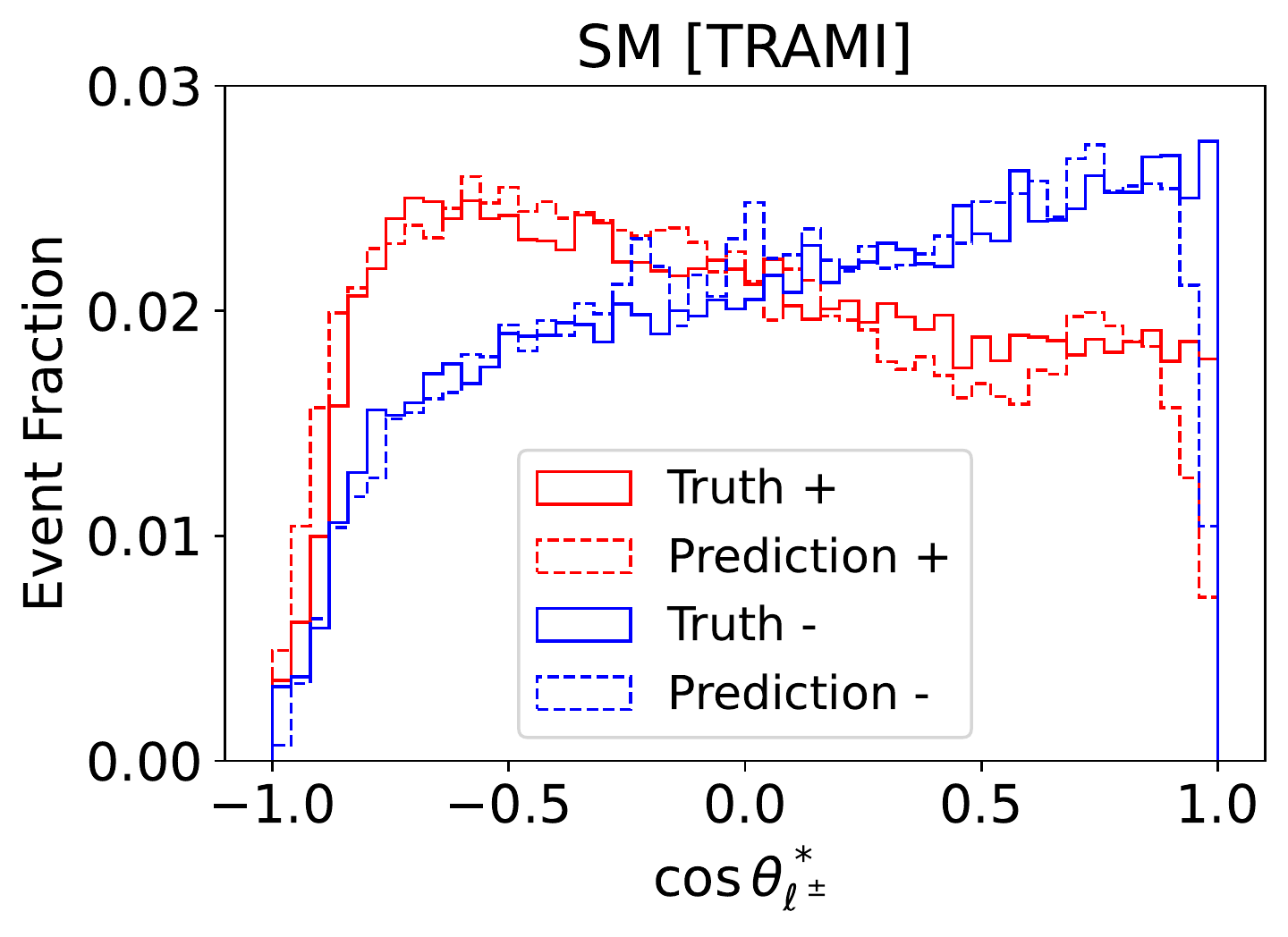}
\includegraphics[width=0.3\textwidth]{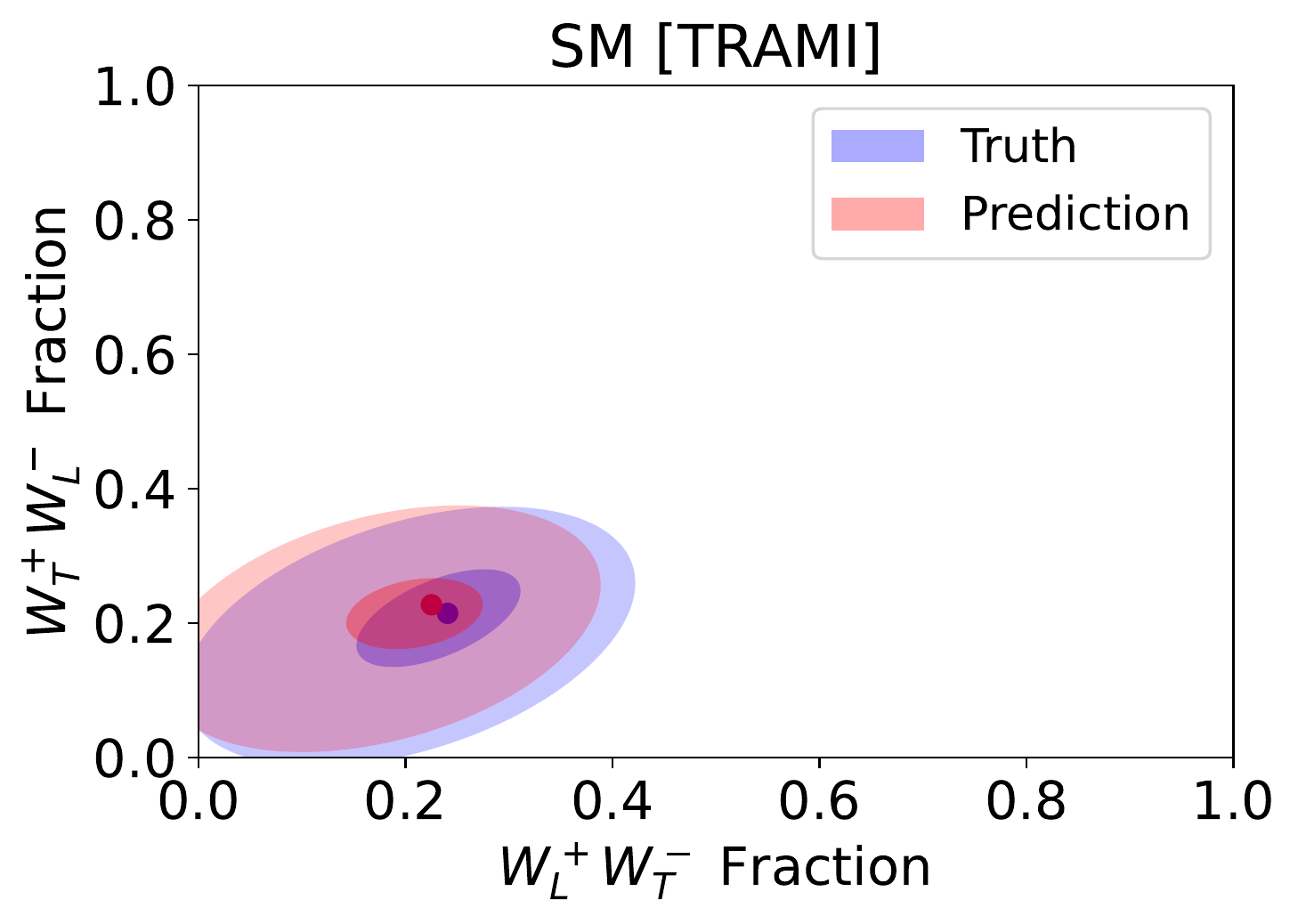}
\includegraphics[width=0.3\textwidth]{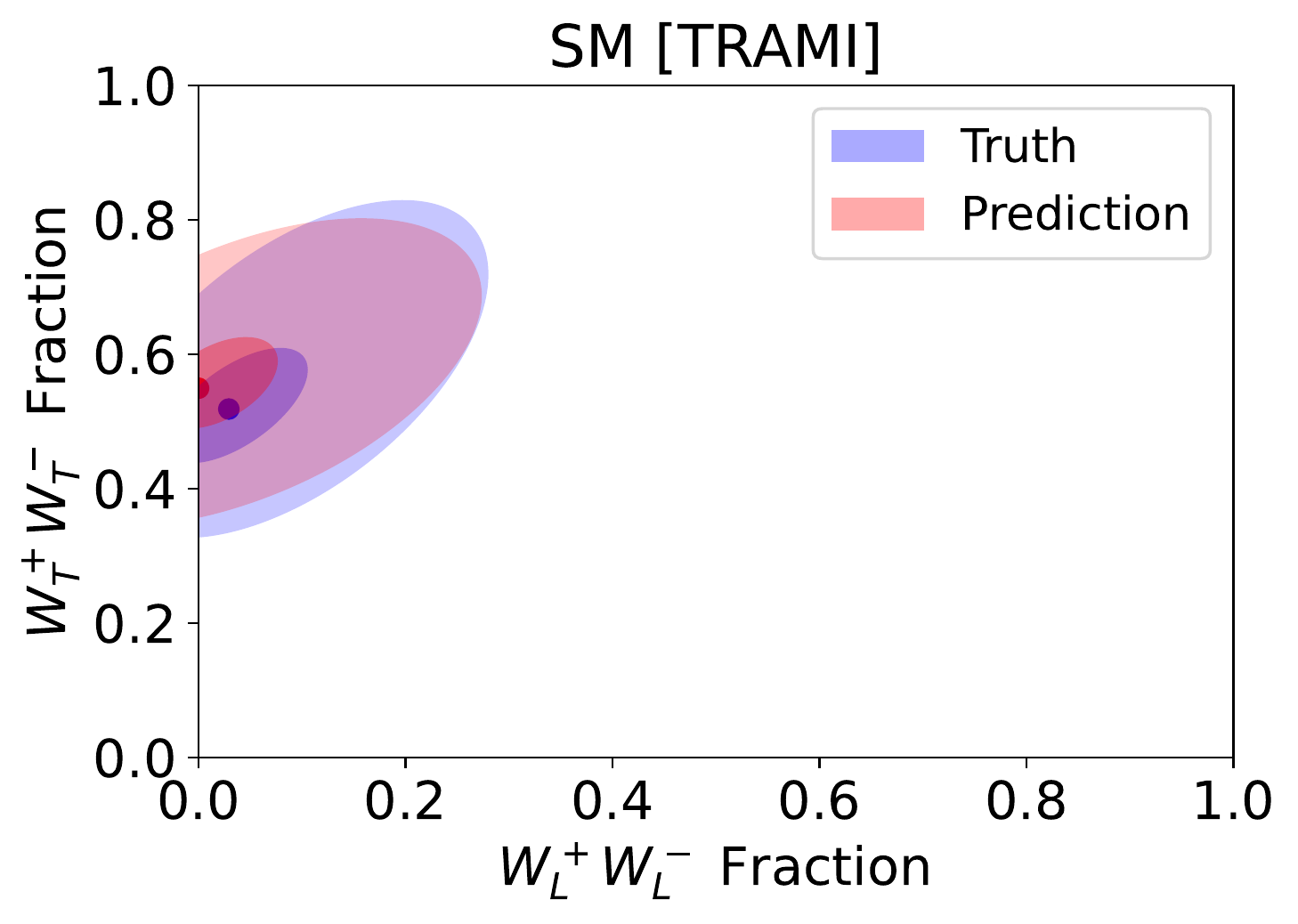}\\ 
\includegraphics[width=0.3\textwidth]{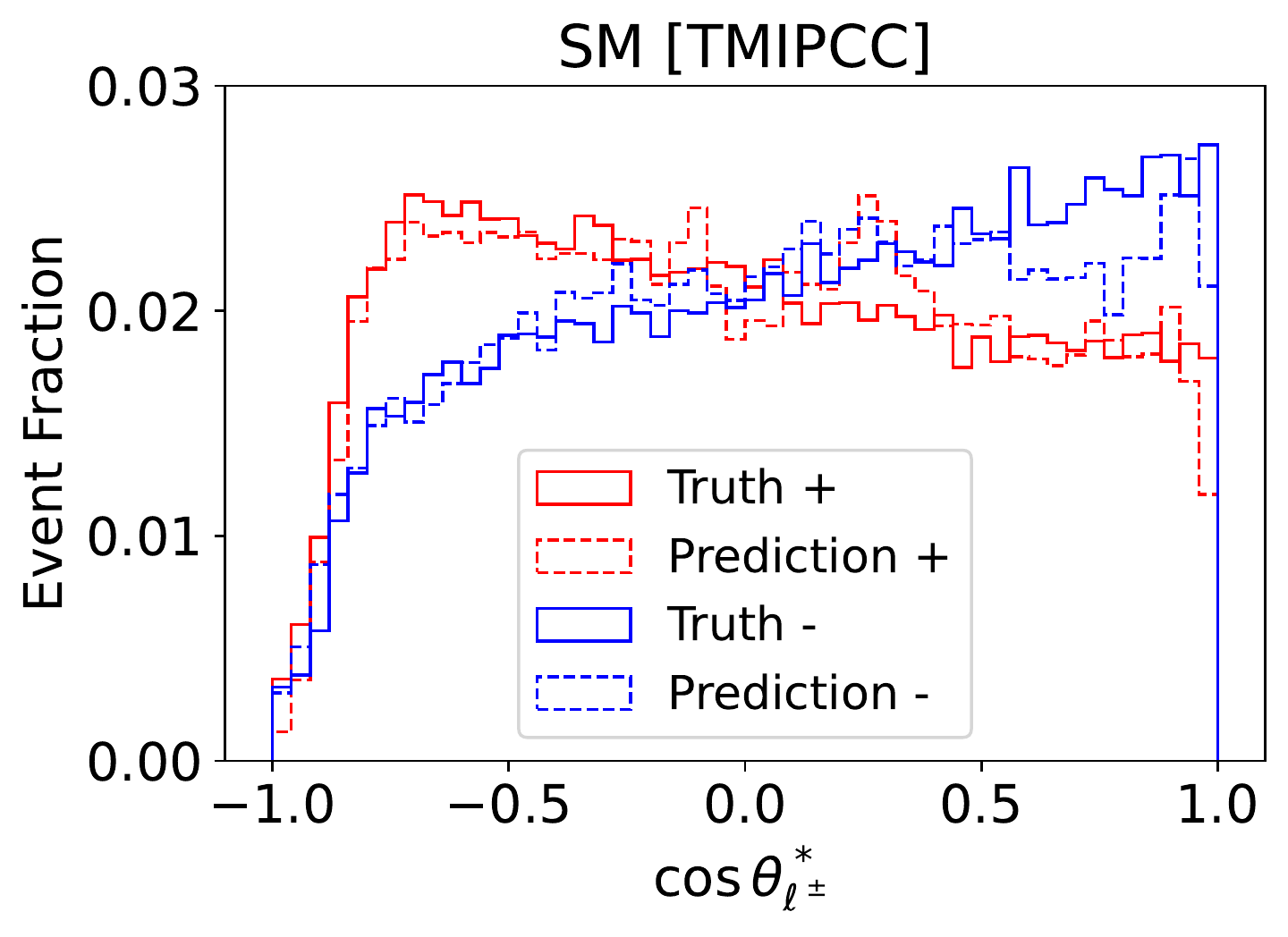}
\includegraphics[width=0.3\textwidth]{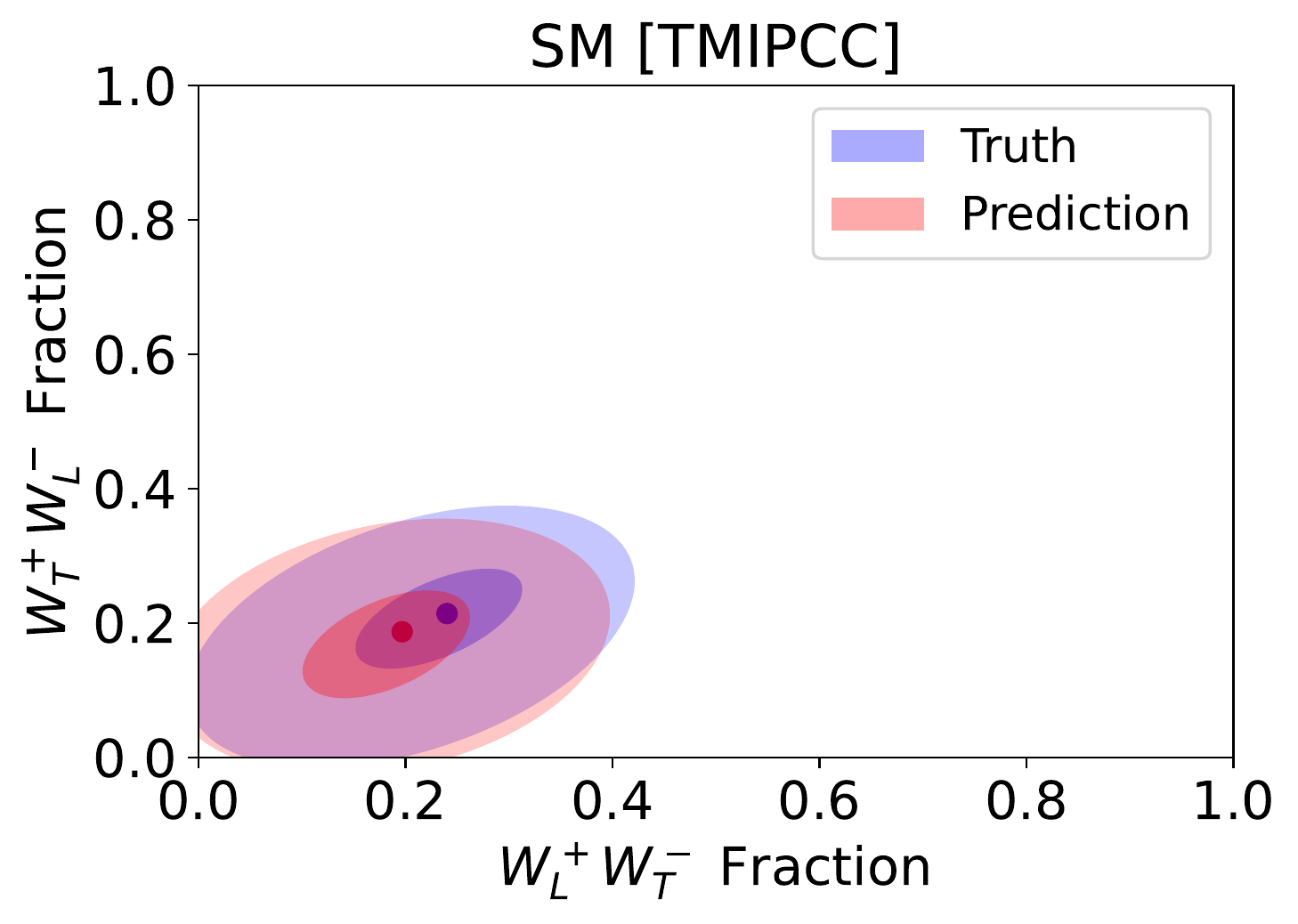}
\includegraphics[width=0.3\textwidth]{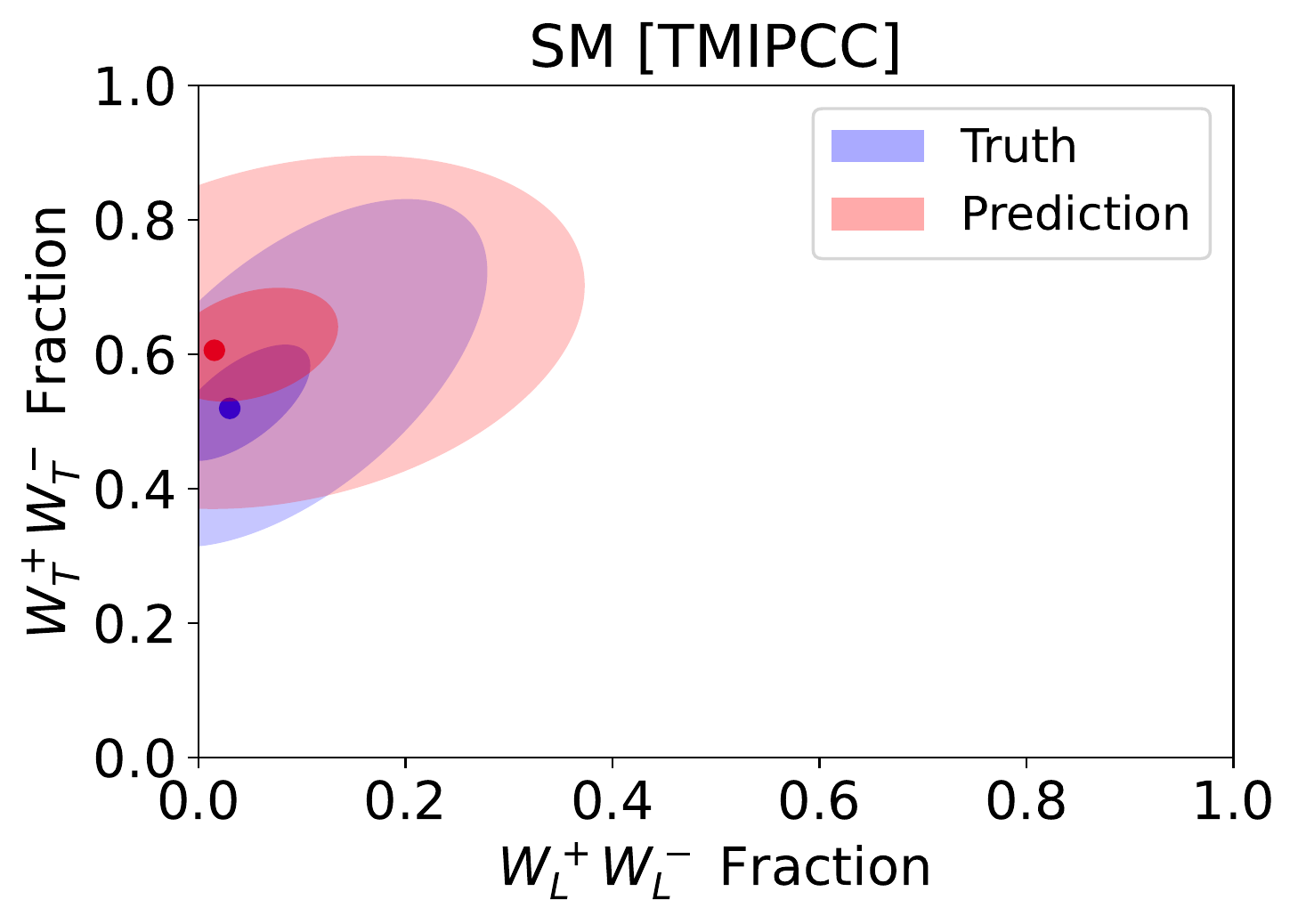}
\caption{The projected lepton angle ($ \cos \theta^*_{\ell^\pm}$) distributions and the $\Delta \chi^2$ contours on the polarization fraction planes for the SM $W^+W^-$ scattering at the 13 TeV LHC. 
Different shades from inside out correspond to $\Delta \chi^2=1$ calculated on datasets with integrated luminosities of 30 and 3 ab$^{-1}$.
Plots from top to bottom are obtained with the TRANS, TRAMI, and TMIPCC networks, respectively. The corresponding results for truth level $ \cos \theta^*_{\ell^\pm}$ are also presented for comparison.
\label{fig:smfit}}
\end{figure}

We first apply our network to the SM $W^+ W^-$ scattering at the 13 TeV LHC. 
The one-dimensional lepton angle distributions and the fitted polarization fractions obtained from three networks are shown in Fig.~\ref{fig:smfit}. 
The left panels show the comparison of the truth level $\cos \theta^*_{\ell^\pm}$ and the network output distributions, where we have projected the two dimensional $\cos \theta^*_{\ell^+} - \cos \theta^*_{\ell^-}$ distributions into each component for visibility. 
In the middle and right panels, the $\Delta \chi^2=1$ contours on the $f_{LT}-f_{TL}$ plane and $f_{LL}-f_{TT}$ plane for integrated luminosities 30 ab$^{-1}$ (may not realistic) and 3 ab$^{-1}$ are shown. 
We can find that the networks reproduce the distributions of the truth level lepton angle well\footnote{Note that they are not so accurate in reproducing the lepton angle for a single event as discussed in Sec.~\ref{sec::network}. For example, the RMSE for the $\cos \theta^*_{\ell^+}$ ($\cos \theta^*_{\ell^-}$) of TRANS, TRAMI, and TMIPCC are 0.548 (0.546), 0.474 (0.470), and 0.515 (0.483), respectively.}.
In particular, the TRAMI network, which makes the features focus on the $\cos \theta^*_{\ell^\pm}$ and decorrelate with the momentum of the $W$ boson pair, has almost the same reconstruction precisions as the truth $\cos \theta^*_{\ell^\pm}$; {\it i.e.}, the sizes of the contours are similar. 
As for the TMIPCC network, while the precision of the $f_{LT}$ and $f_{TL}$ fractions is similar to that obtained with the truth $\cos \theta^*_{\ell^\pm}$, both the $f_{LL}$ and $f_{TT}$ precision is worse, partly because of the difficulty in training the network with a more complex loss function. 
Overall, giving the cross section (after preselection cuts) of SM $W^+W^-$ scattering as 4.36 fb, each fraction of polarization can be resolved with deviation of $\sim 0.2$ at the LHC for an integrated luminosity of 3 ab$^{-1}$. 

\begin{figure}[thb]
\includegraphics[width=0.3\textwidth]{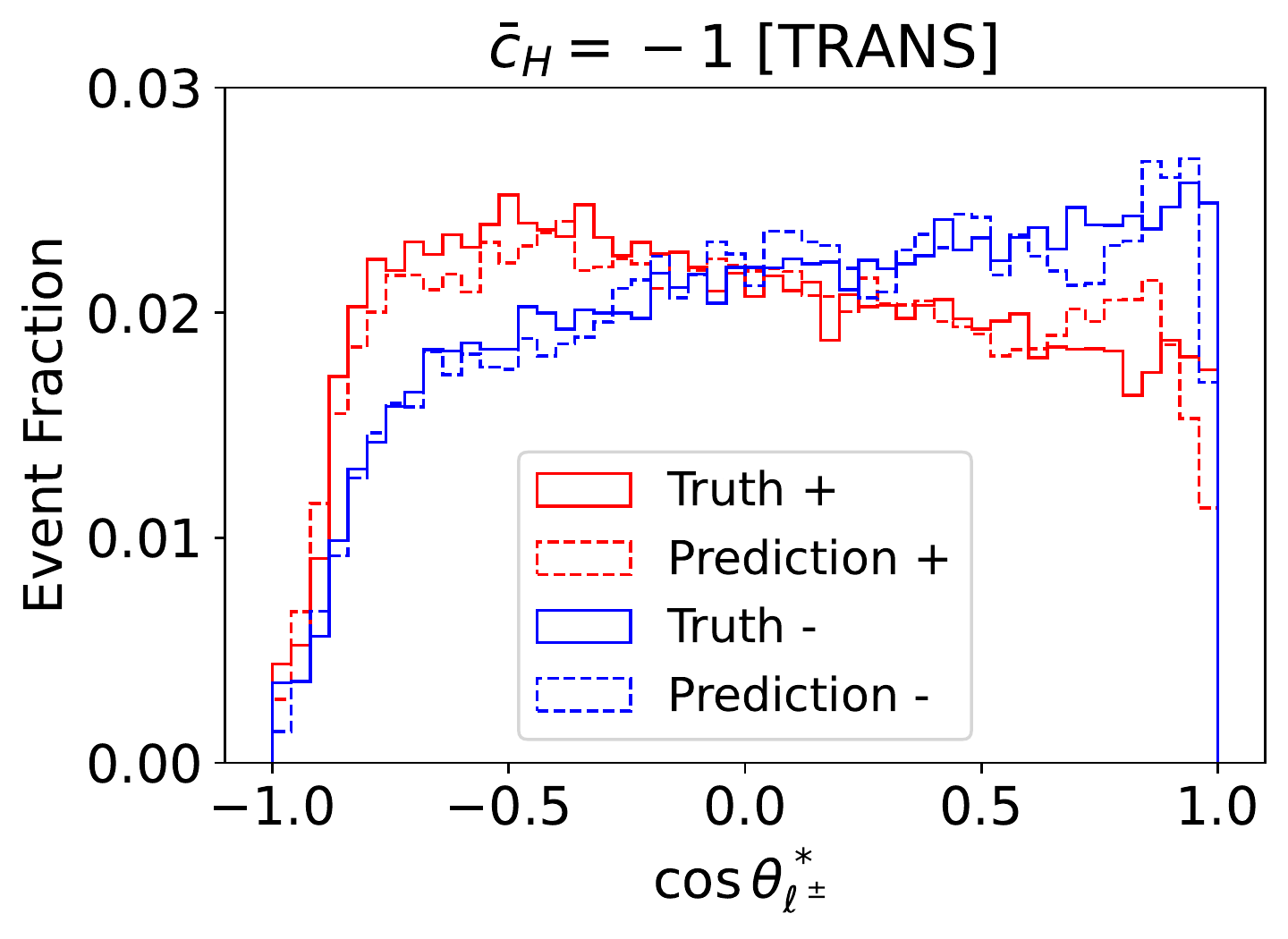}
\includegraphics[width=0.3\textwidth]{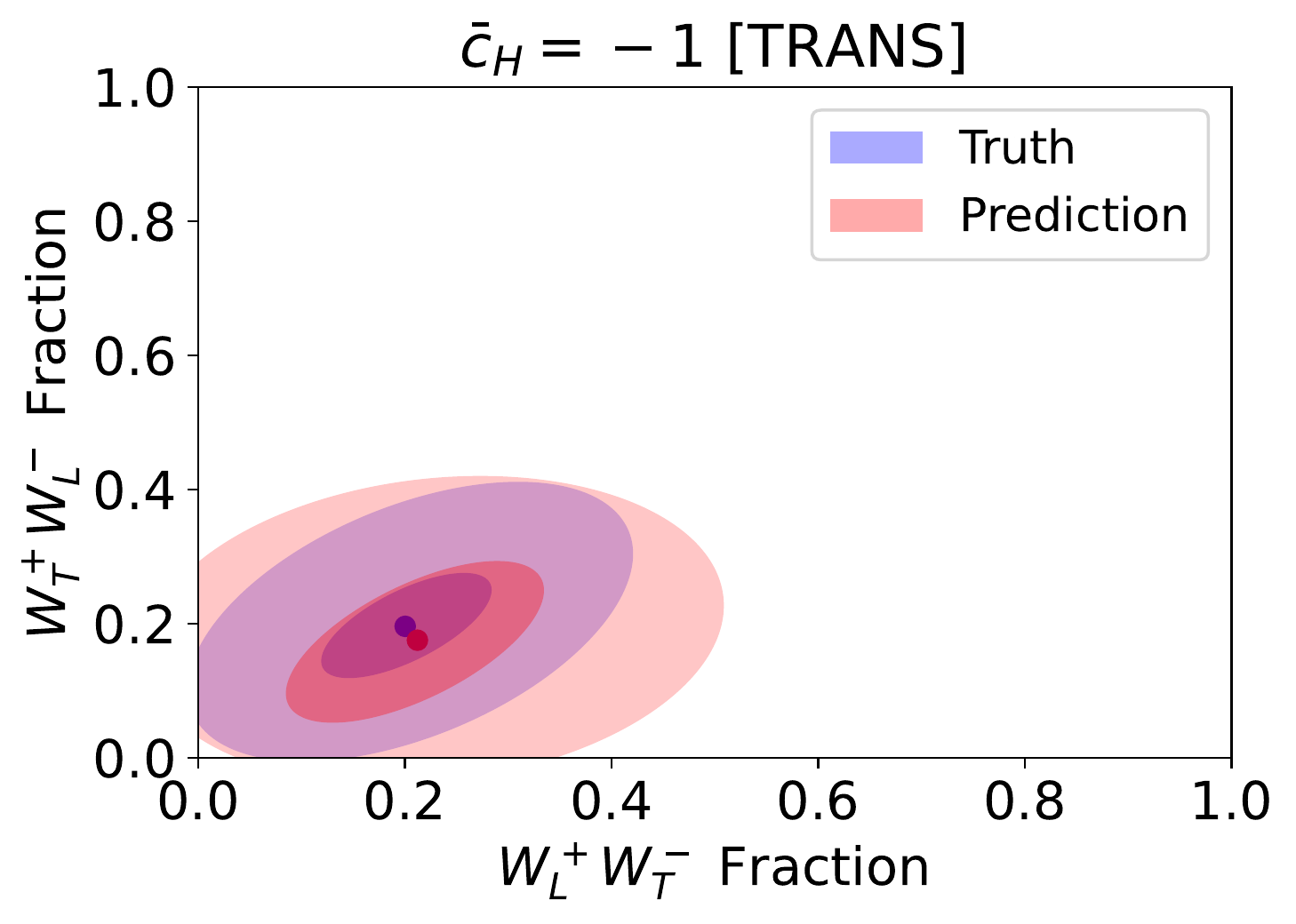}
\includegraphics[width=0.3\textwidth]{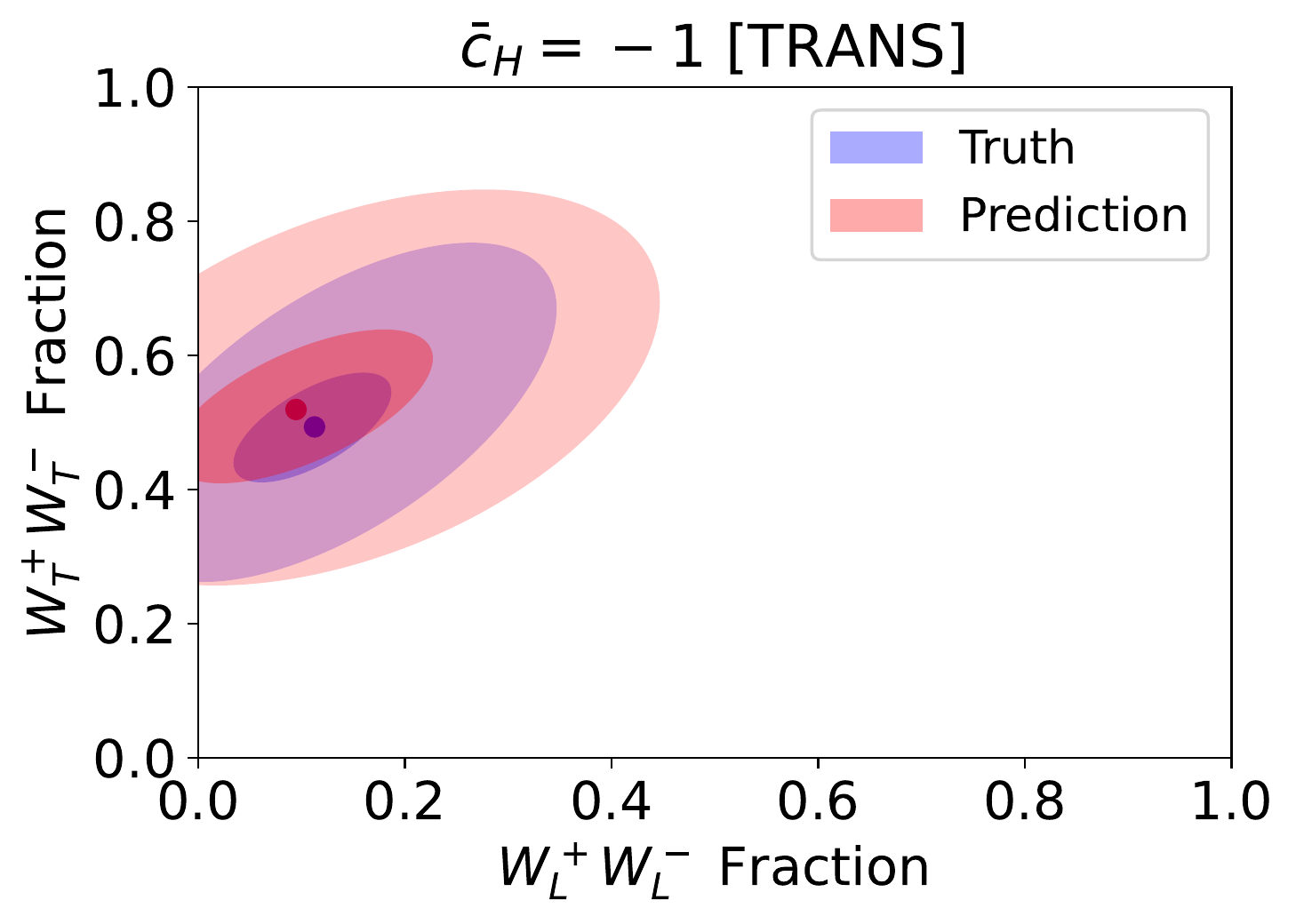}\\
\includegraphics[width=0.3\textwidth]{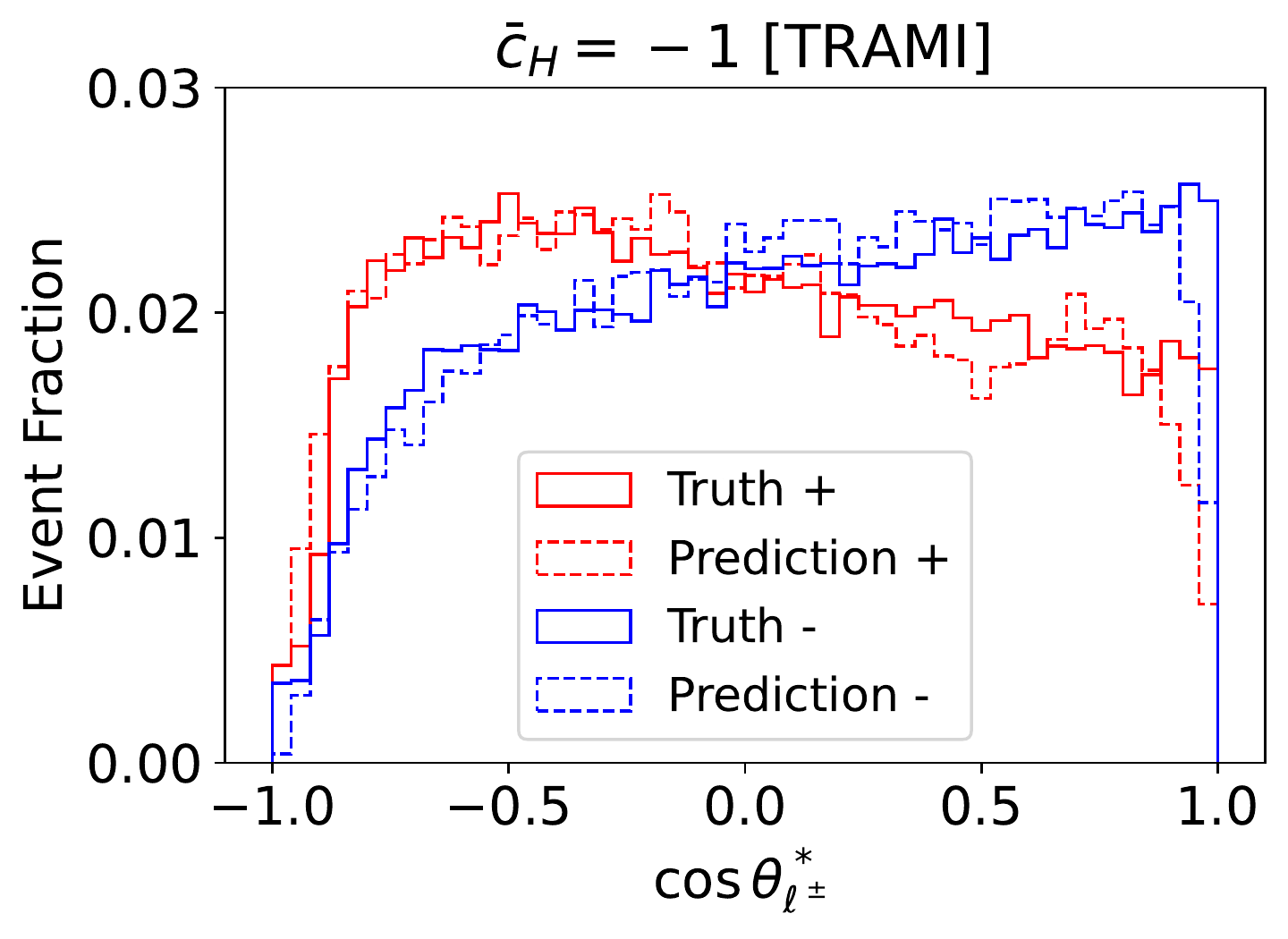}
\includegraphics[width=0.3\textwidth]{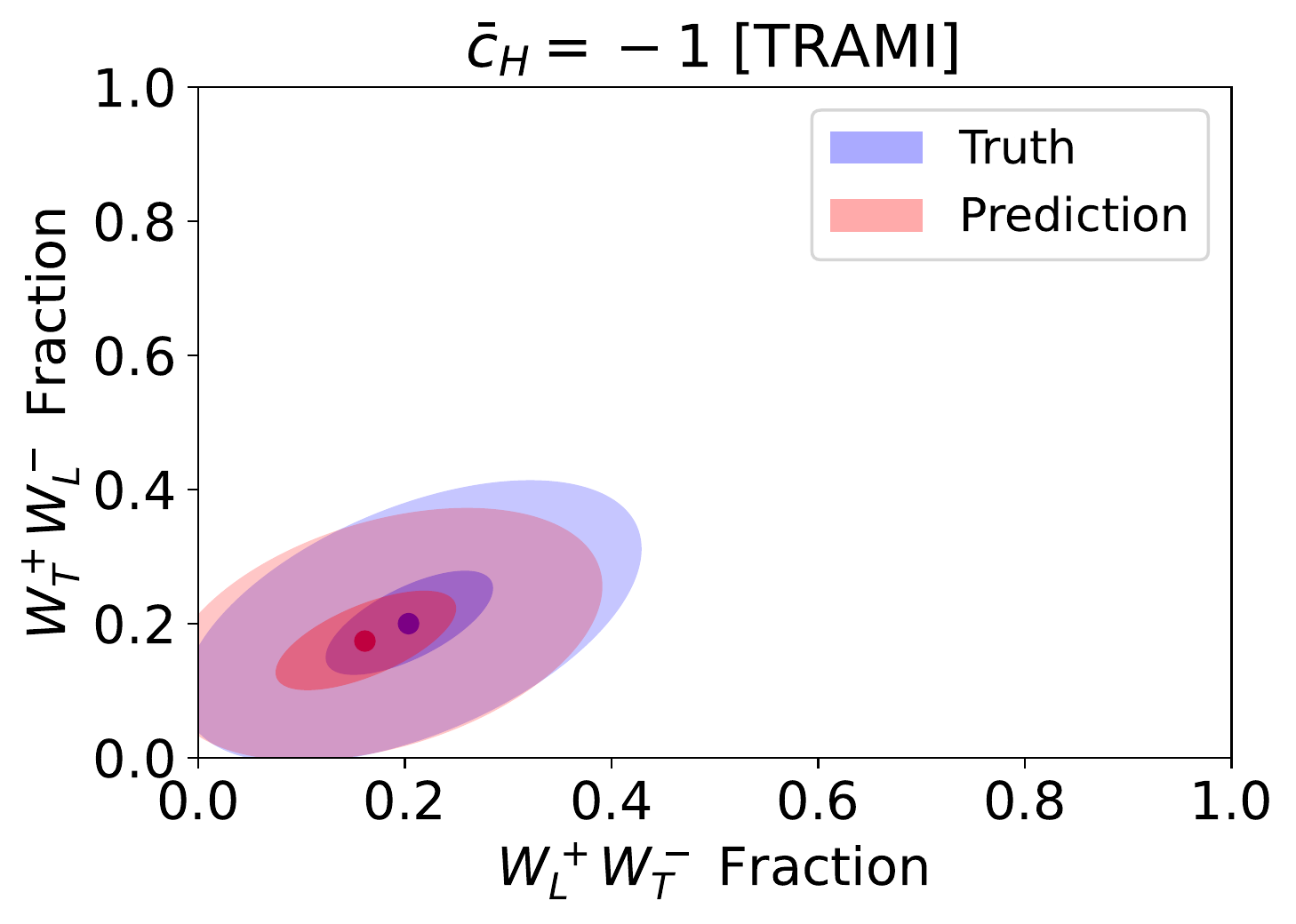}
\includegraphics[width=0.3\textwidth]{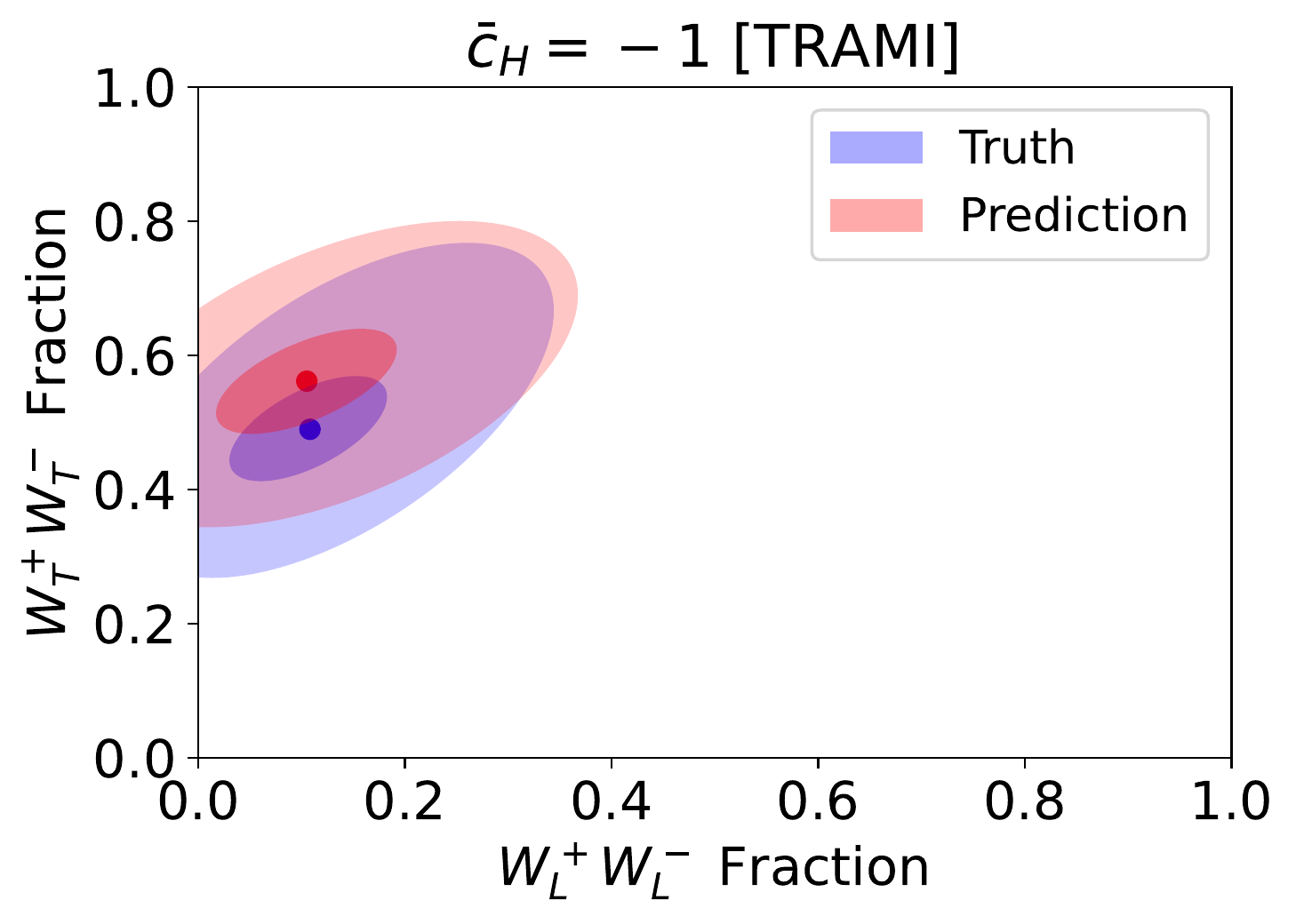}
\includegraphics[width=0.3\textwidth]{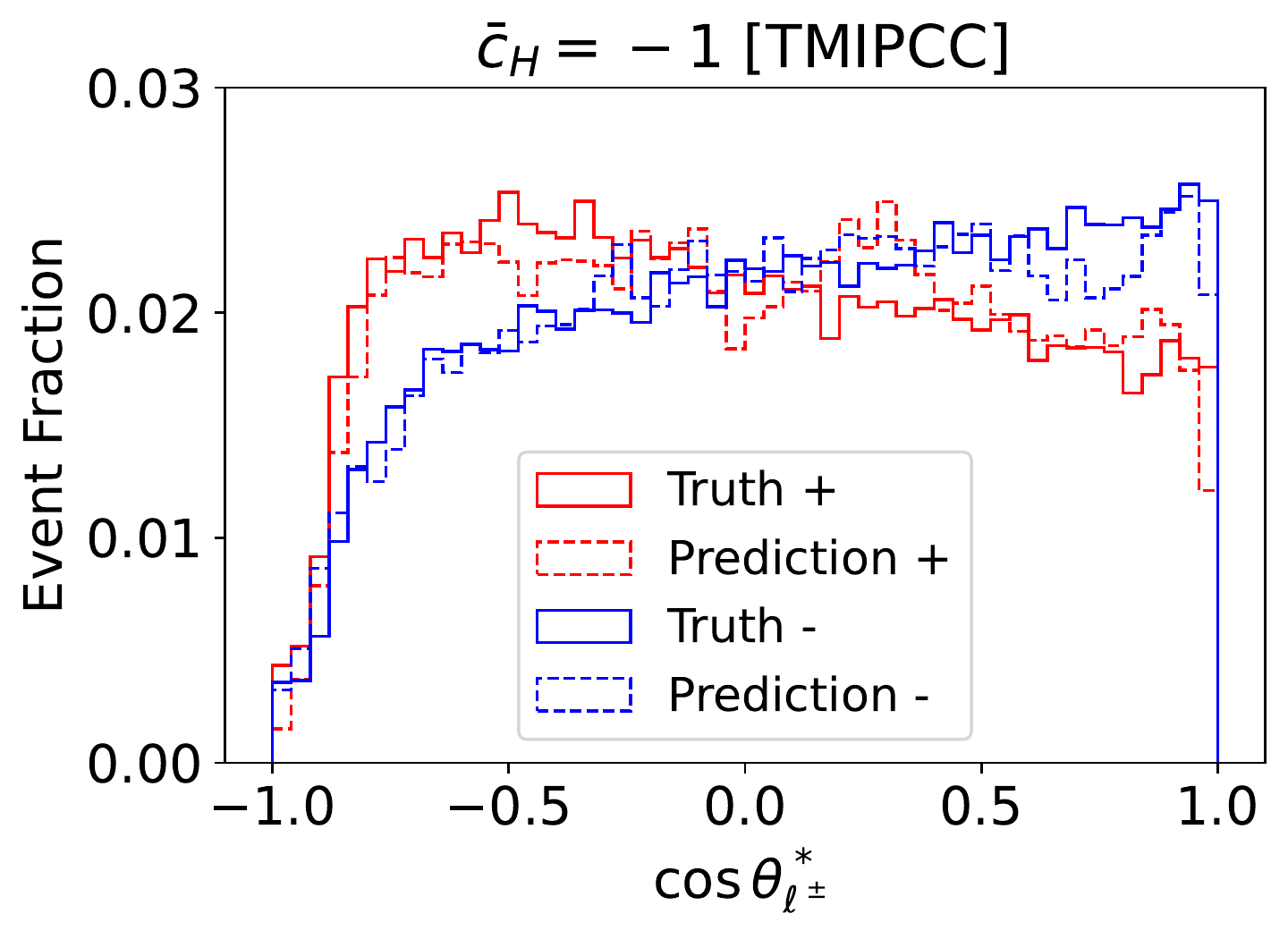}
\includegraphics[width=0.3\textwidth]{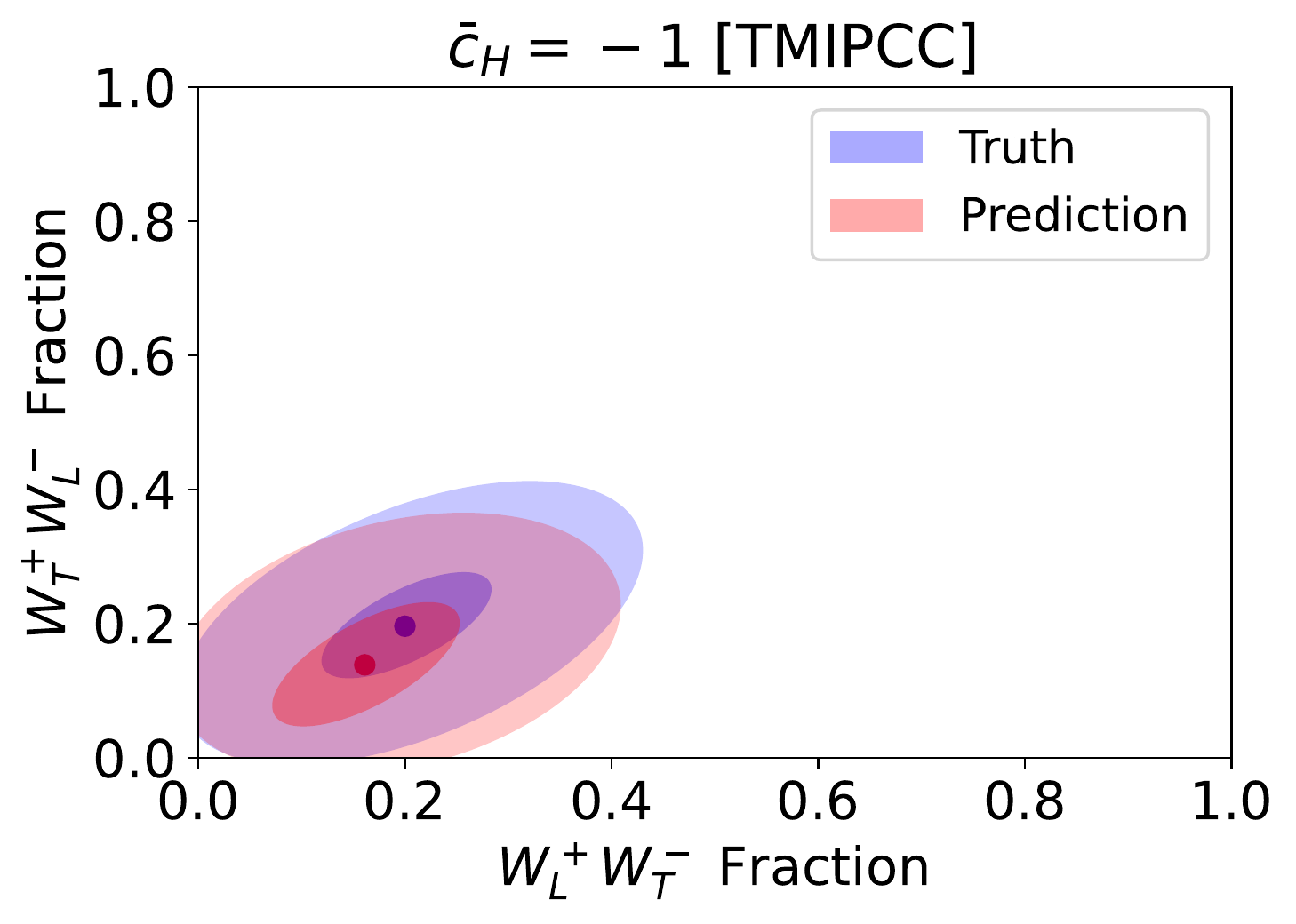}
\includegraphics[width=0.3\textwidth]{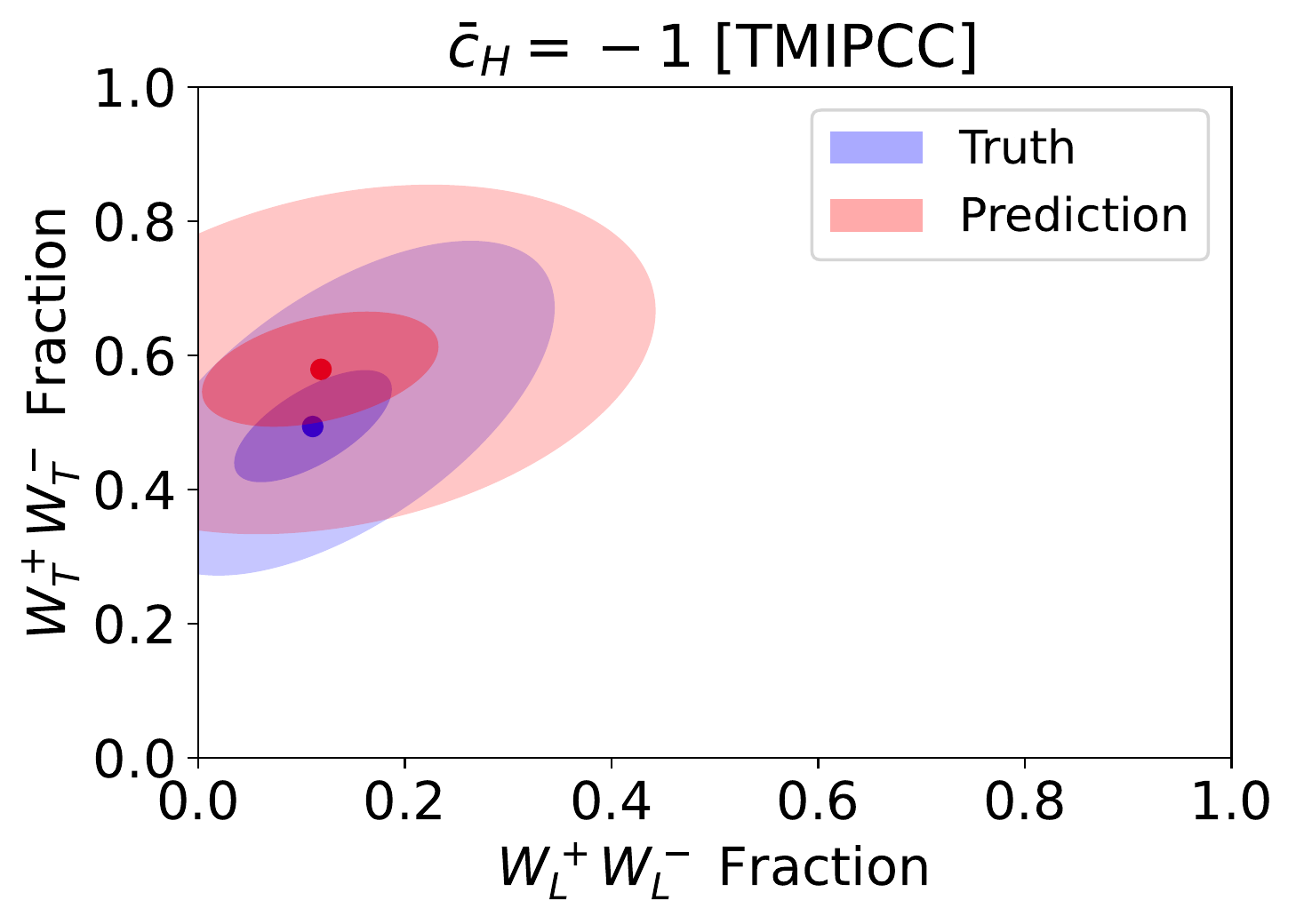}
\caption{For the $W^+W^-$ scattering in EFT with $\bar{c}_H=-1$ at the 13 TeV LHC. Meanings of the plots are the same as Fig.~\ref{fig:smfit}.  \label{fig:chm1fit}}
\end{figure}

The $W^+W^-$ scattering could be affected by any new physics that is related to the electroweak symmetry breaking of the SM. 
A general framework to describe the new physics effects is the effective field theory (EFT). To study the network performances in new physics models, following the strategy as discussed in Ref.~\cite{Li:2020fna}, we consider the following operator~\cite{Giudice:2007fh,Contino:2013kra}:
\begin{align}
\mathcal{O}_H = \frac{\bar{c}_H}{2 v^2} \partial^\mu [\Phi^\dagger \Phi]  \partial_\mu [\Phi^\dagger \Phi] \Rightarrow  \frac{\bar{c}_H}{2} \partial^\mu h \partial_\mu h
\end{align}
where the $\Phi$ field is the Higgs doublet and $h$ denotes the SM Higgs boson field. 
This operator leads to the following changes to the Higgs couplings:
\begin{align}
\mathcal{L}_{H}  \supset & \frac{g m_W}{c^2_W} [1- \frac{1}{2} \bar{c}_H]  Z_\mu Z^\mu h  + g m_W [1-\frac{1}{2}\bar{c}_H ] W^\dagger_\mu W^\mu h  \nonumber \\
 & + [ \frac{y_f}{\sqrt{2}} [1-\frac{1}{2}\bar{c}_H] \bar{f} P_R f h +h.c.]~. \label{eq:hcouplings}
\end{align}
Although the updated global fit requires $\bar{c}_H \lesssim 0.4$~\cite{Dawson:2020oco}, we apply our network to the case with $\bar{c}_H = -1$ to illustrate the network performance when the new physics contributions are sizable. The results are given in Fig.~\ref{fig:chm1fit}. Compared to the SM case, the fraction of the longitudinal $W$ boson is greatly enhanced due to the incomplete cancellation in $W_L W_L \to W_L W_L$ scattering. 
However, the overall kinematic properties and the total production cross section (which is 4.82 fb after preselection for $\bar{c}_H = -1$) of the $W^+W^-$ scattering in the EFT with non-zero $\bar{c}_H$ are similar to the SM ones, so the performances of the three networks on the EFT are similar to that on the SM as shown in Fig.~\ref{fig:smfit}. 
The one-dimensional lepton angle distributions predicted by all three networks match the truth lepton angle distributions well. 
Among the three networks, the TRAMI performs the best---the discrimination power of which is quite close to the truth $ \cos \theta^*_{\ell^\pm}$. 
On the other hand, in the TMIPCC network, only the precision of $f_{LT}$ and $f_{TL}$ is comparable to those obtained with the truth $ \cos \theta^*_{\ell^\pm}$.

\subsection{The $W^+ W^-$ polarization in the 2HDM and from resonant production}

Many new physics models predict light states mediating the $W^+W^-$ scattering, the effects of which can not be fully described in the EFT. 
We consider the type-II 2HDM~\cite{Aoki:2009ha,Branco:2011iw} as a benchmark model, as it is featured by both the Higgs coupling modification and the existence of another scalar mediator (besides the SM Higgs) in $W^+W^-$ scattering. 
There are six parameters: masses of scalar bosons ($m_{H_1}, m_{H_2}$, $m_A$, and $m_{H^\pm}$), the mixing angle between two $CP$-even scalars $\alpha$, and the ratio between two vacuum expectation values $\tan \beta$.
The $m_{H_1}$ has been measured to be around 125 GeV (we only consider the case of $H_1$ being the SM-like Higgs boson). The $m_A$ and $m_{H^\pm}$ are not relevant in the $W^+ W^- j j$ production, assuming they are much larger than $m_{H_2}$. 
The couplings of $CP$-even scalars to the $W$ boson are given by
\begin{align}
\mathcal{L} \supset  \frac{2 m^2_W}{v} \sin (\alpha-\beta) H_1 W^{+}_{\mu} W^{\mu -} + \frac{2 m^2_W}{v} \cos (\alpha-\beta) H_2 W^{+}_{\mu} W^{\mu -}~.
\end{align}
The values of $\tan \beta$ alone are not related to the $HWW$ coupling, although it can influence the $WW$ scattering indirectly via changing the decay width of $H_2$. 
We fix $\tan \beta =5$ without loss of generality, and only need to deal with two free parameters: $m_{H_2}$ and $\sin (\alpha -\beta)$. 

\begin{figure}[thb]
\includegraphics[width=0.3\textwidth]{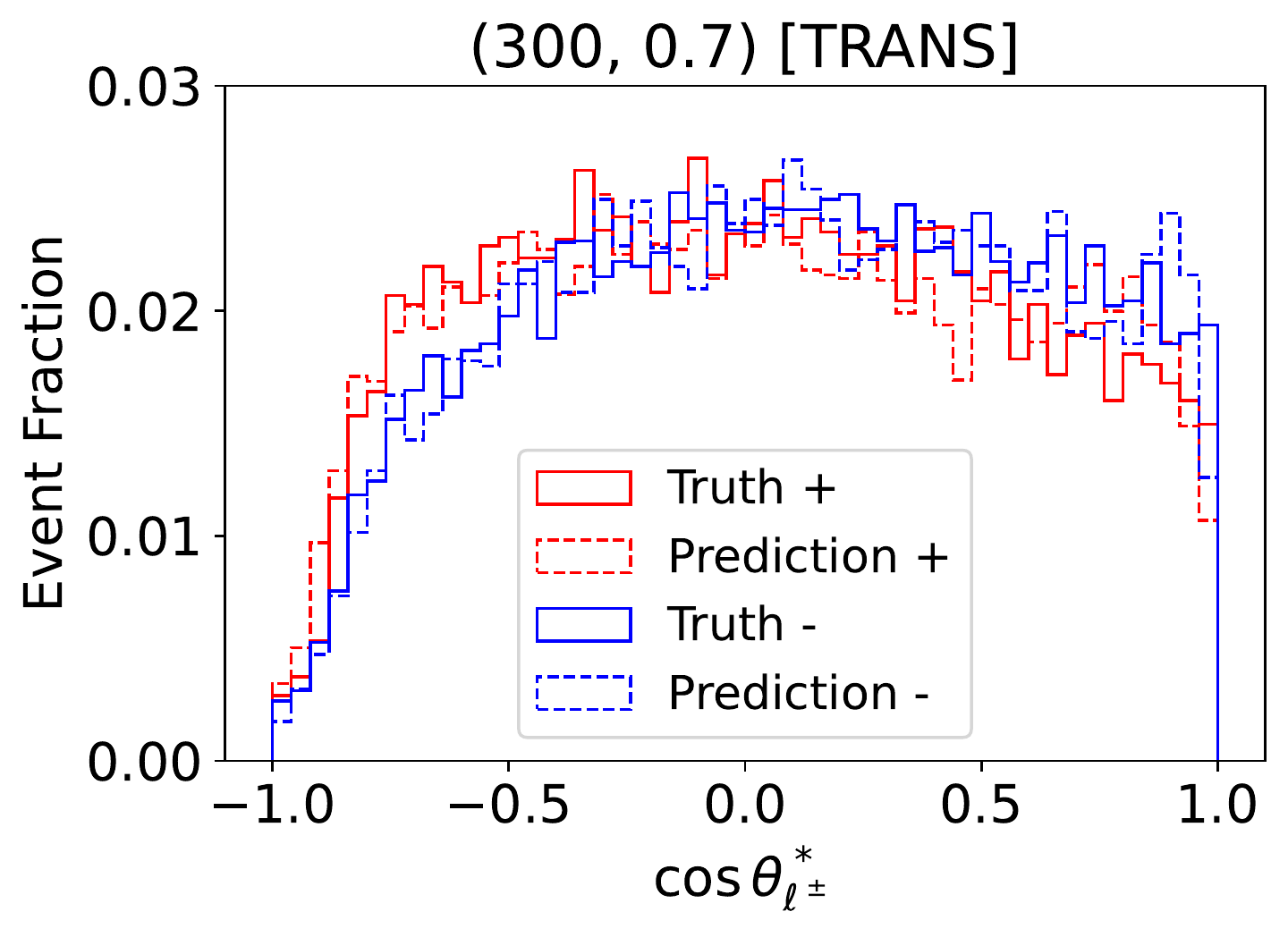}
\includegraphics[width=0.3\textwidth]{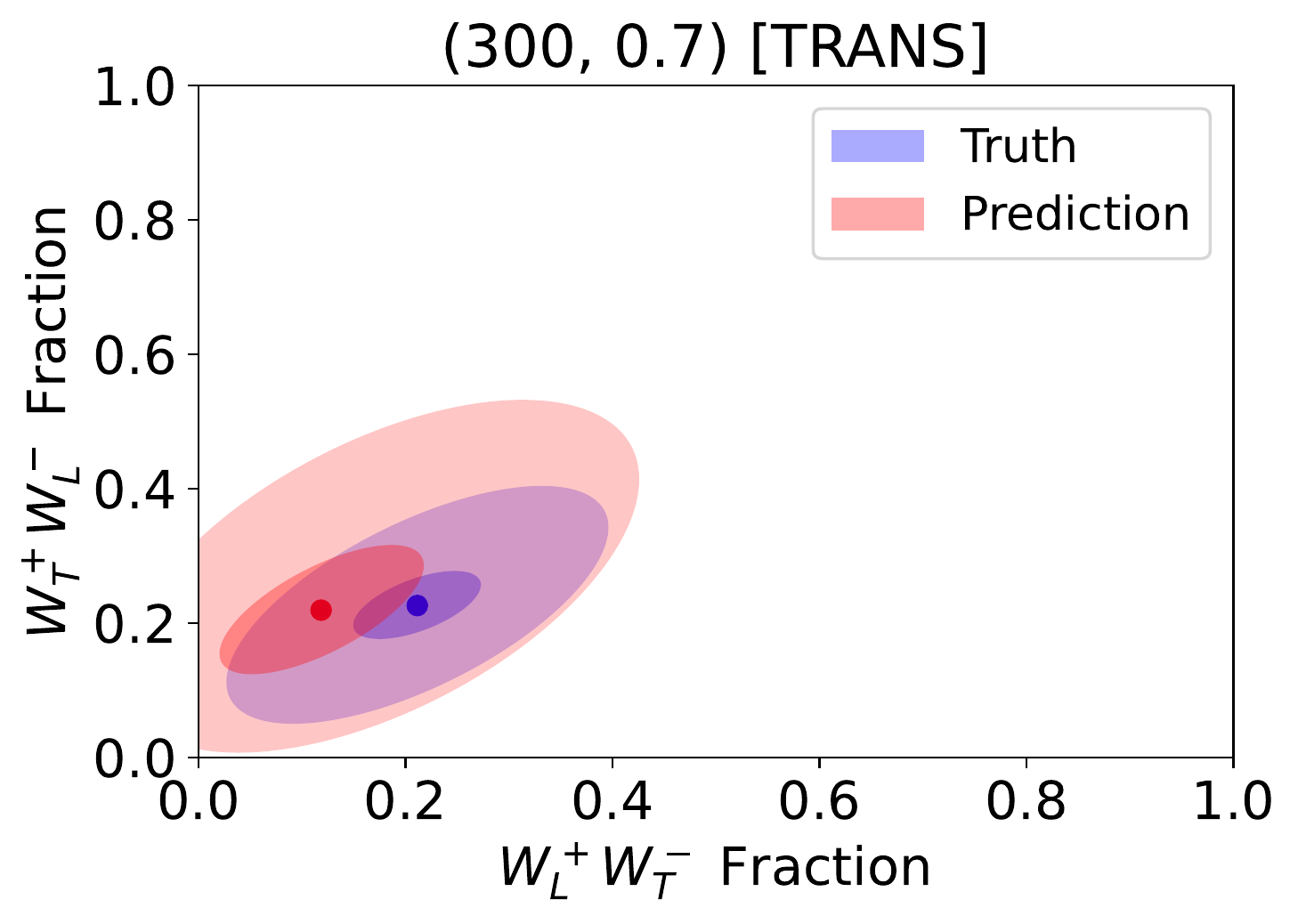}
\includegraphics[width=0.3\textwidth]{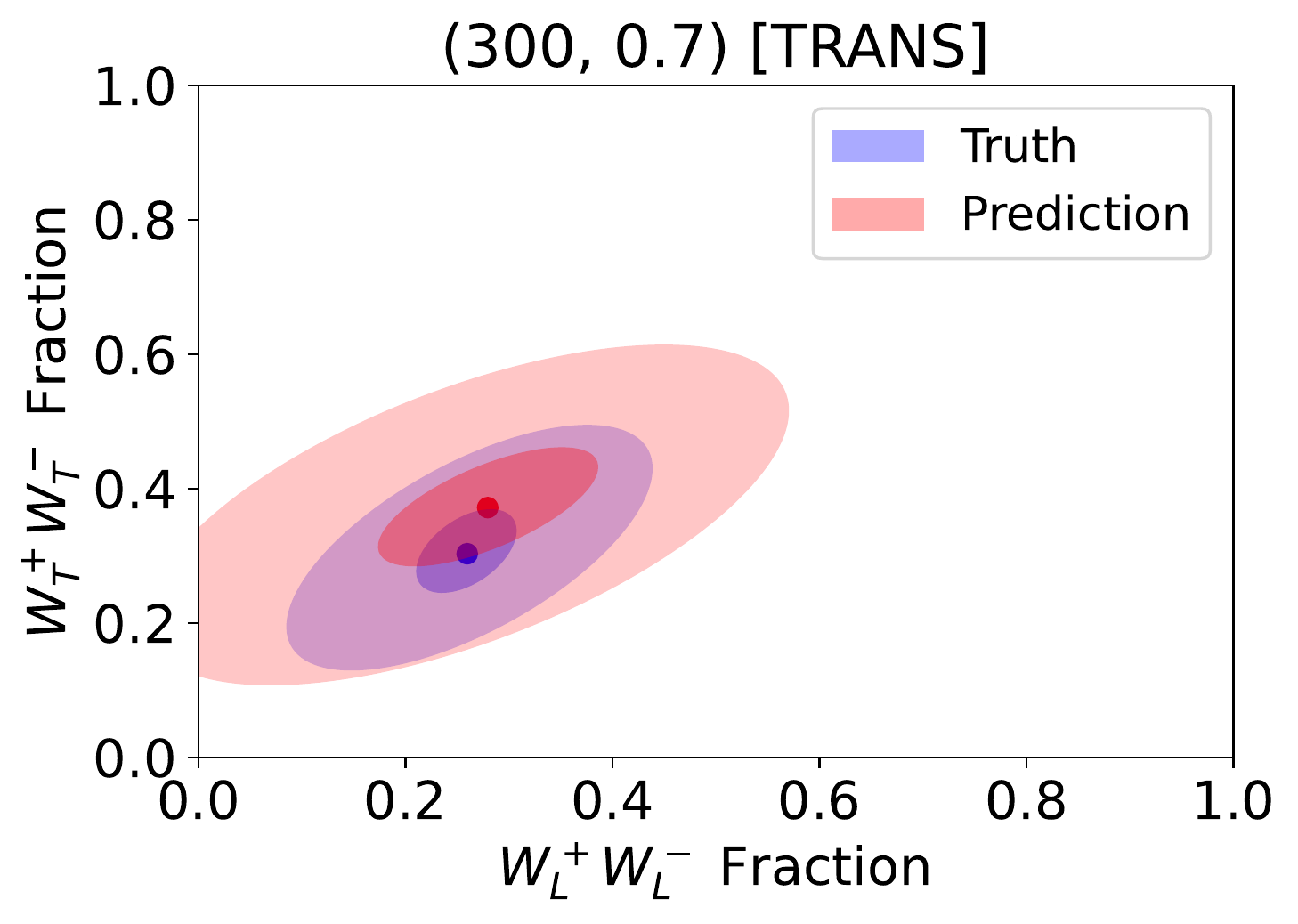}\\
\includegraphics[width=0.3\textwidth]{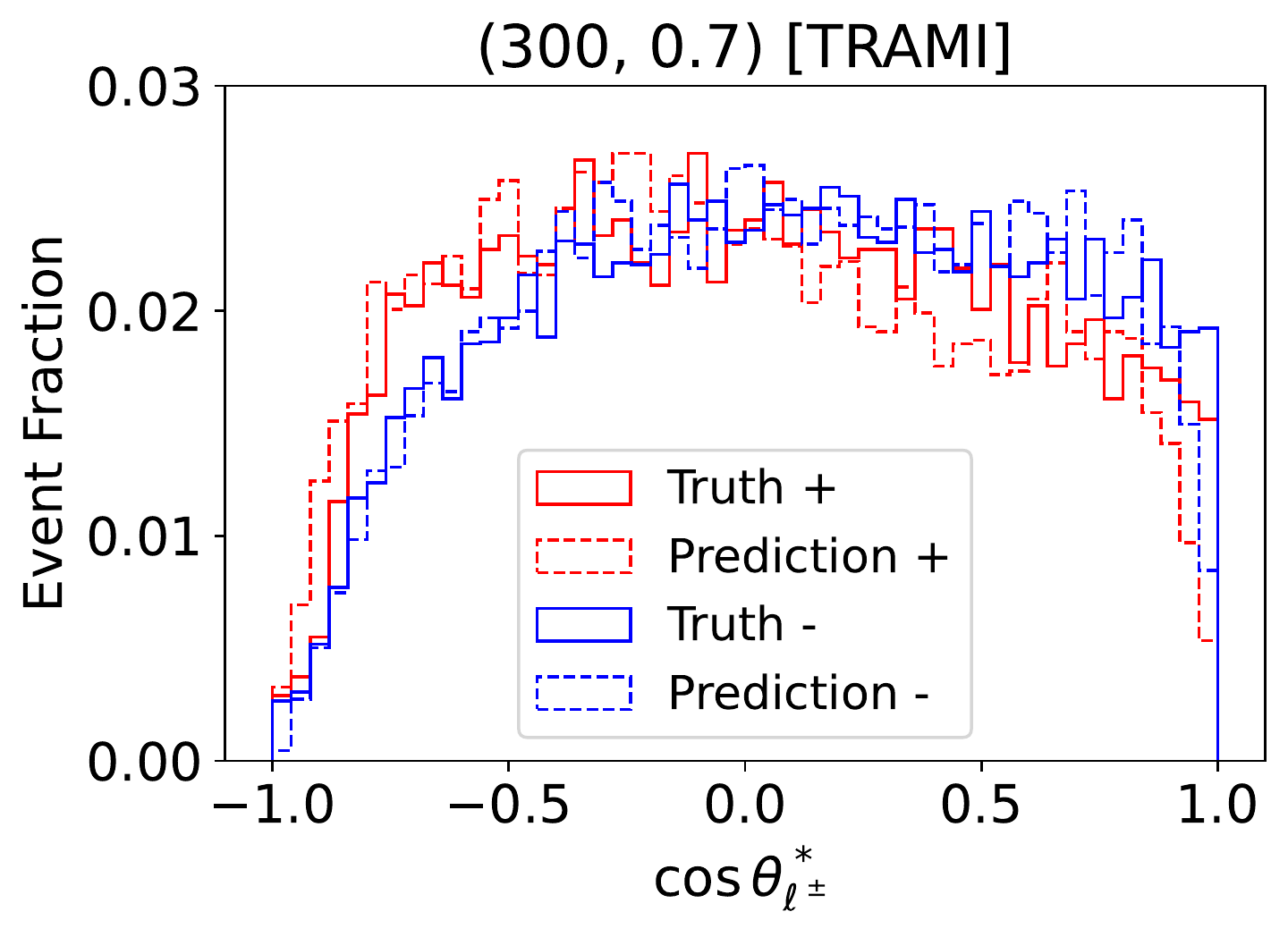}
\includegraphics[width=0.3\textwidth]{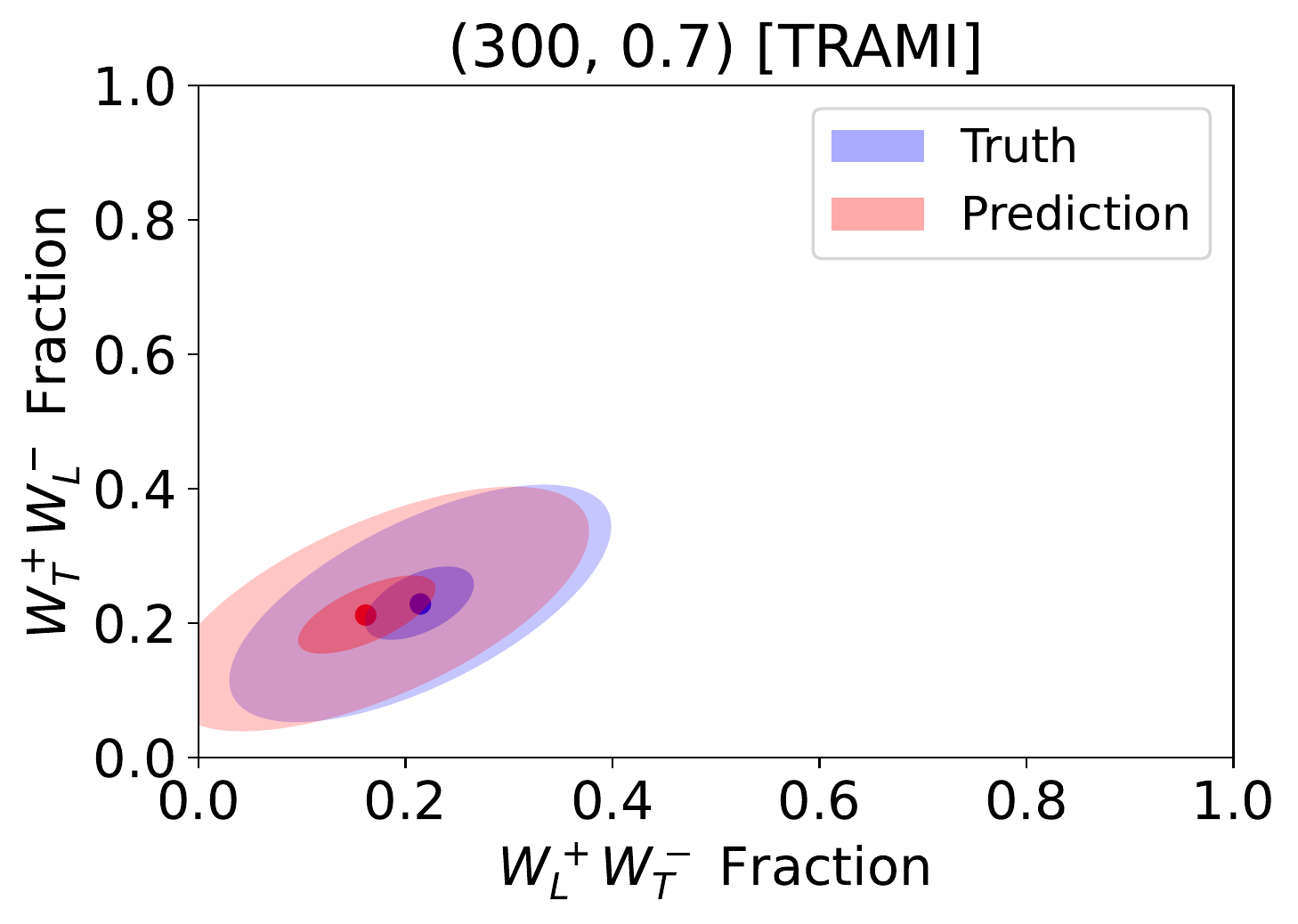}
\includegraphics[width=0.3\textwidth]{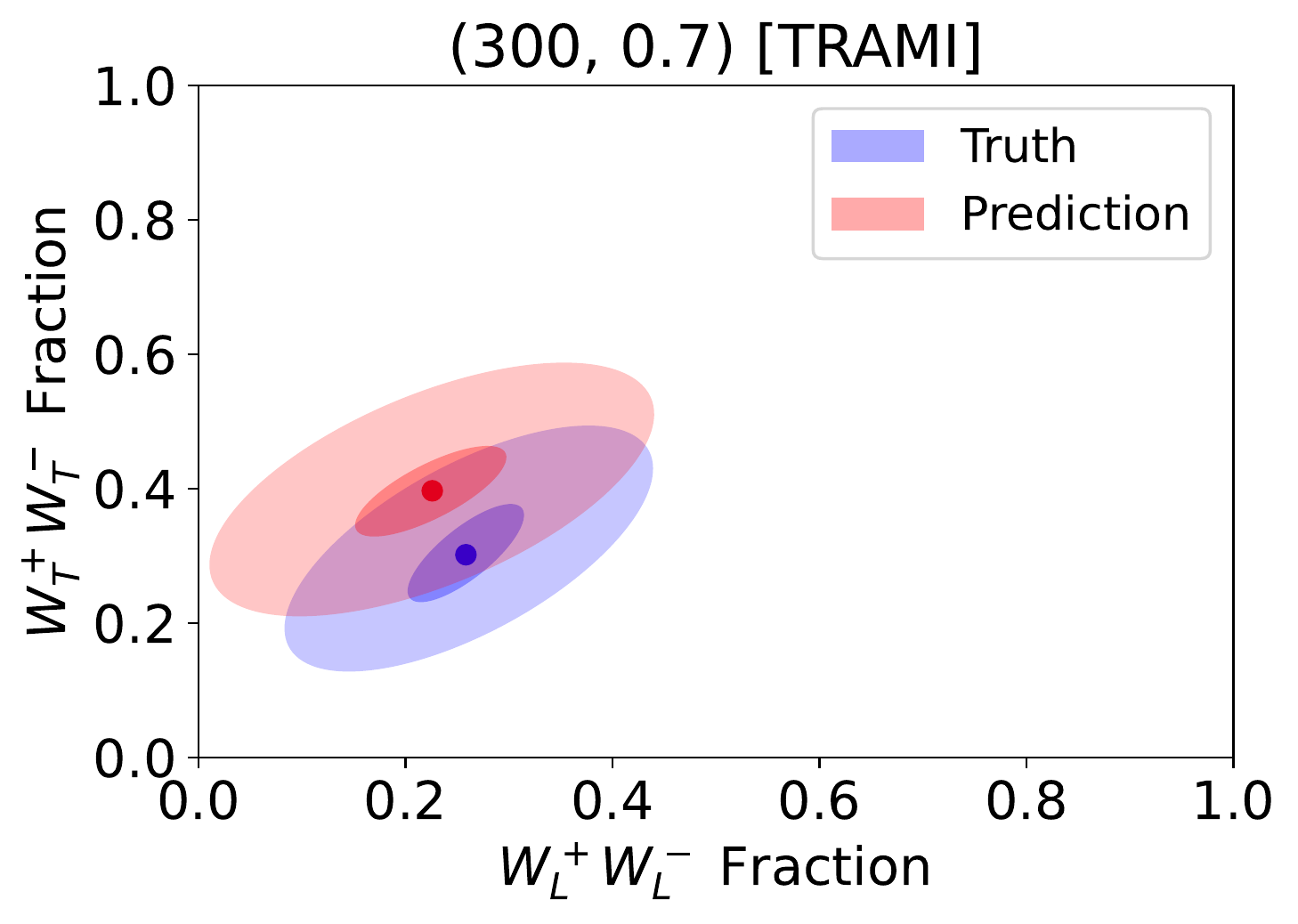}\\
\includegraphics[width=0.3\textwidth]{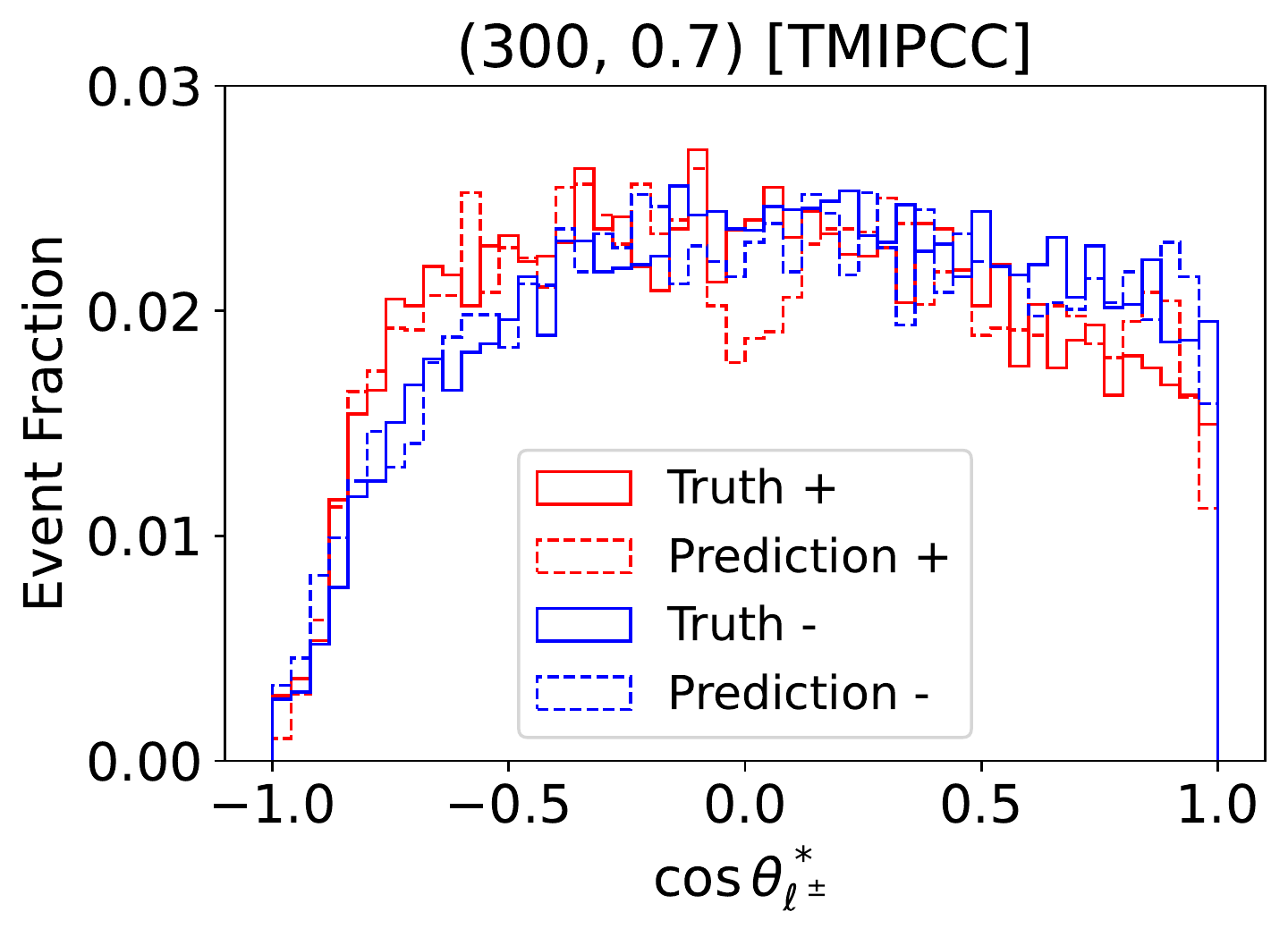}
\includegraphics[width=0.3\textwidth]{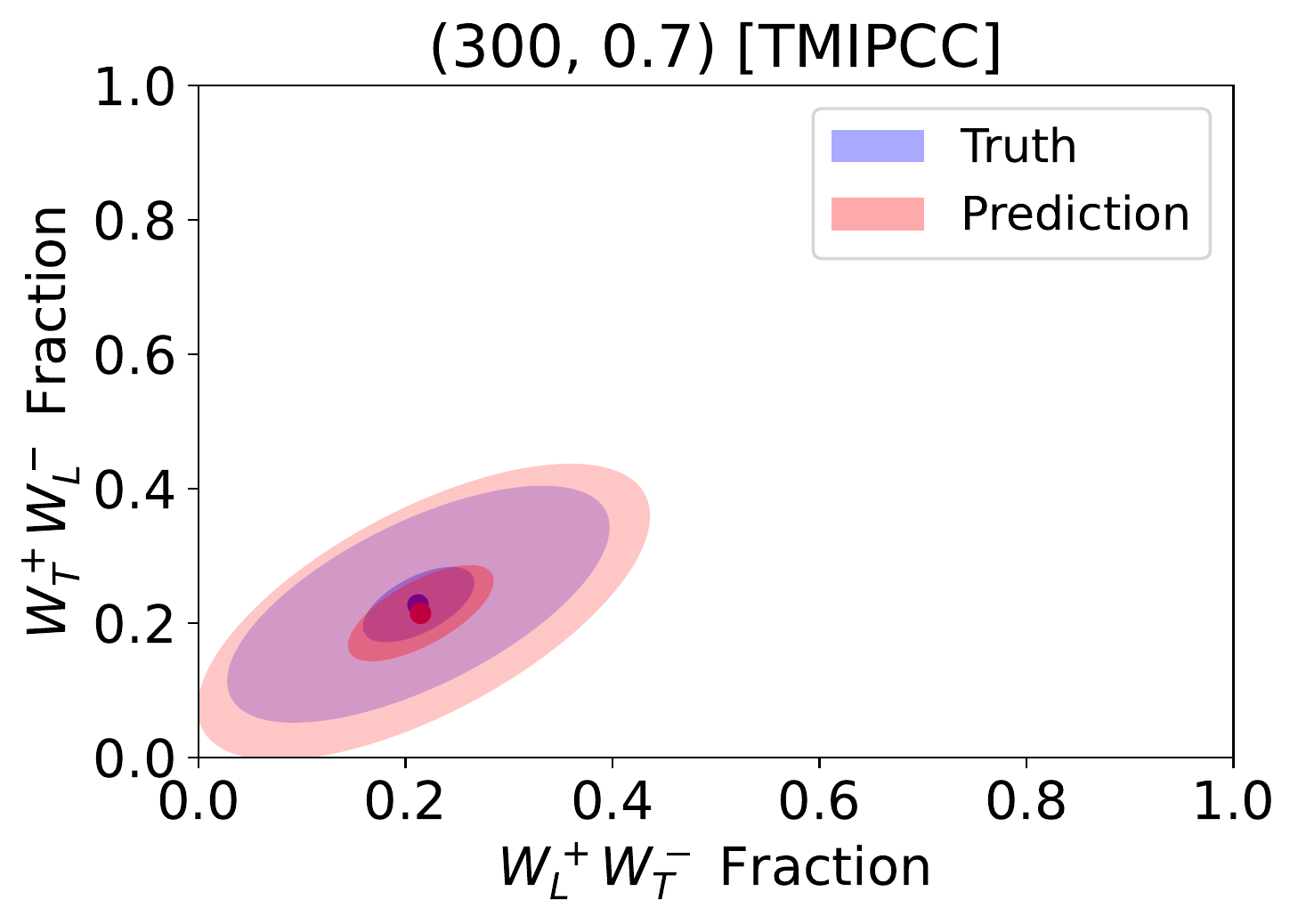}
\includegraphics[width=0.3\textwidth]{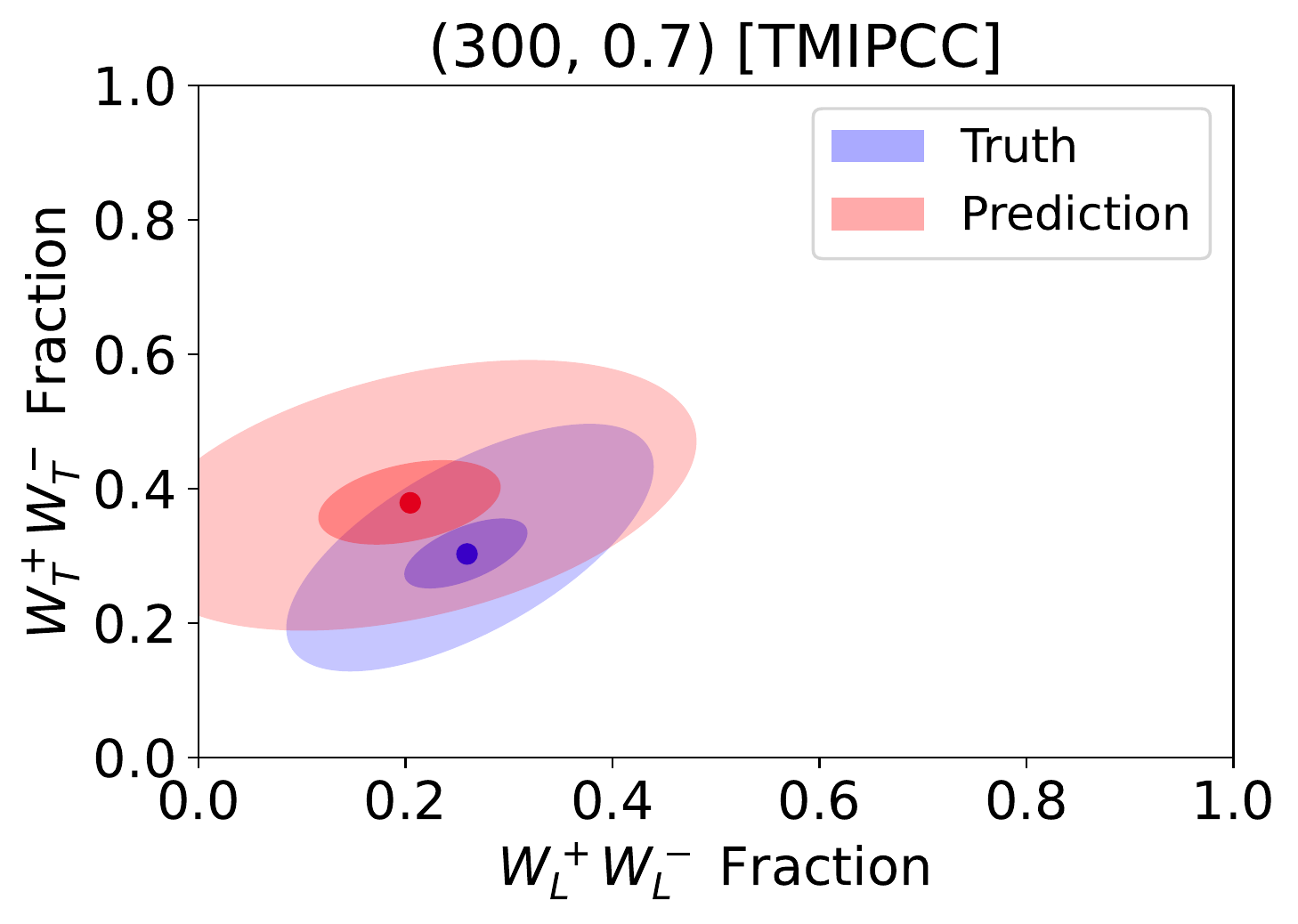}
\caption{ For the $W^+W^-$ scattering in the 2HDM with $m_{H_2}=300$ GeV and $\sin (\alpha) = 0.7$ at the 13 TeV LHC. Meanings of the plots are the same as Fig.~\ref{fig:smfit}.  \label{fig:2hdmfit}}
\end{figure}

In Fig.~\ref{fig:2hdmfit}, the projected one-dimensional lepton angle ($ \cos \theta^*_{\ell^\pm}$) distributions and the $\Delta \chi^2$ contours on the polarization fraction planes for the $W^+ W^-$ scattering in 2HDM with $m_{H_2}=300$ GeV and $\sin (\alpha) = 0.7$ at the 13 TeV LHC are shown. 
Because of the resonant contribution from the $H_2$, the cross section of the $W^+ W^-$ scattering is increased to 8.362 fb. 
As in the SM and EFT, the lepton angle distributions can be reproduced well by all three networks. 
However, the precision of the polarization fractions obtained from the TRANS network are not as good as those obtained from the truth lepton angles. 
This is because the features in the TRANS network (which is trained only with the SM events) contain the SM kinematic information (in particular, the invariant mass of the $W$ boson pair). The differences between the kinematic properties of the 2HDM and the SM degrade the performance. 
The situation is much improved for the TRAMI network, in which the information of $W$ boson pair momentum is decorrelated from the features.


\begin{figure}[htb]
\includegraphics[width=0.3\textwidth]{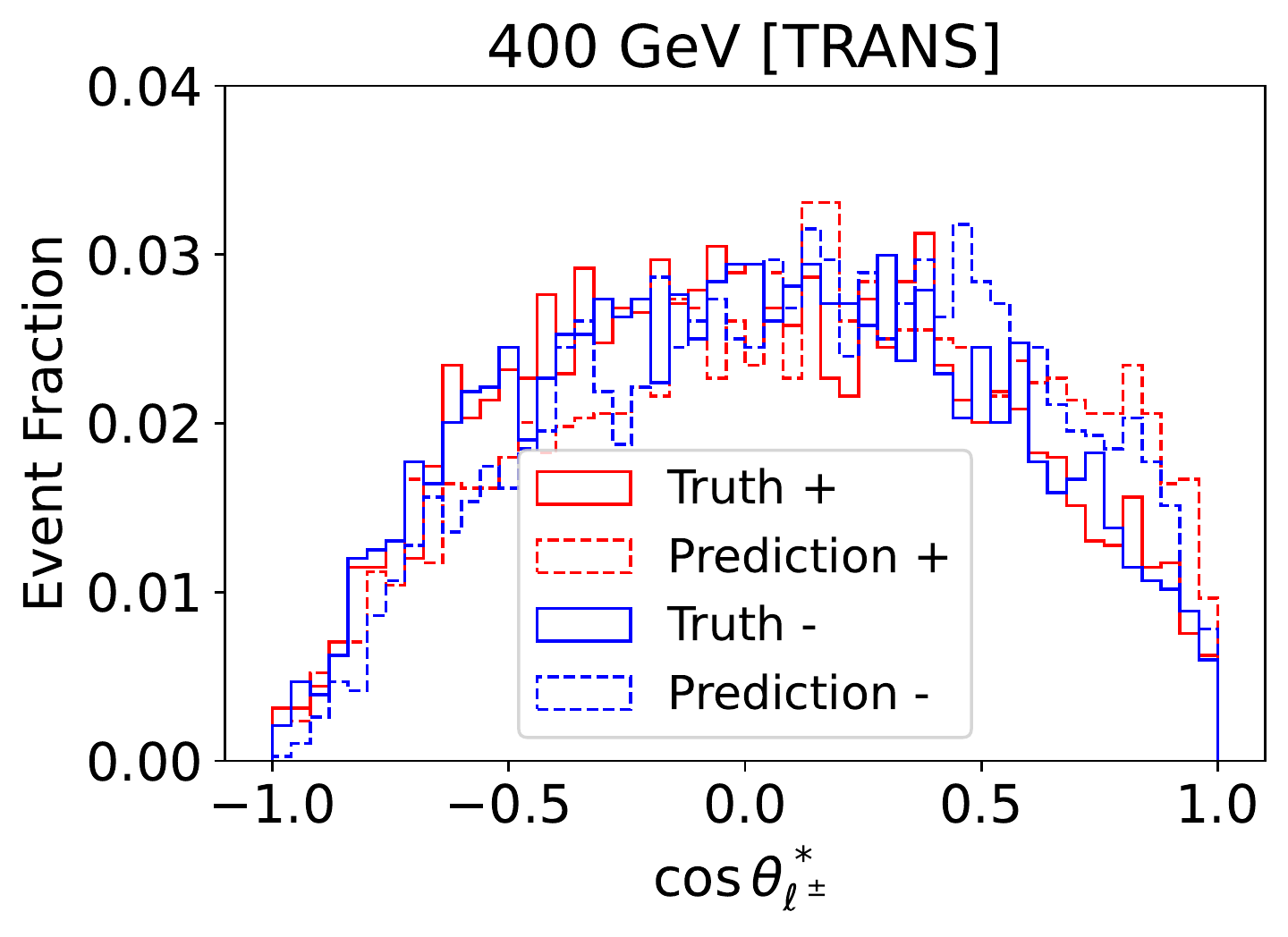}
\includegraphics[width=0.3\textwidth]{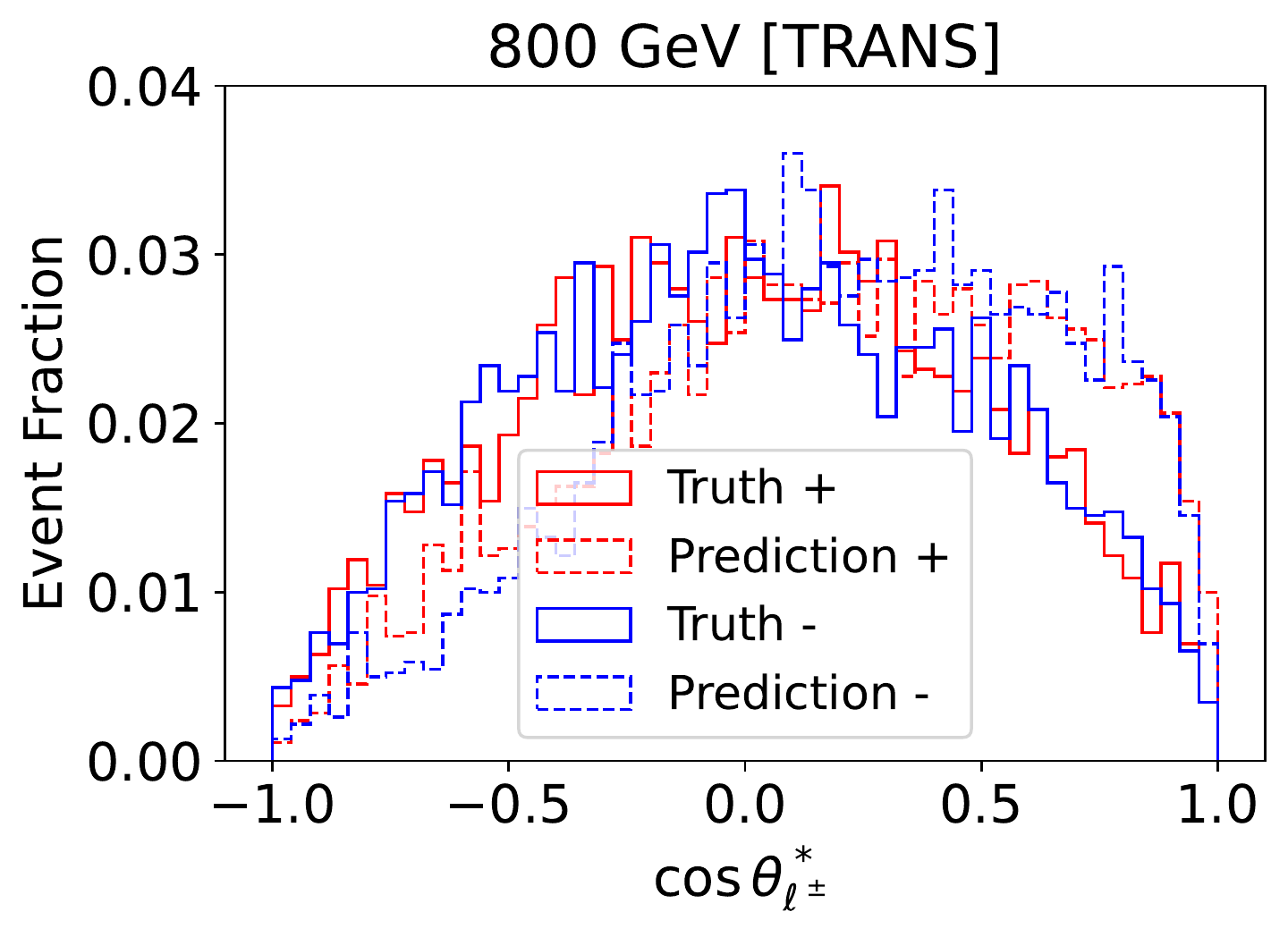}
\includegraphics[width=0.3\textwidth]{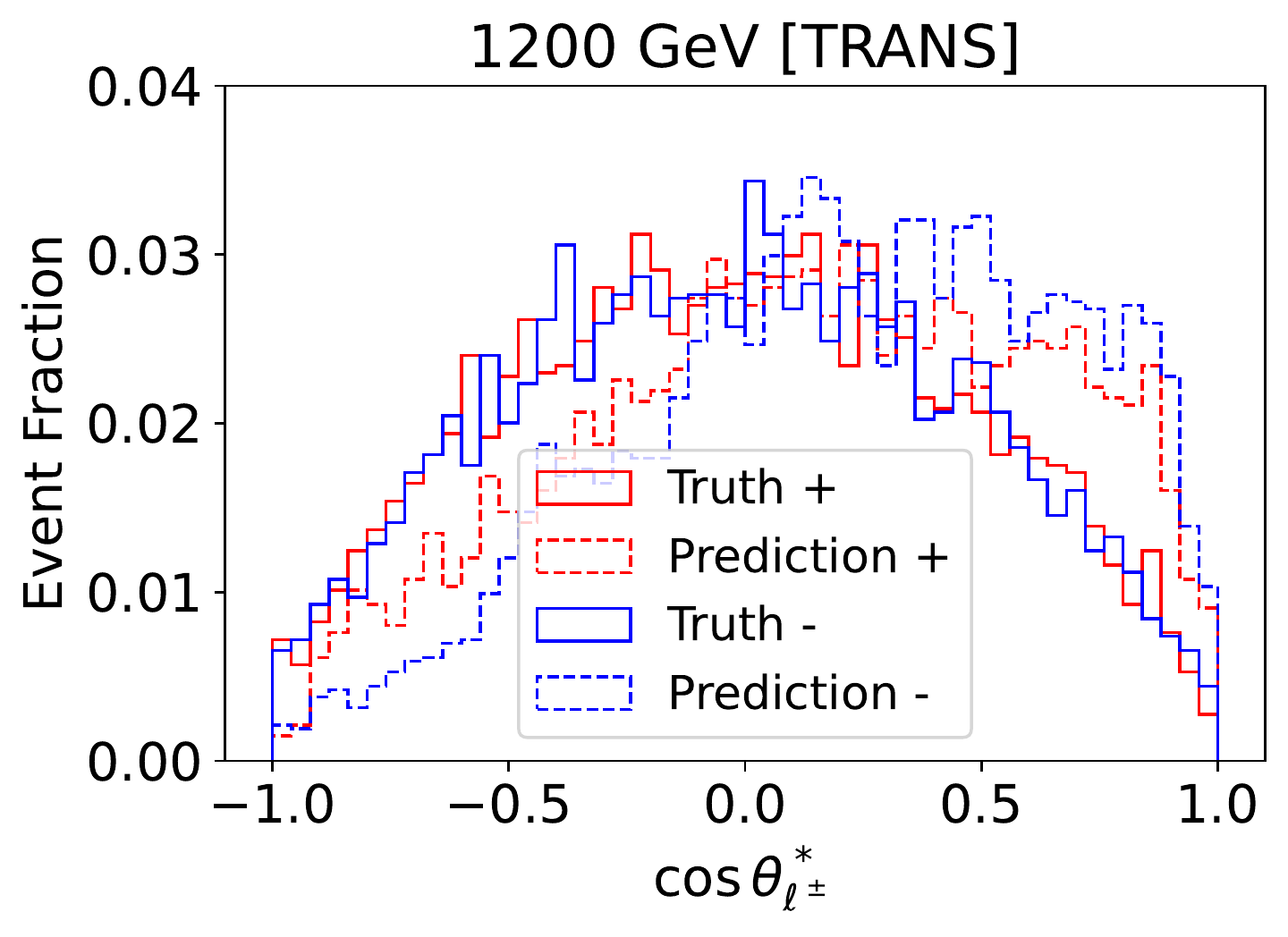} \\ 
\includegraphics[width=0.3\textwidth]{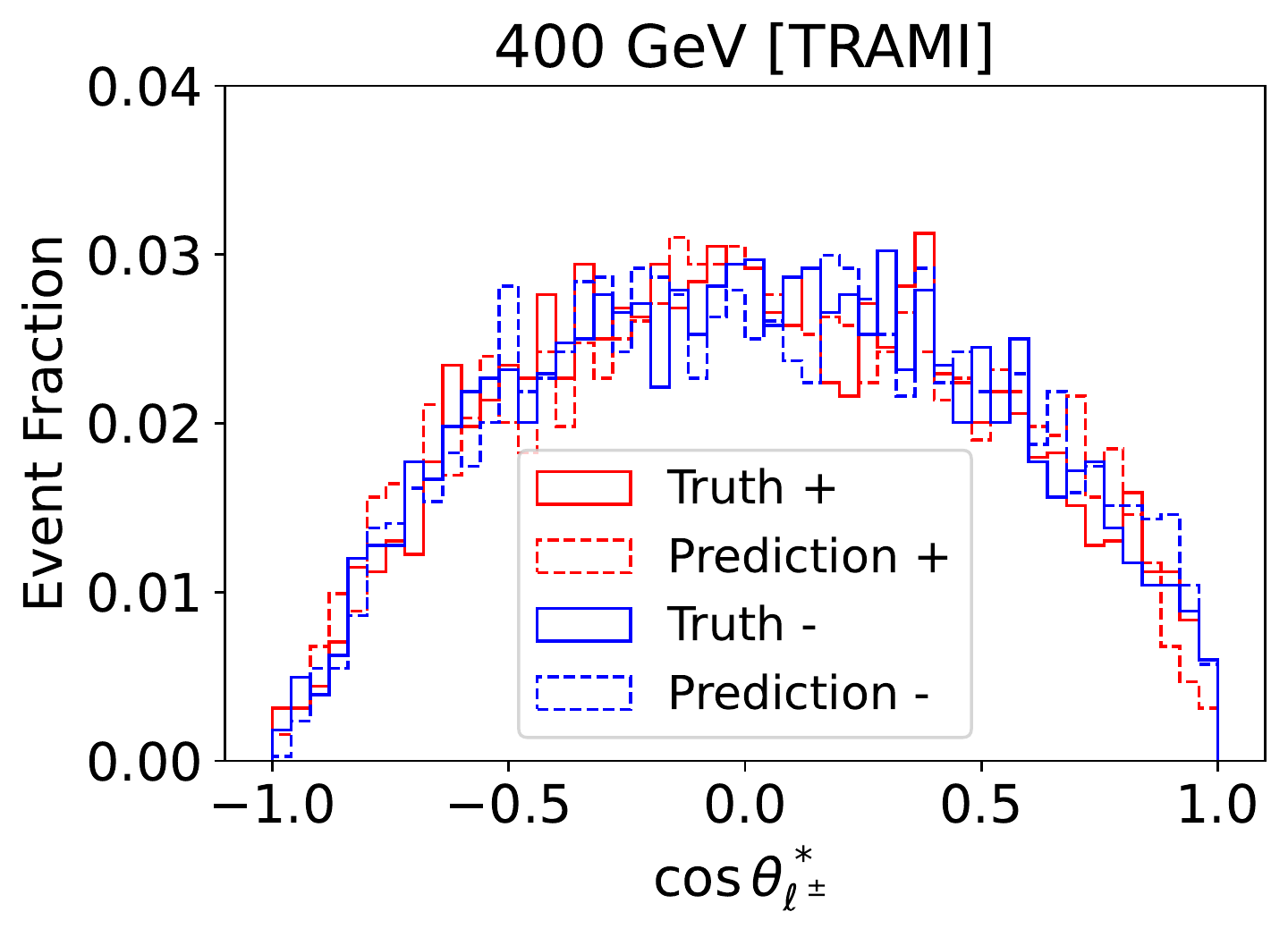}
\includegraphics[width=0.3\textwidth]{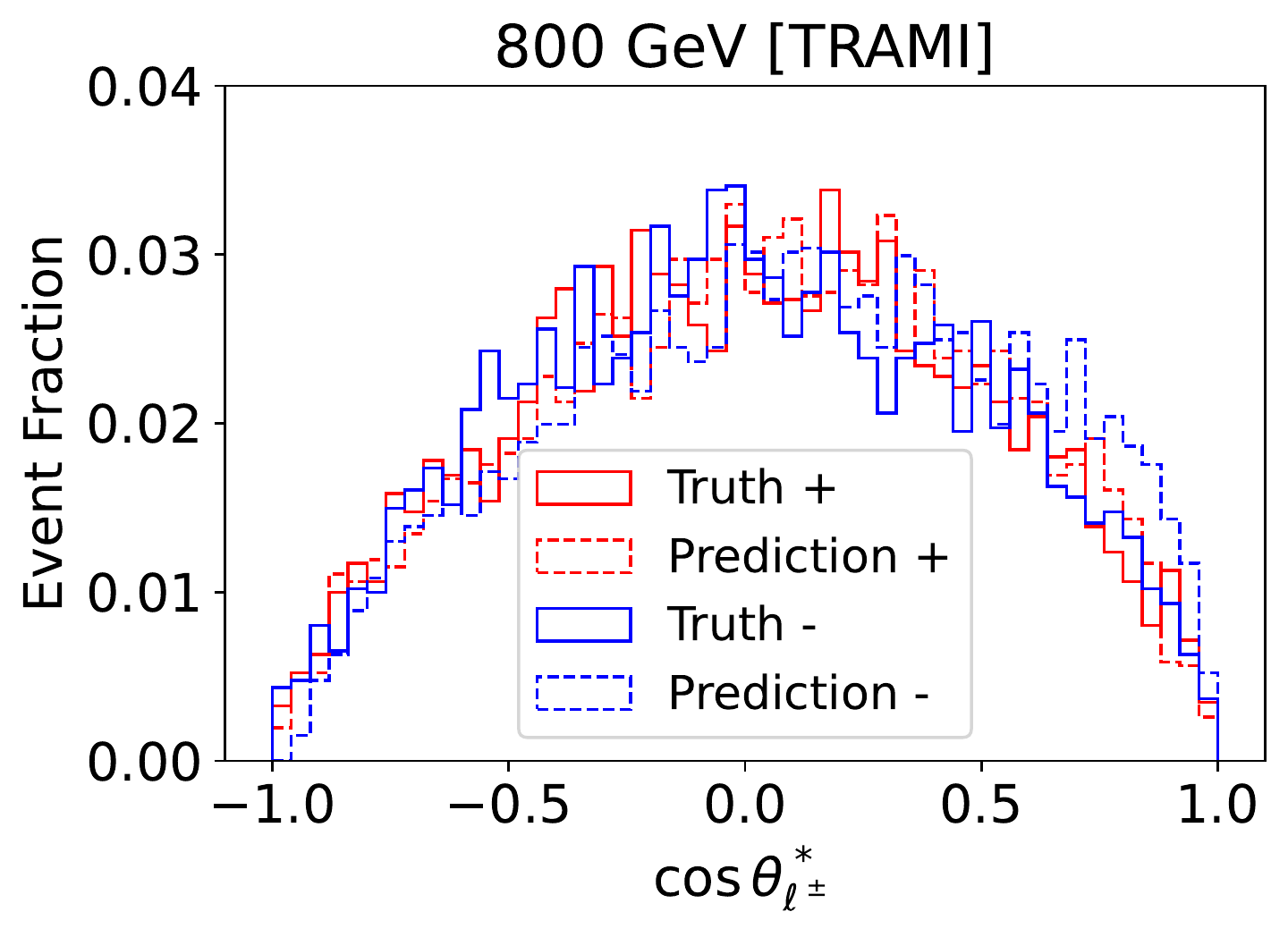}
\includegraphics[width=0.3\textwidth]{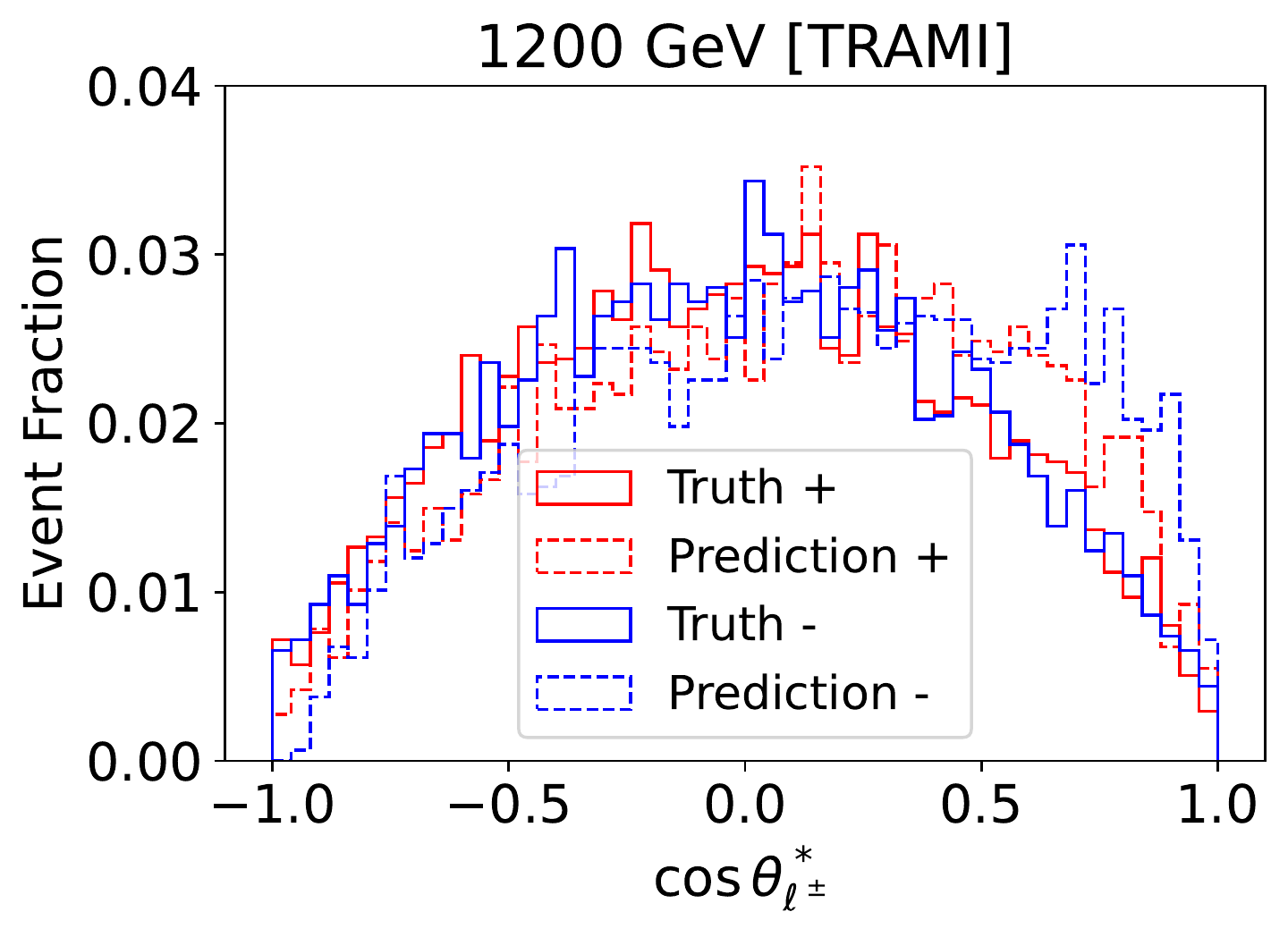} \\
\includegraphics[width=0.3\textwidth]{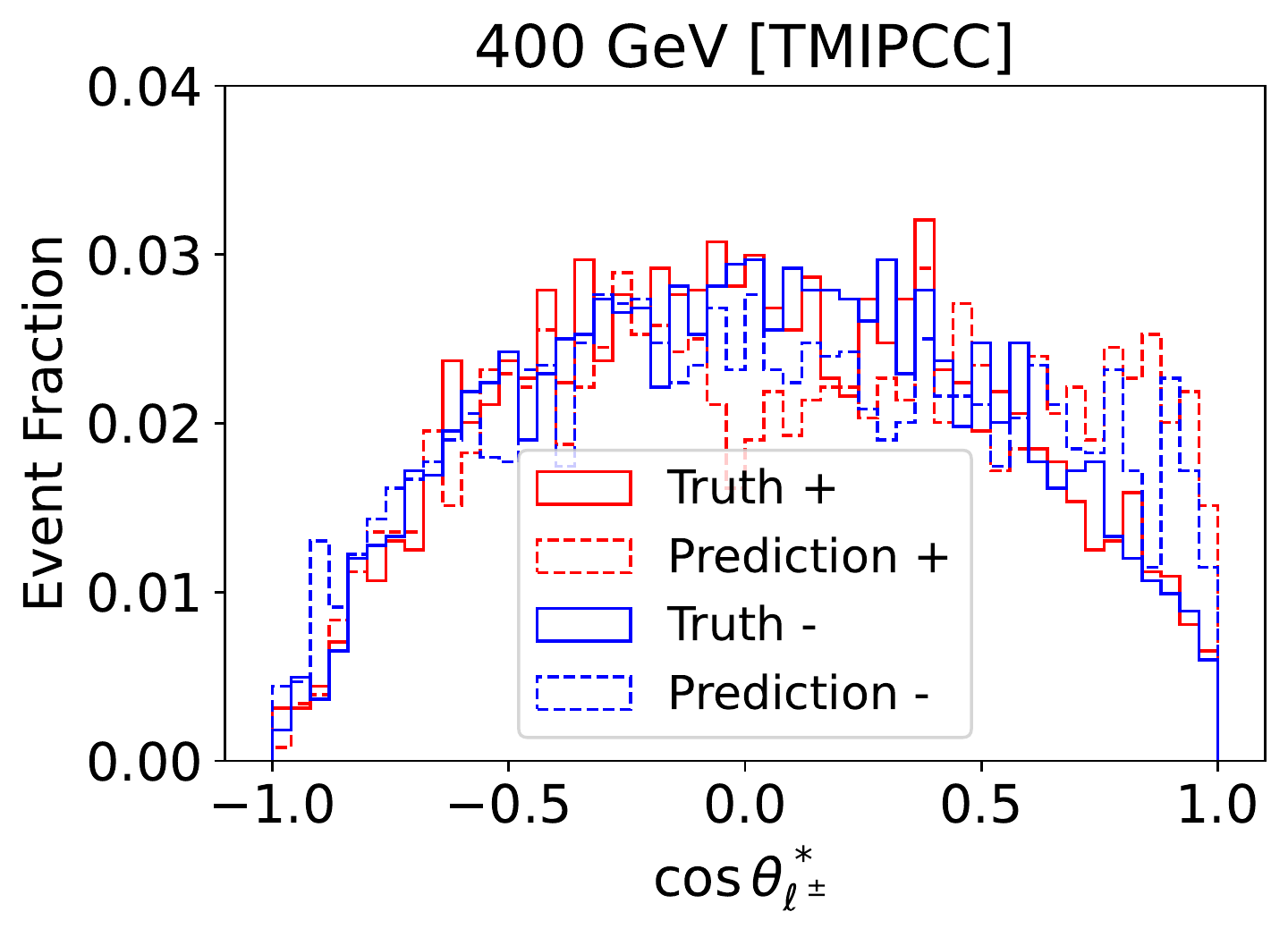}
\includegraphics[width=0.3\textwidth]{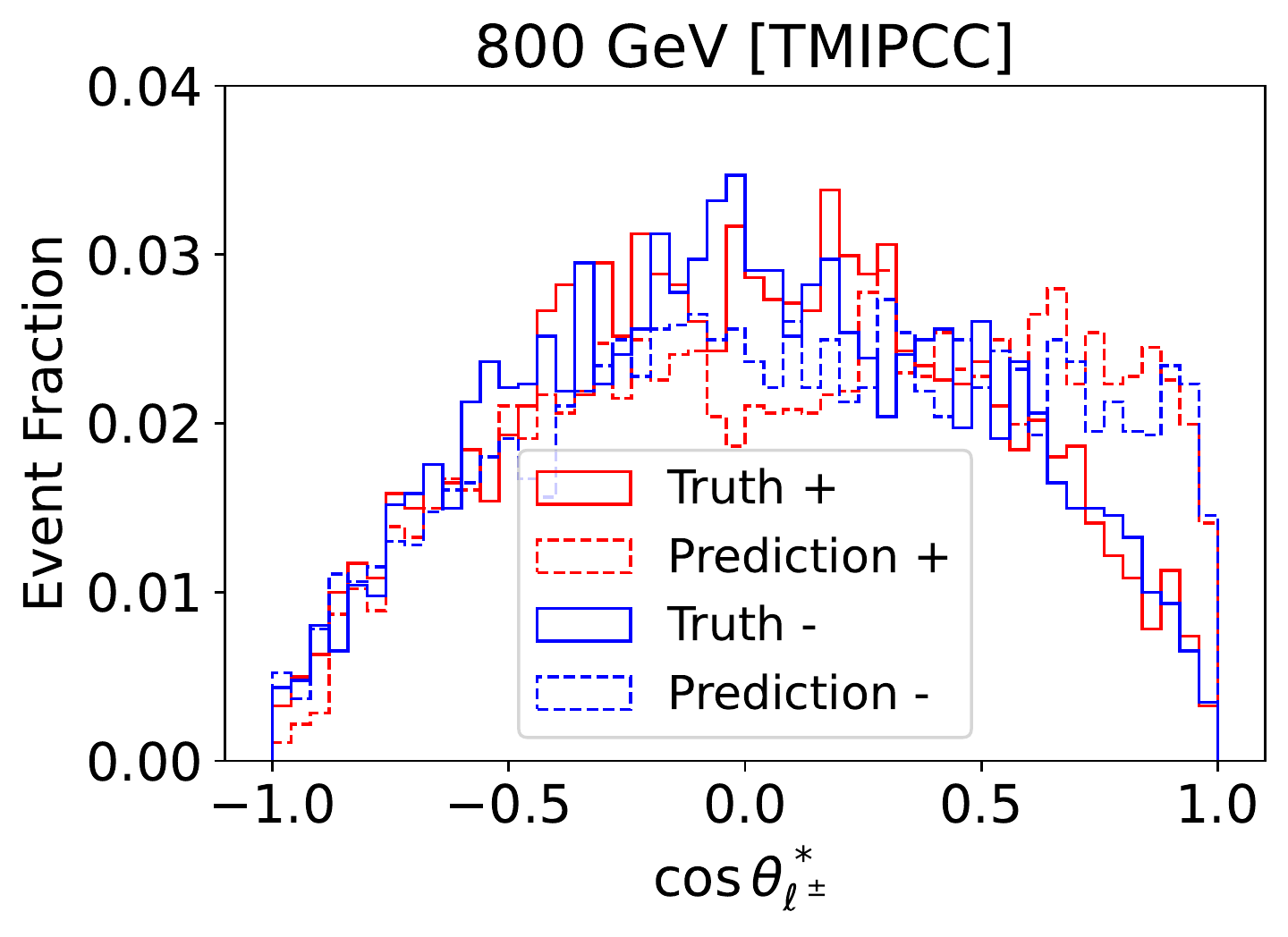}
\includegraphics[width=0.3\textwidth]{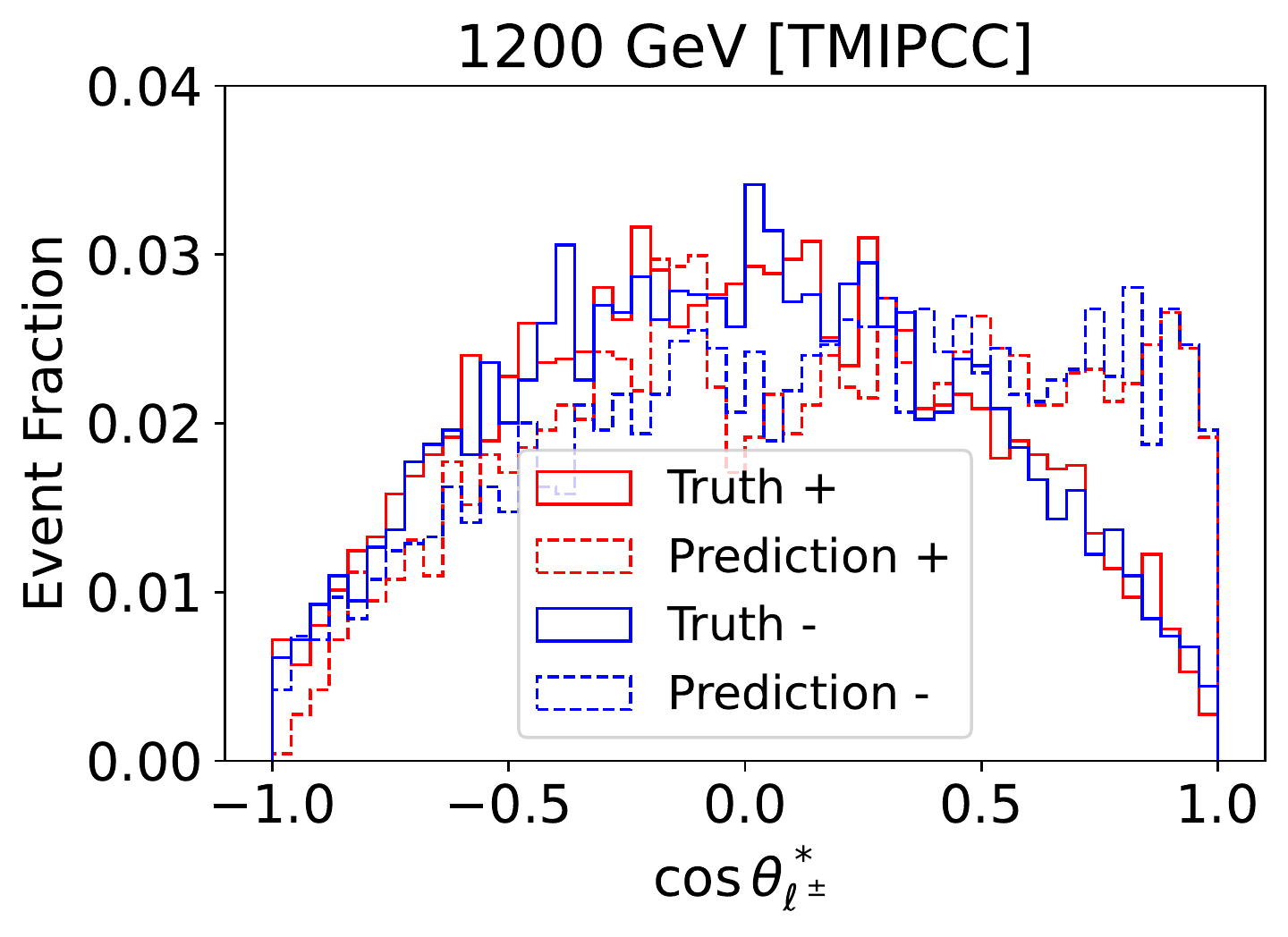}
\caption{A comparison of truth level $\cos \theta^*_{\ell^\pm}$ and the network output distributions for the resonant $W^+W^-$ scattering at the 13 TeV LHC with resonance mass $400$, 800, and 1200 GeV (from left to right). Plots from top to bottom are obtained with the TRANS, TRAMI, and TMIPCC networks, respectively. \label{fig:peakangle}}
\end{figure}

To illustrate the performance in a more extreme case, we apply those networks to the process of $W^+W^-$ scattering solely through a heavy resonance. 
Both the polarization pattern and the kinematic features are dramatically different from the SM ones. 
Assuming the mediator is a scalar boson, only the polarization modes $W_L W_L$ and $W_T W_T$ are allowed. The ratio $f_{LL}/f_{TT} \sim 67$ for the scalar mass around 400 GeV and is increasing fast for a heavier scalar. 
In Fig.~\ref{fig:peakangle}, we show the comparison of the truth level $\cos \theta^*_{\ell^{\pm}}$ and the network output distributions for the $W^+ W^-$ scattering through a heavy scalar resonance at the 13 TeV LHC. 
As expected, the TRAMI network performs much better than the TRANS network, since the TRAMI is not supposed to be sensitive to the $WW$ production mechanism. 
The reconstructed lepton angle distributions from the TRAMI network remain close to the truth ones for resonance mass less than $\sim 1$ TeV. 
The heavier the resonance is, the larger the deviation between the network predictions and truth values.

Up to this point, we have not found it necessary to add the PCC to the loss function, as the performance of the TMIPCC network is not comparable to that of the TRAMI network in all cases. 
This is mainly because the training samples for the Transformer network are generated at 13 TeV, and the decorrelation of collision energy will not be necessary if we are extracting the $W$ polarization fractions at the same collision energy. 
In this case, the TRAMI network performs the best and is recommended to use. 
However, if we want to apply the same network to processes at different collision energy, the subtraction of collision energy dependence becomes essential.  
In the next section, we further study the network performance at different collision energy by taking the $W^+W^-$ scattering in the 2HDM at 100 TeV as an example. 

\section{The $W^+W^-$ polarization in 100 TeV $p$-$p$ collision} \label{sec:model100}

As we discussed above, although the distributions of $\theta^*_{\ell^{\pm}}$ are supposed to be only related to the $W$ boson polarization, the effects of preselection cuts which distort the lepton angle distribution, depend on the collision energy.
The same preselection cuts as proposed in Sec.~\ref{sec:event} for 13 TeV are also adopted here. 

\begin{figure}[thb]
\includegraphics[width=0.24\textwidth]{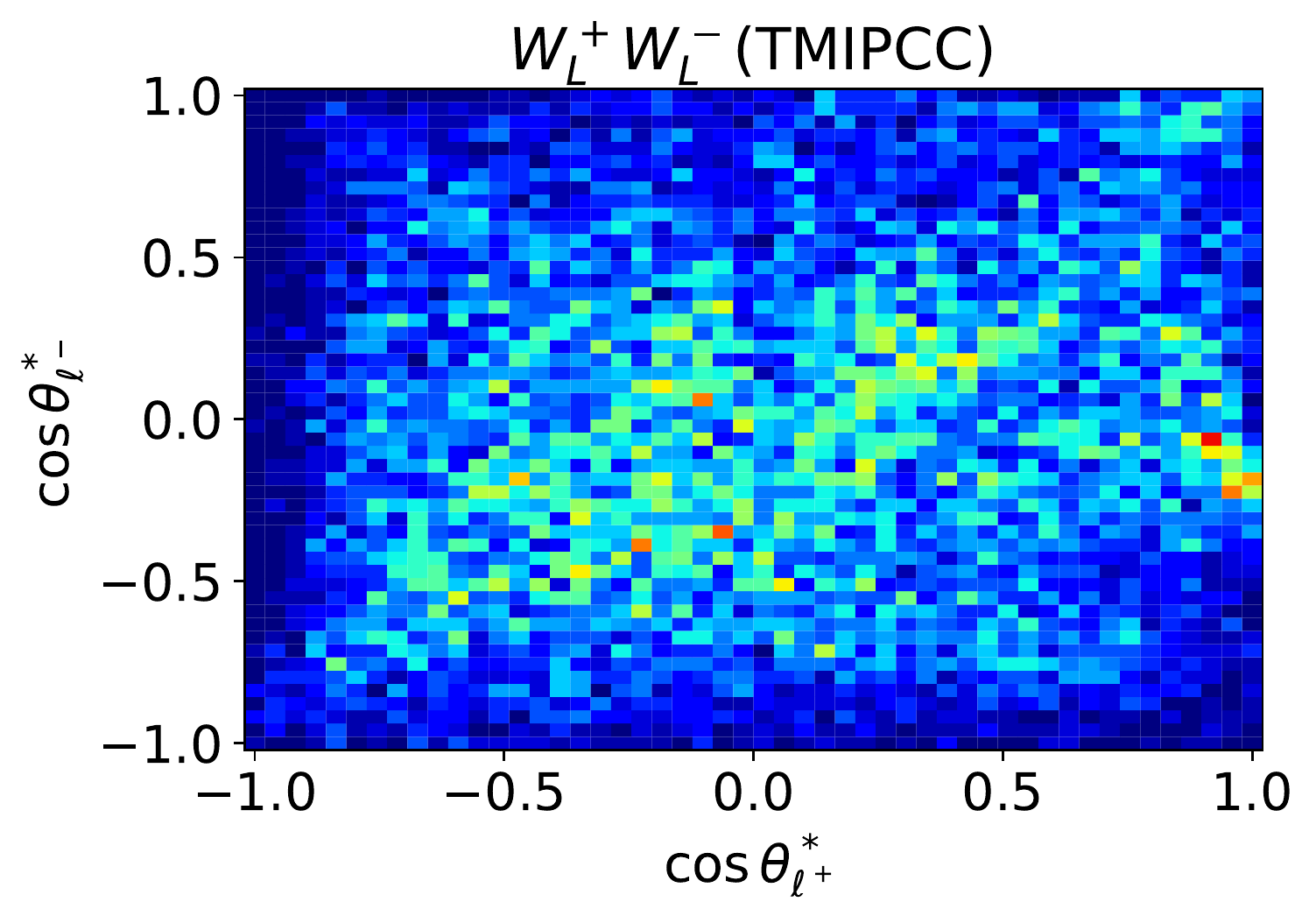}
\includegraphics[width=0.24\textwidth]{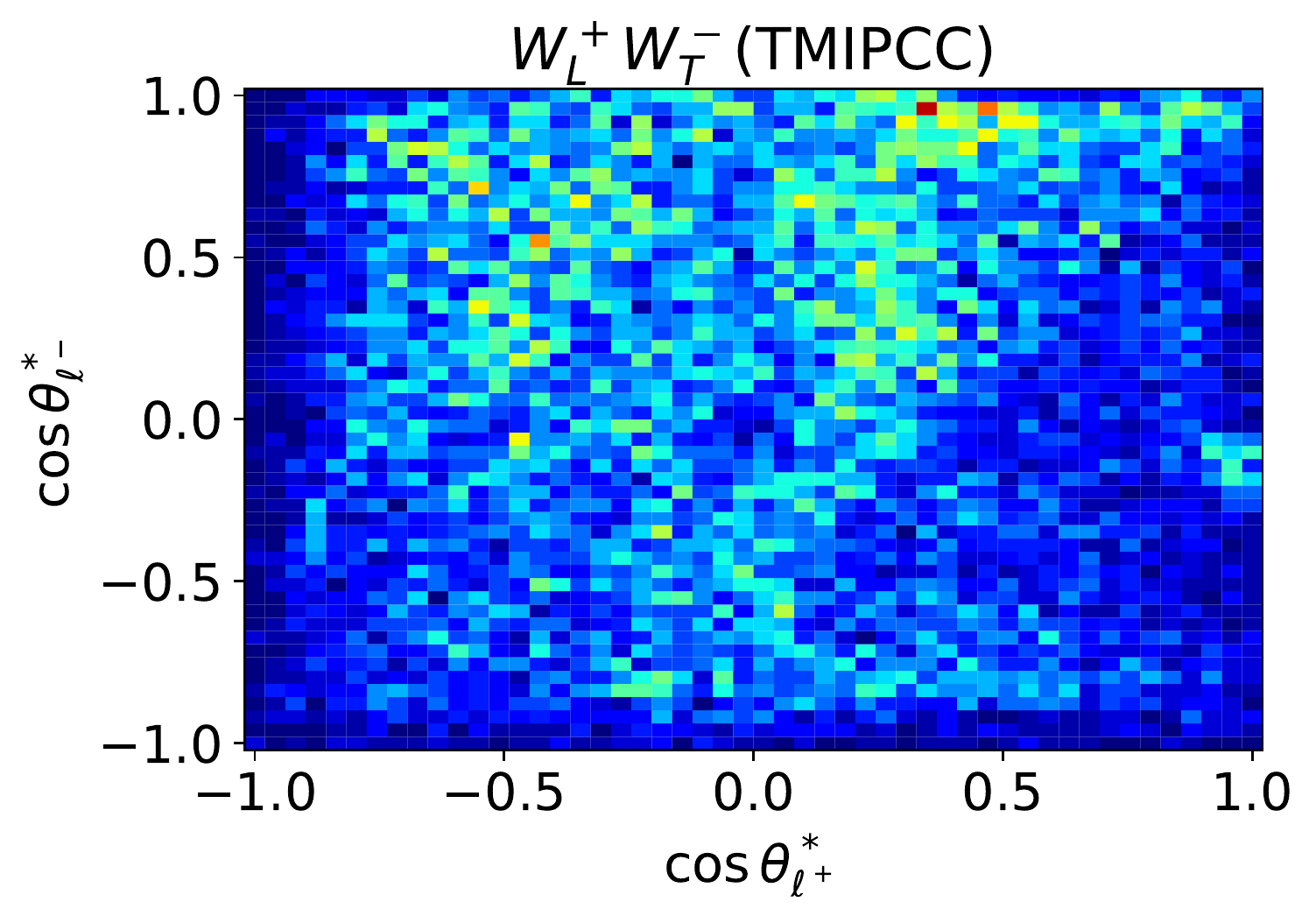}
\includegraphics[width=0.24\textwidth]{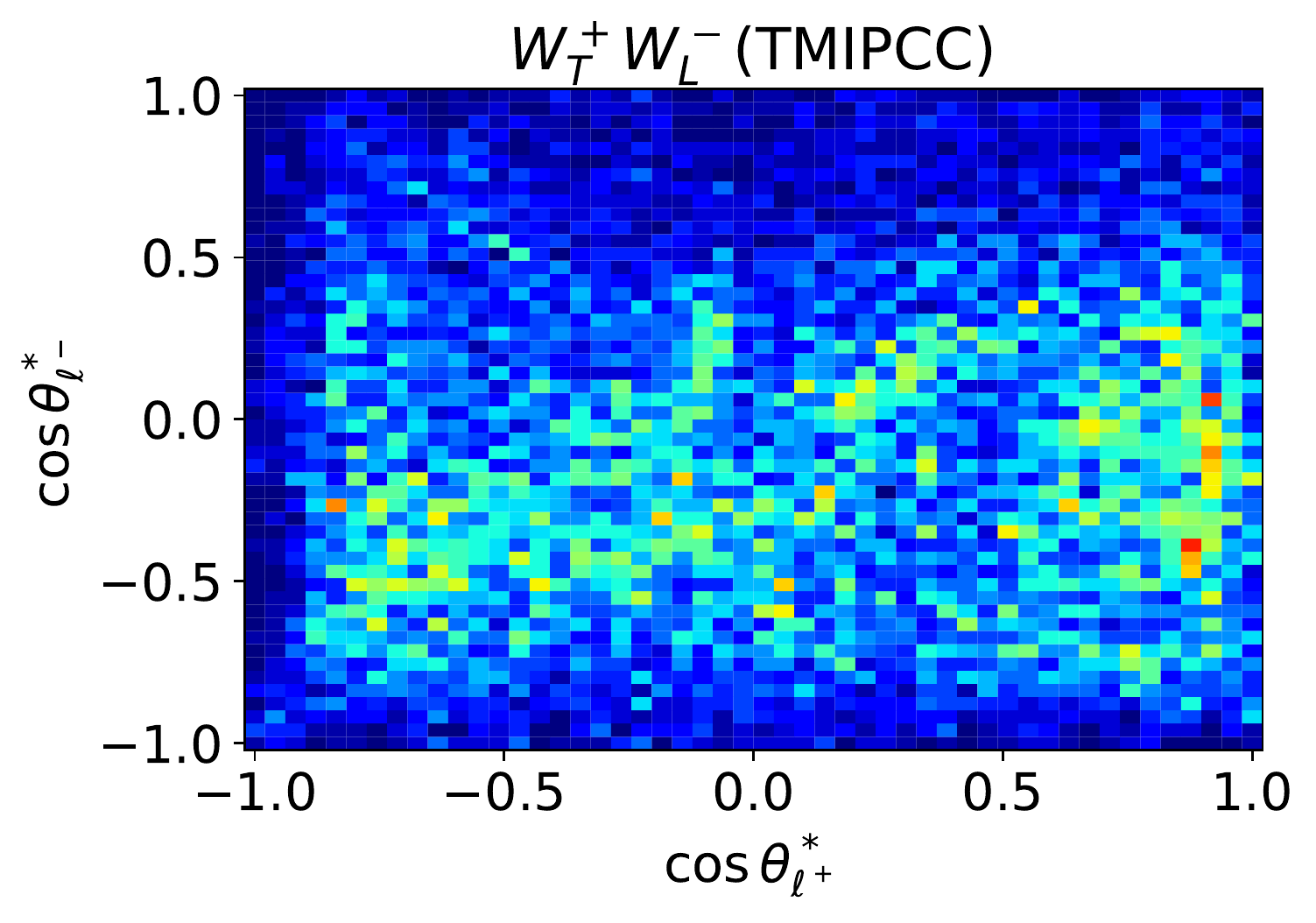}
\includegraphics[width=0.24\textwidth]{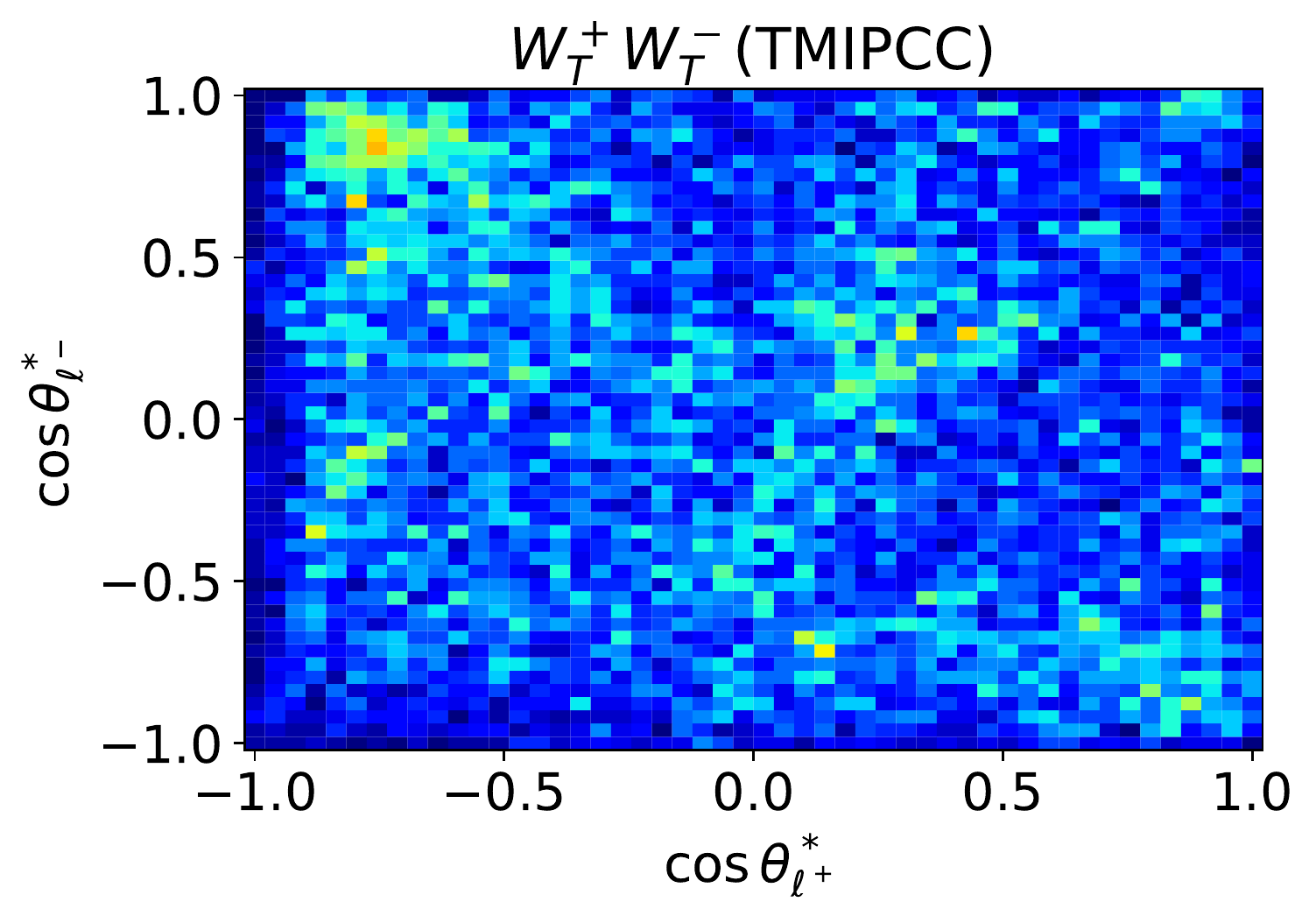}\\
\includegraphics[width=0.24\textwidth]{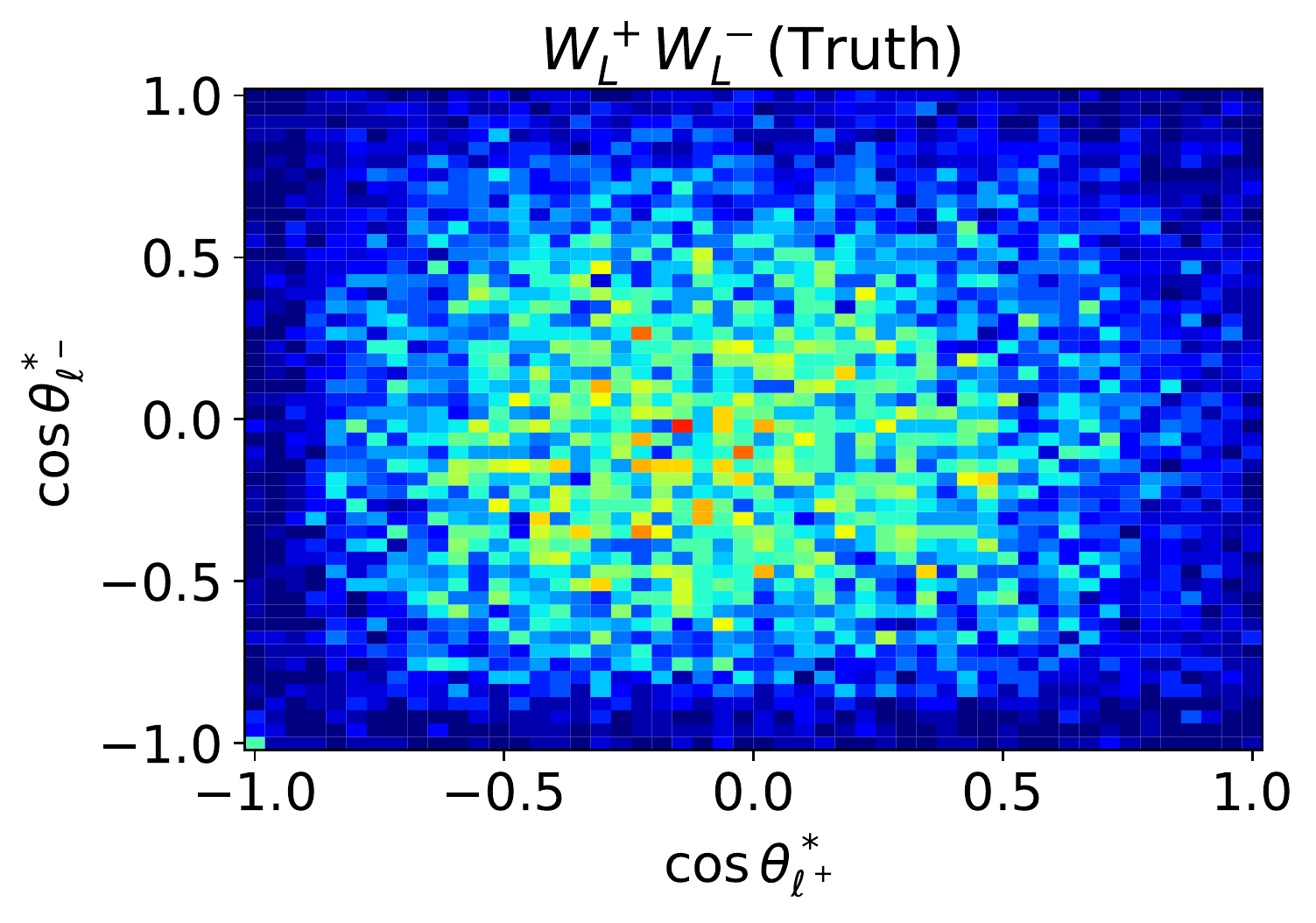}
\includegraphics[width=0.24\textwidth]{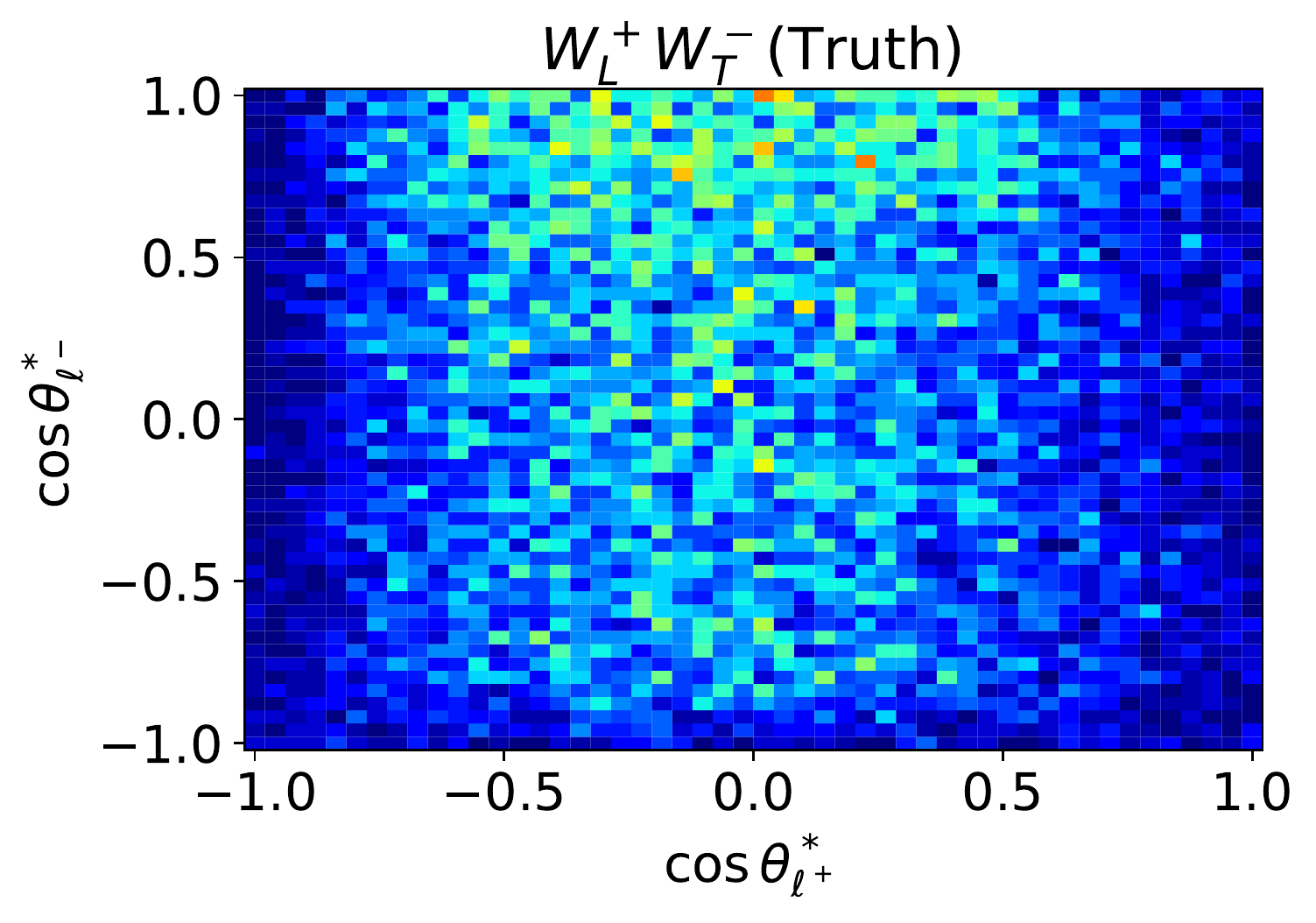}
\includegraphics[width=0.24\textwidth]{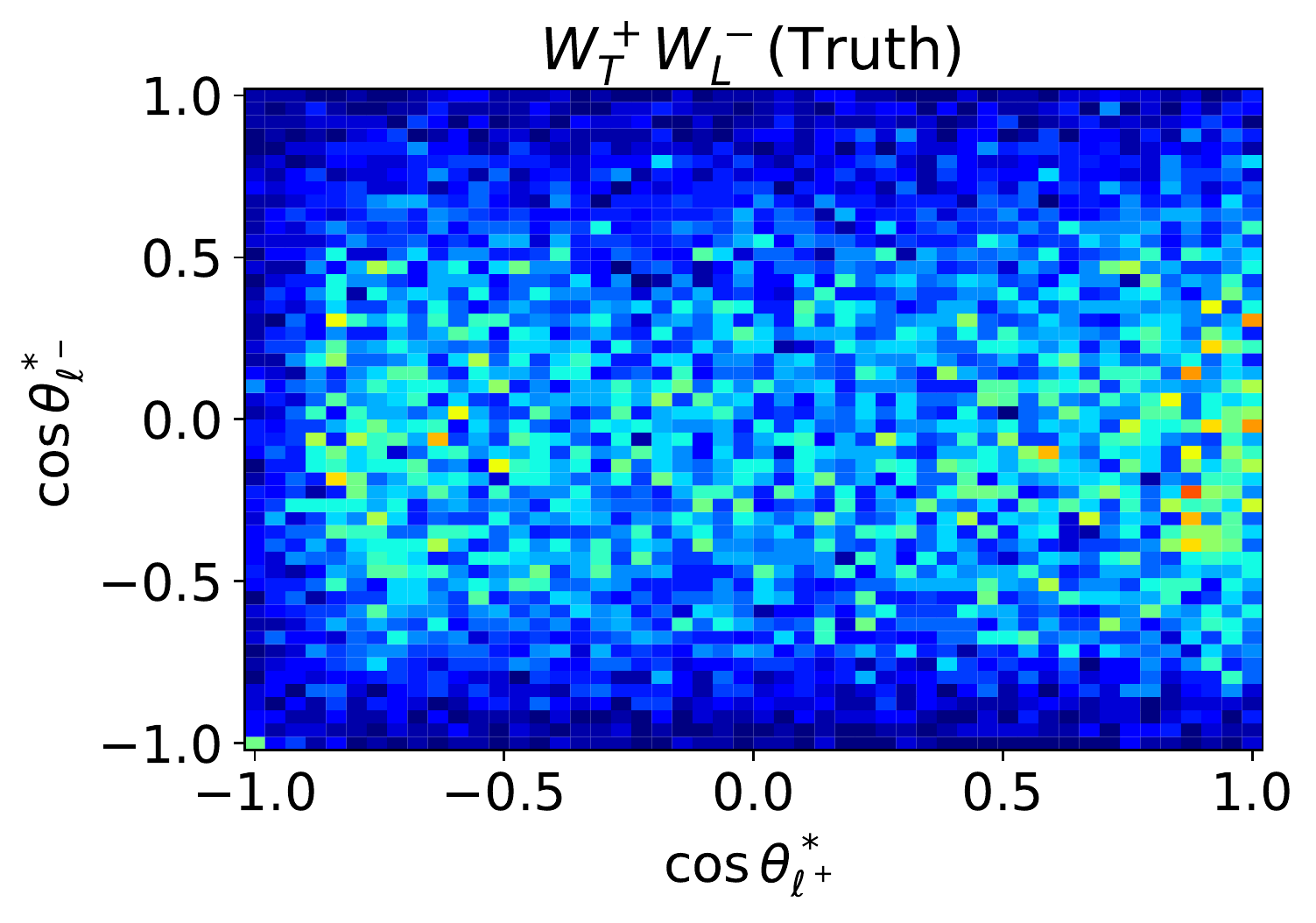}
\includegraphics[width=0.24\textwidth]{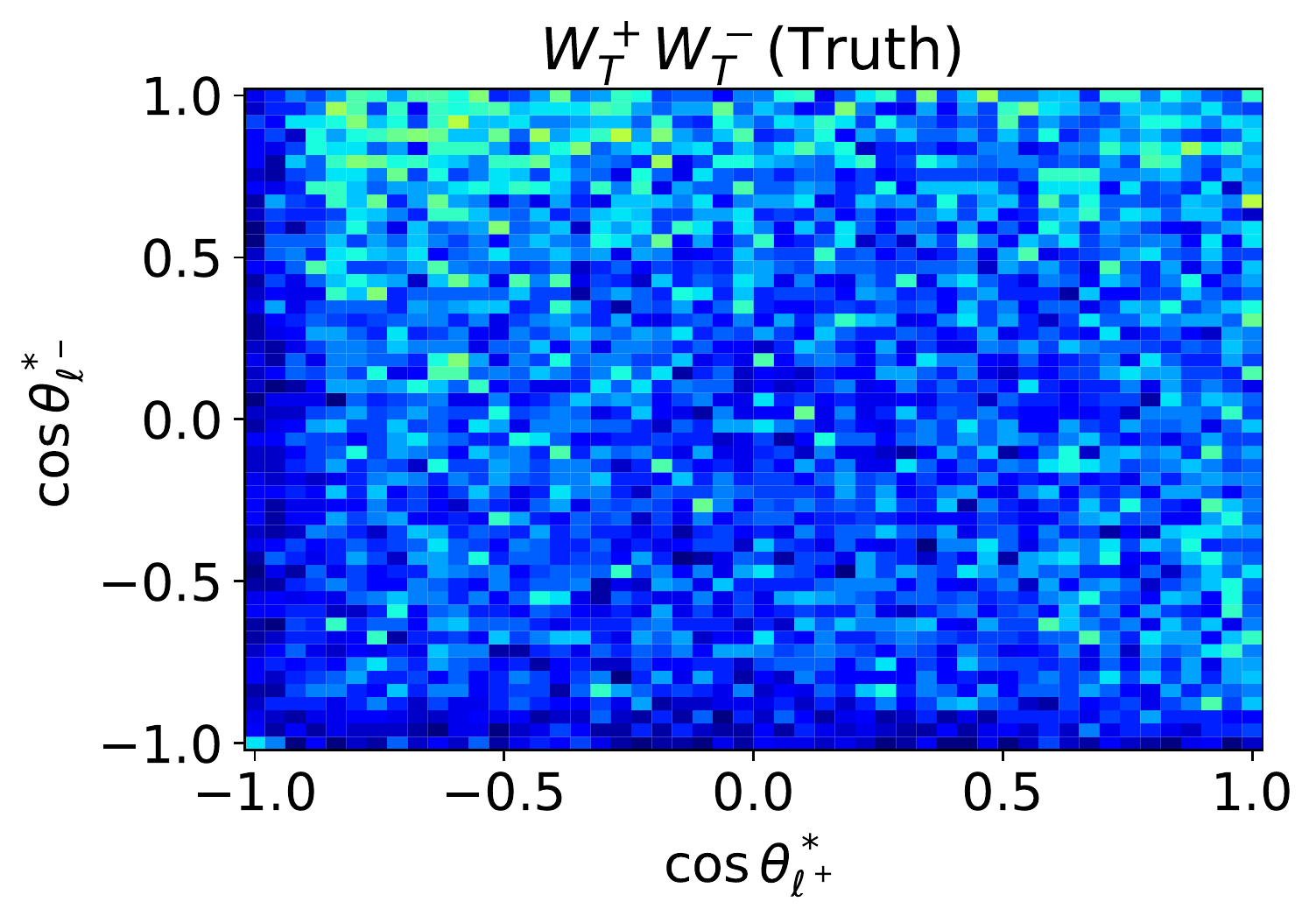}
\caption{Two dimensional distributions of $\cos \theta^*_{\ell^+}$-$\cos \theta^*_{\ell^-}$ for different polarization modes of the SM $W^+W^-$ scattering at 100 TeV. Upper panels: TMIPCC network prediction; Lower panels: truth $\cos \theta^*_{\ell^\pm}$.
\label{fig:2dtemp100}}
\end{figure}

Fig.~\ref{fig:2dtemp100} shows the two-dimensional templates on the $\cos \theta^*_{\ell^+}$-$\cos \theta^*_{\ell^-}$ plane for the TMIPCC network prediction and the truth lepton angles at 100 TeV. 
Compared with Fig.~\ref{fig:2dtemp}, we can find that the distributions for the truth lepton angle vary with collision energy.  
In particular, the effects of the preselection cuts are milder at the 100 TeV collision, leading to slightly sharper lepton angle distributions at the truth level. 
On the other hand, the templates from the network predictions become less precise for the 100 TeV case even after applying both MI and PCC in the loss function. 
This is attributed to the fact that the Transformer networks are only trained on events at the 13 TeV LHC. 
It is possible that one can optimize the results at 100 TeV by using 100 TeV event samples to train the Transformer networks.

\begin{figure}[htbp]
\includegraphics[width=0.3\textwidth]{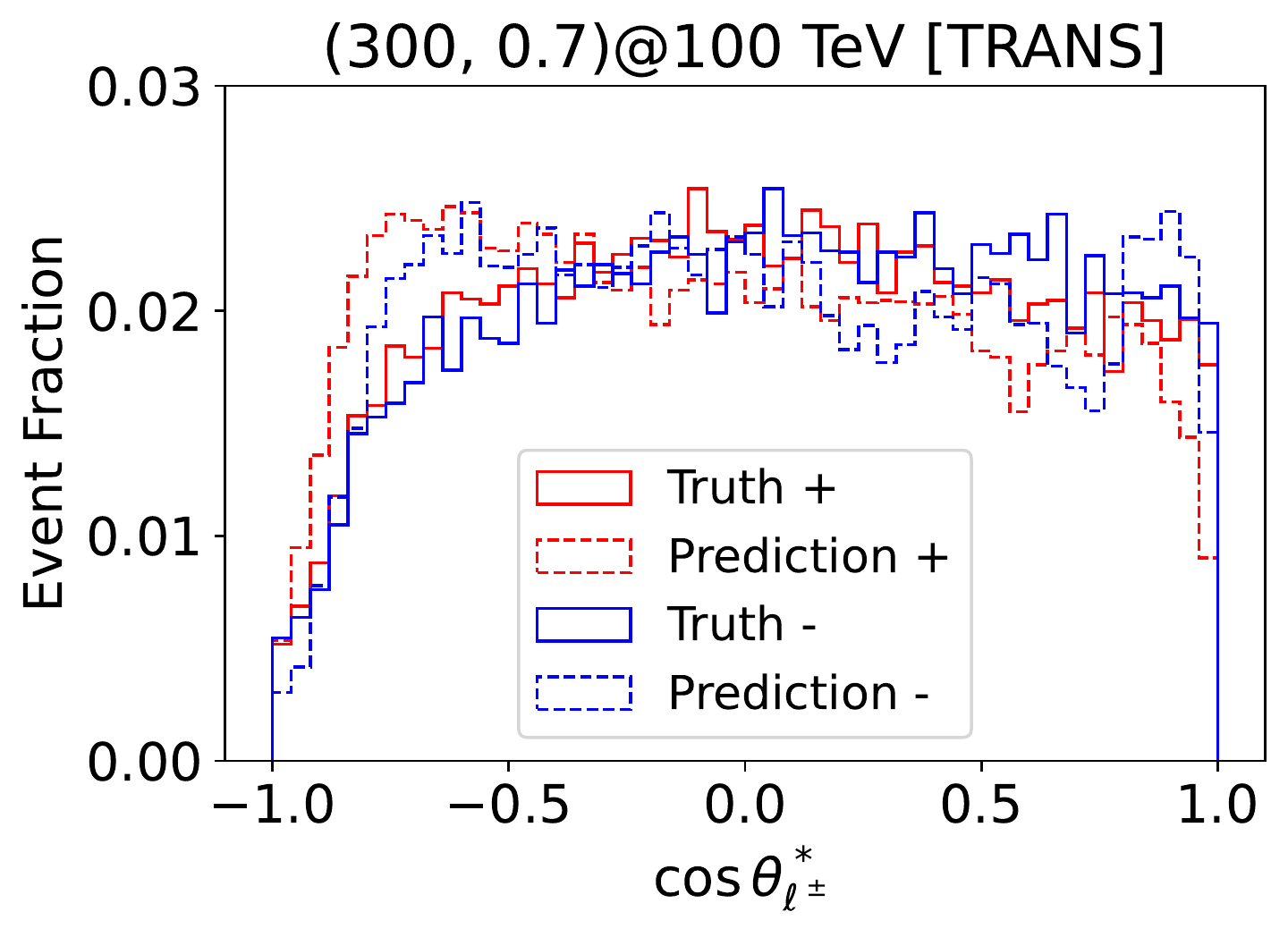}
\includegraphics[width=0.3\textwidth]{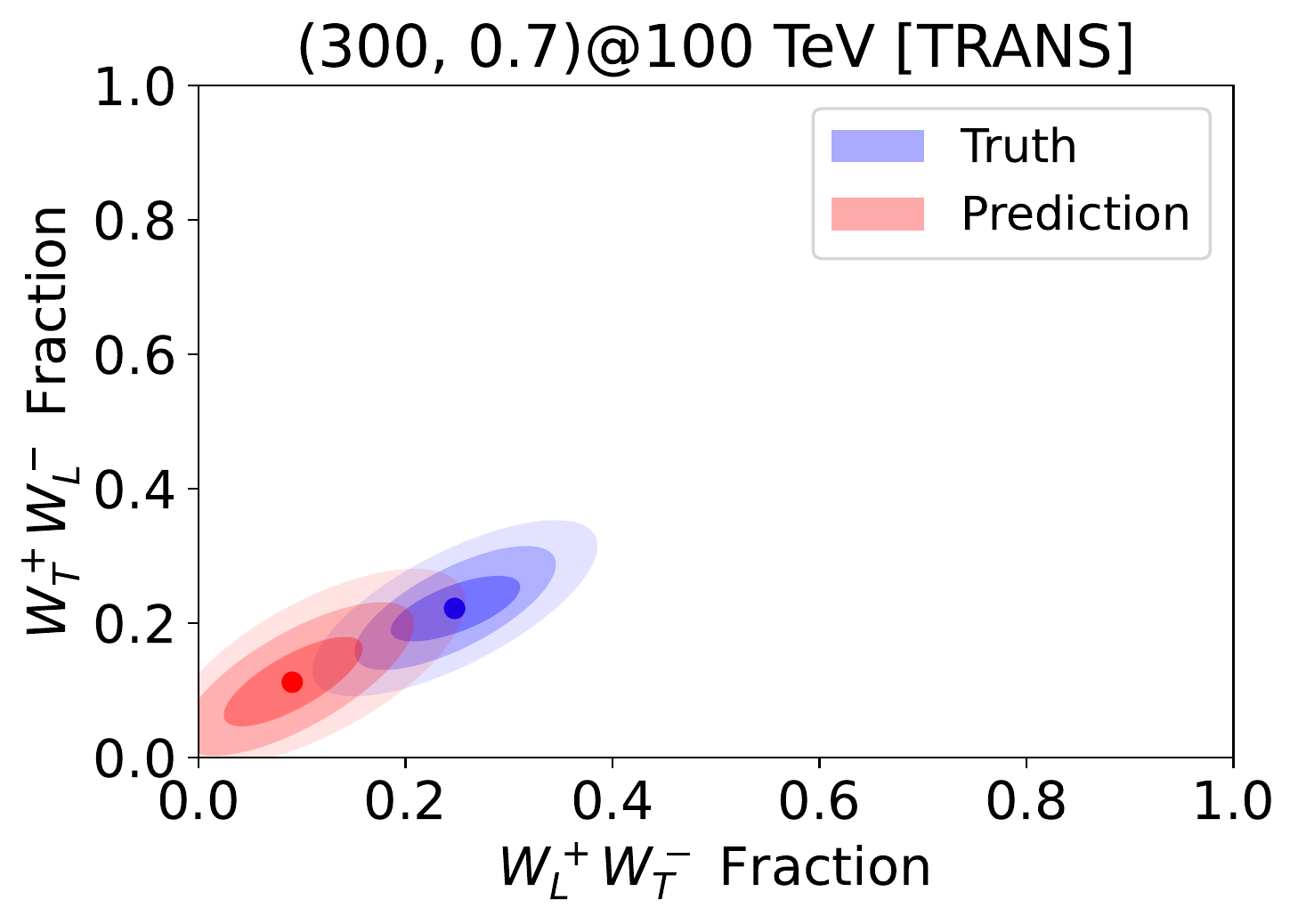}
\includegraphics[width=0.3\textwidth]{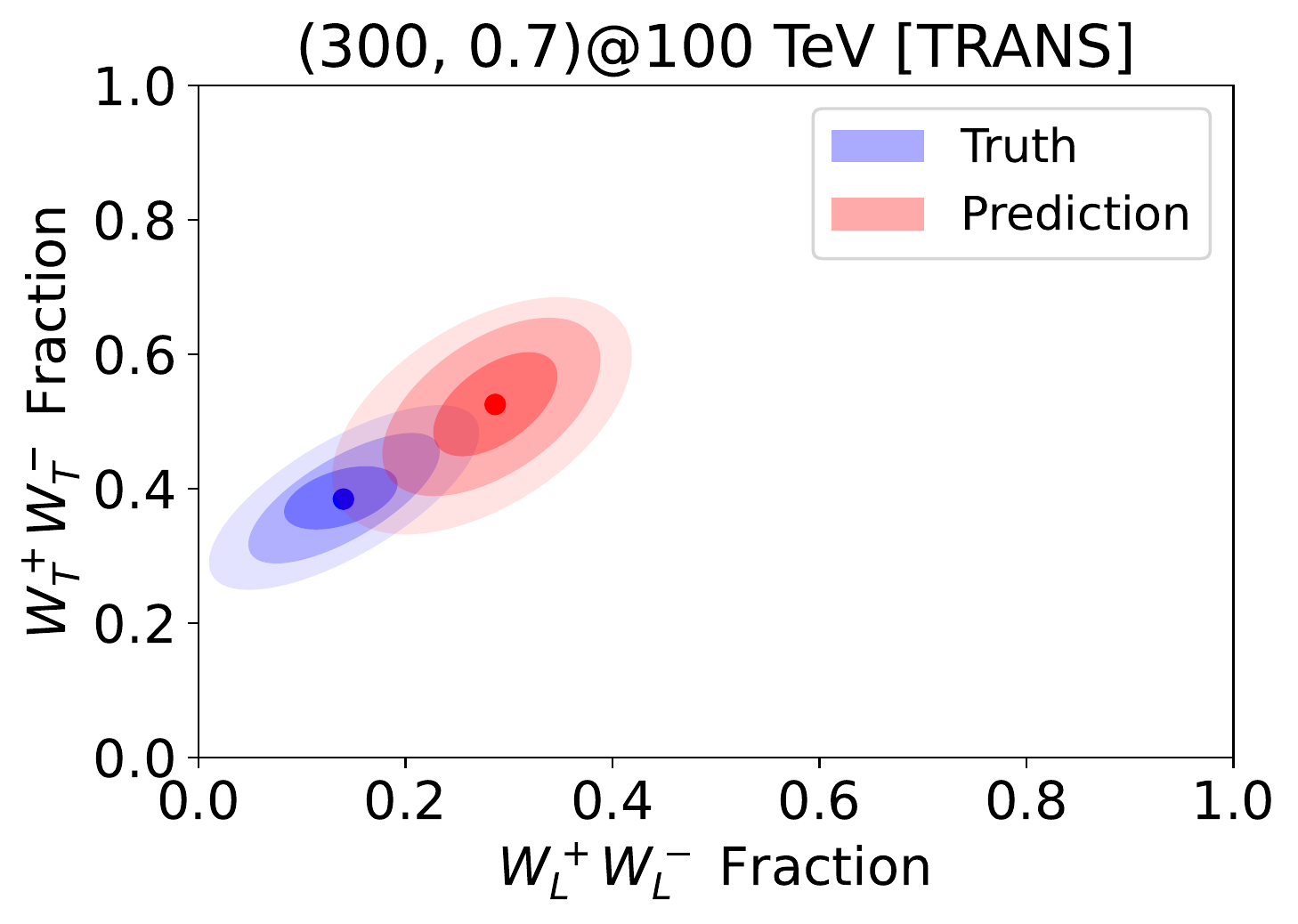}\\
\includegraphics[width=0.3\textwidth]{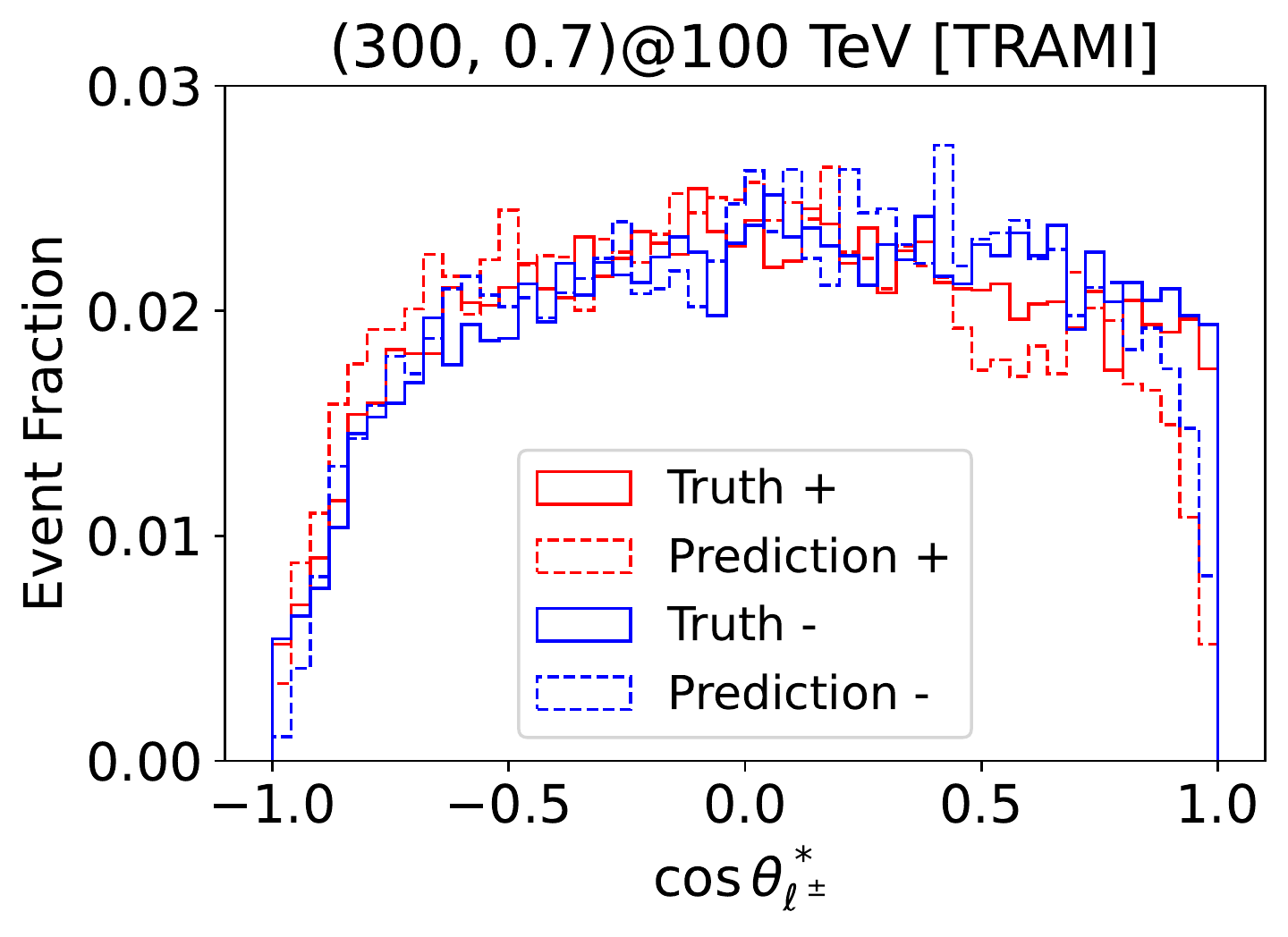}
\includegraphics[width=0.3\textwidth]{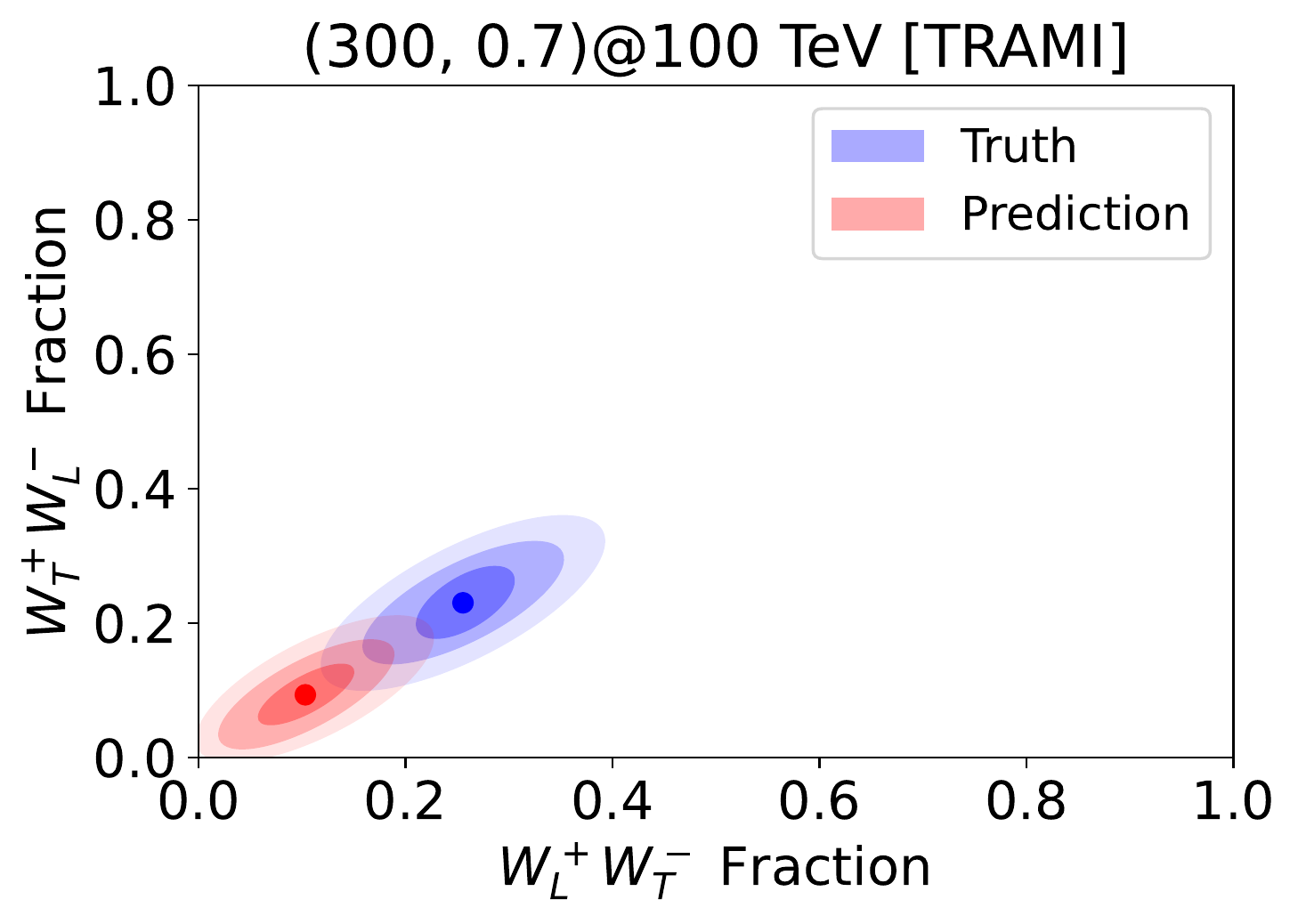}
\includegraphics[width=0.3\textwidth]{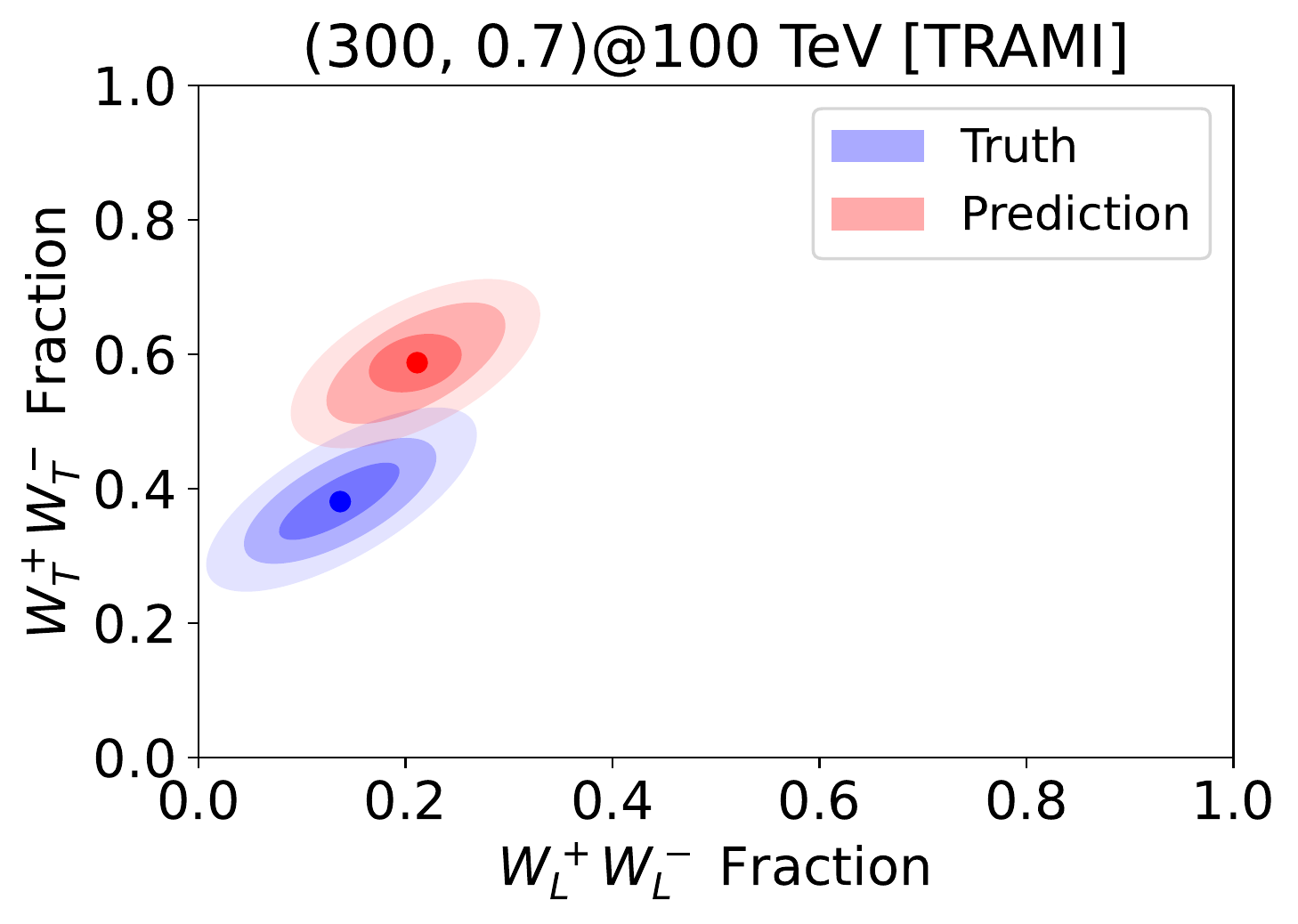}\\
\includegraphics[width=0.3\textwidth]{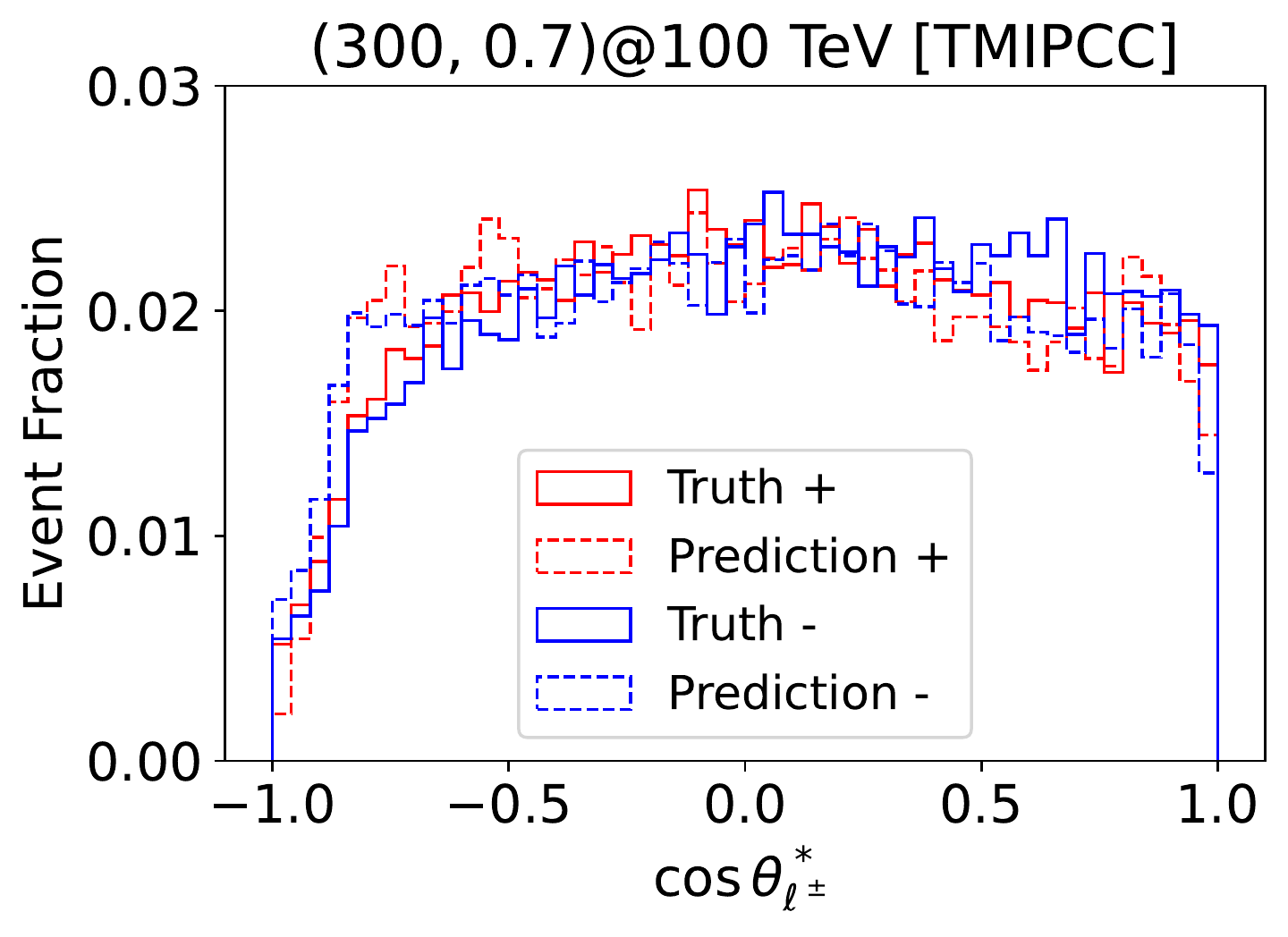}
\includegraphics[width=0.3\textwidth]{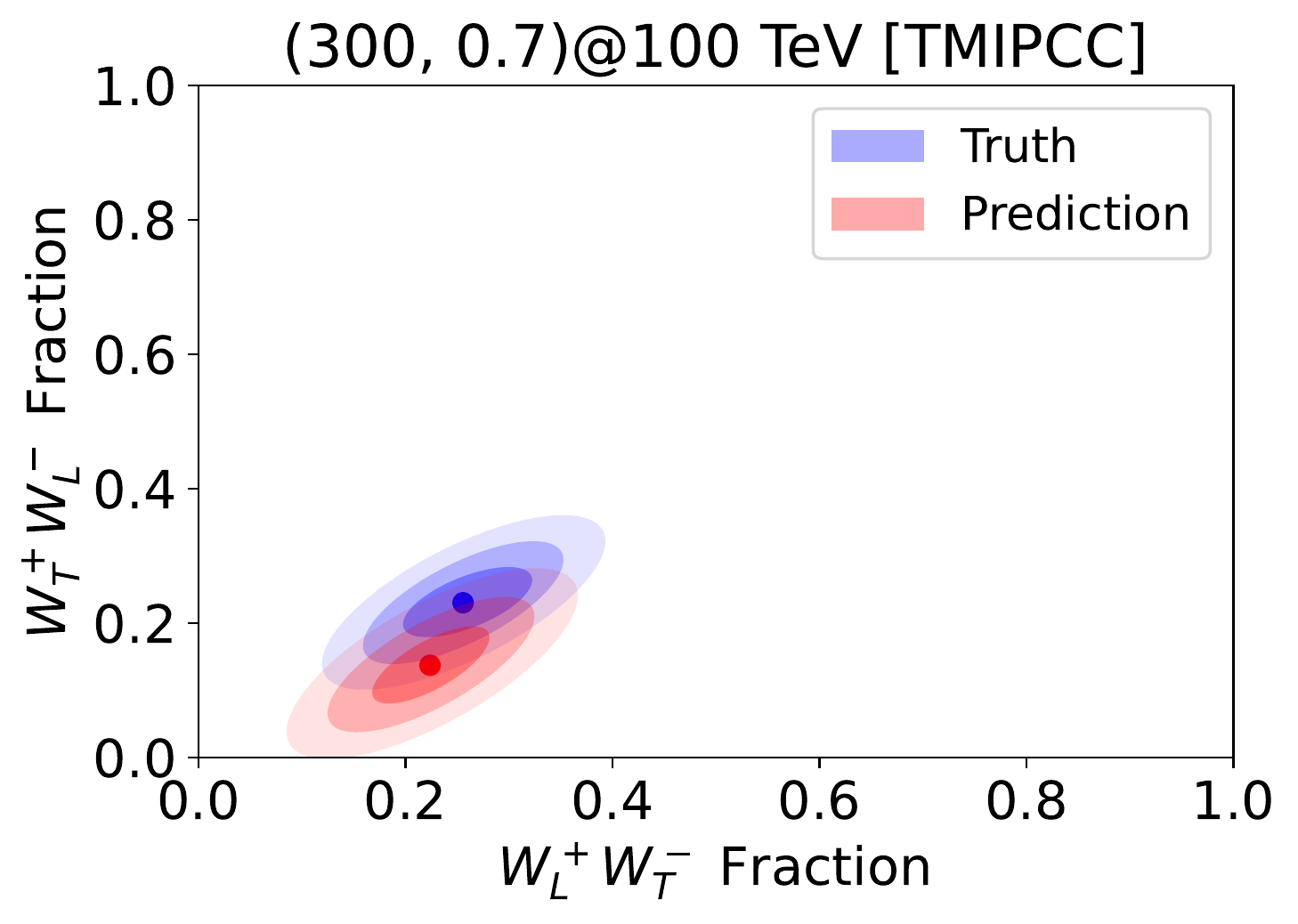}
\includegraphics[width=0.3\textwidth]{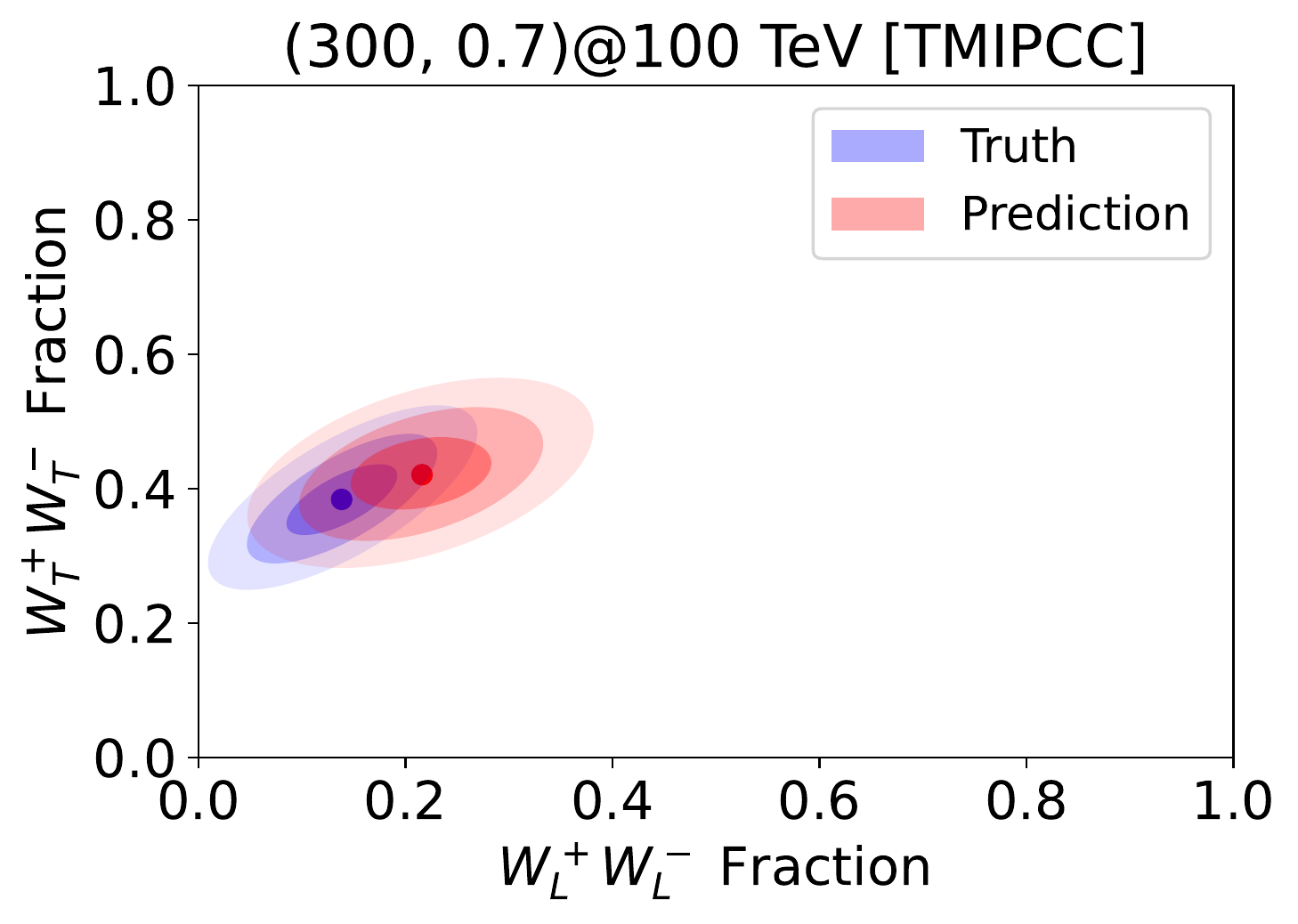}
\caption{For the $W^+W^-$ scattering in the 2HDM with $m_{H_2}=300$ GeV and $\sin (\alpha) = 0.7$ at 100 TeV. Meanings of the plots are the same as Fig.~\ref{fig:smfit}, except that the different shades of the $\Delta \chi^2$ contours
from inside out correspond to $\Delta \chi^2=1$ calculated on datasets with integrated luminosities 3 ab$^{-1}$, 1 ab$^{-1}$ and 500 fb$^{-1}$, respectively.  \label{fig:2hdmfit100}}
\end{figure}

The projected one-dimensional $\cos \theta^*_{\ell^\pm}$ distributions, as well as the fitted polarization fractions from the network predictions and truth lepton angles, are presented in Fig.~\ref{fig:2hdmfit100}. 
The $W^+W^-$ scattering in the 2HDM with $m_{H_2}=300$ GeV and $\sin (\alpha) = 0.7$ at 100 TeV has been taken as an example. After preselection, the production cross section for the process is 148.74 fb. 
Unlike the 13 TeV case, the reduced performance of the TRANS network is visible in the $\cos \theta^*_{\ell^\pm}$ distribution this time. 
As for the polarization fraction, the TRAMI network can no longer work well, and decorrelating the collision energy dependence is essential. 
We can see that the TMIPCC network outperforms TRAMI in the 100 TeV case, although there is still a certain amount of deviation between the predicted ones and the truth ones. 

\section{Subtracting the backgrounds: SM as a case study} \label{sec:bkg}
So far, we have only considered the application of the networks to the $W^+W^-$ scattering processes. 
In practice, there will be events of non-VBS processes that pass the preselections, behaving as backgrounds in our analysis. 
As a result, we can only obtain the superposed distribution of $\cos \theta^*_{\ell^+}$-$\cos \theta^*_{\ell^-}$, from which the contributions from the background processes need to be subtracted out before applying the fit to the templates. 
However, due to the uncertainties in the backgrounds simulation, the background subtraction can not be perfect.
This will lead to reduced precision in extracting the polarization fractions. 

Since we are considering the dileptonic channel of the $W^+W^-$ scattering, the dominant background processes are the dileptonic $t\bar{t}$ and $tW$ processes, mixed electroweak-QCD $W^+W^- j j$ production, as well as the $WZ j j$ production (both at orders of $\mathcal{O}(\alpha_{EW}^4)$ and $\mathcal{O}(\alpha_{EW}^2 \alpha^2_s)$) with gauge bosons decaying leptonically. 
The production cross sections at 13 TeV for the simulated background events before ($\sigma^{\text{fid}}$) and after ($\sigma^{\ell \ell}$) the preselection cuts are listed in Tab.~\ref{tab:xsec}. 
For diboson processes, the transverse momenta of final state jets are required to be greater than 20 GeV.
We will use the measured inclusive cross sections at the LHC for the $t\bar{t}$~\cite{CMS:2016rtp} and $tW$~\cite{Sirunyan:2018lcp} processes,
and use the leading order cross sections which are calculated by \textsc{MG5\_{}aMC@NLO} for the diboson processes.
We note that background events are simulated with at least one lepton in the final state because there could be a misidentified fake lepton due to detector effects. 

\begin{table}[h!]
\begin{center}
\begin{tabular}{|c||c|c|c|c|c|} \hline
  & $t t_\ell$  & $t W_\ell$/$t_\ell W$ & $W_\ell W j j^{\rm QCD}$ & $W_\ell Z j j^{\rm QCD}$ & $W_\ell Z j j^{\rm EW}$  \\\hline 
 $\sigma^{\text{fid}}$ [pb] & 210.3 & 15.9 & 4.68 & 2.20 & 0.487  \\\hline
 $\sigma^{\ell \ell}$ [fb]  & 139.8 &11.6 & 14.7 & 4.49 & 3.68  \\\hline
\end{tabular}
\caption{\label{tab:xsec}  The production cross sections of background processes before and after preselections at the 13 TeV LHC. The superscripts EW and QCD denote the processes at order of $\mathcal{O}(\alpha_{EW}^4)$ and $\mathcal{O}(\alpha_{EW}^2 \alpha^2_s$), respectively.  The subscript $\ell$ denotes the leptonic decay of that particle. }
\end{center}
\end{table}

\begin{figure}[htbp]
\includegraphics[width=0.3\textwidth]{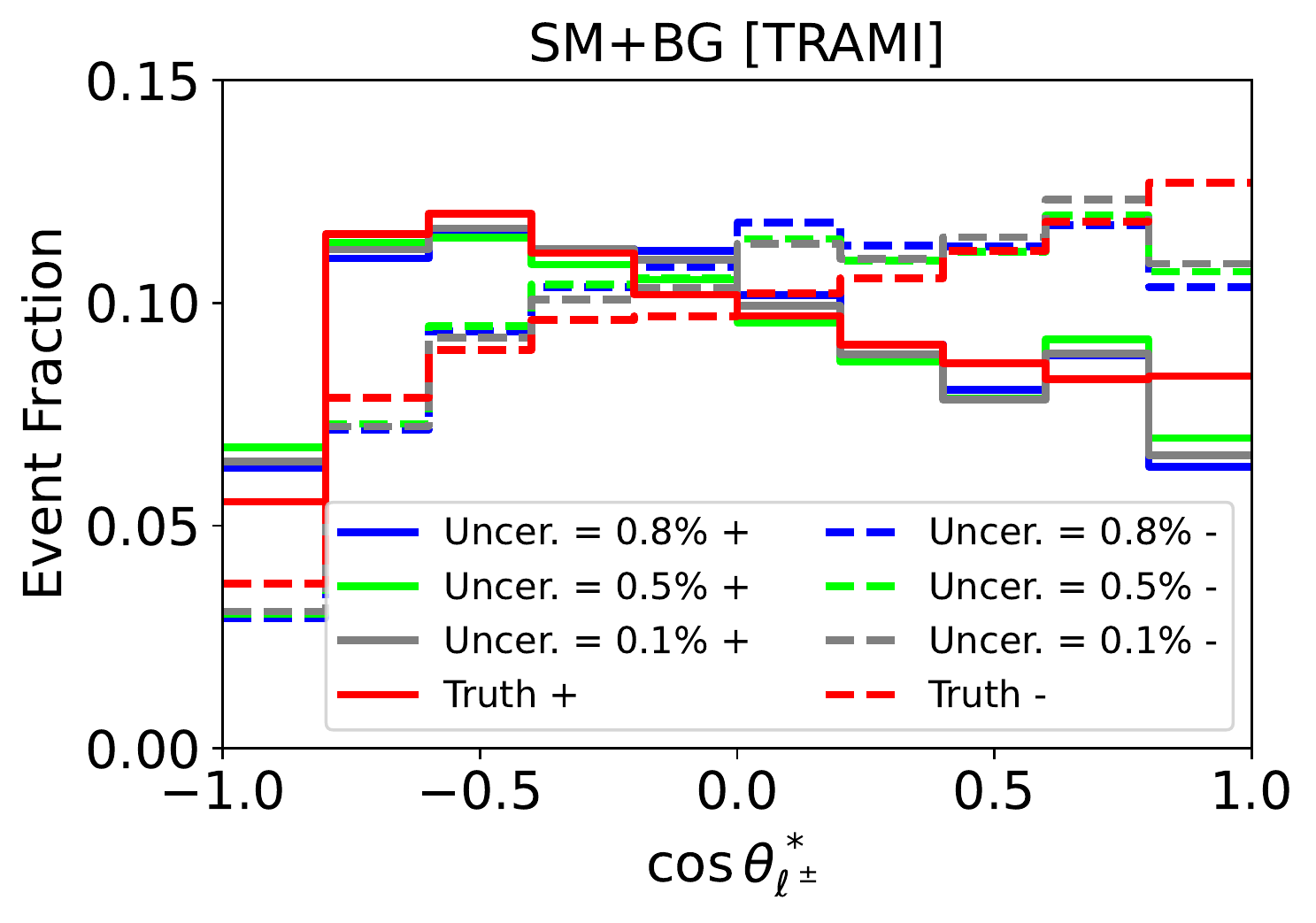}
\includegraphics[width=0.3\textwidth]{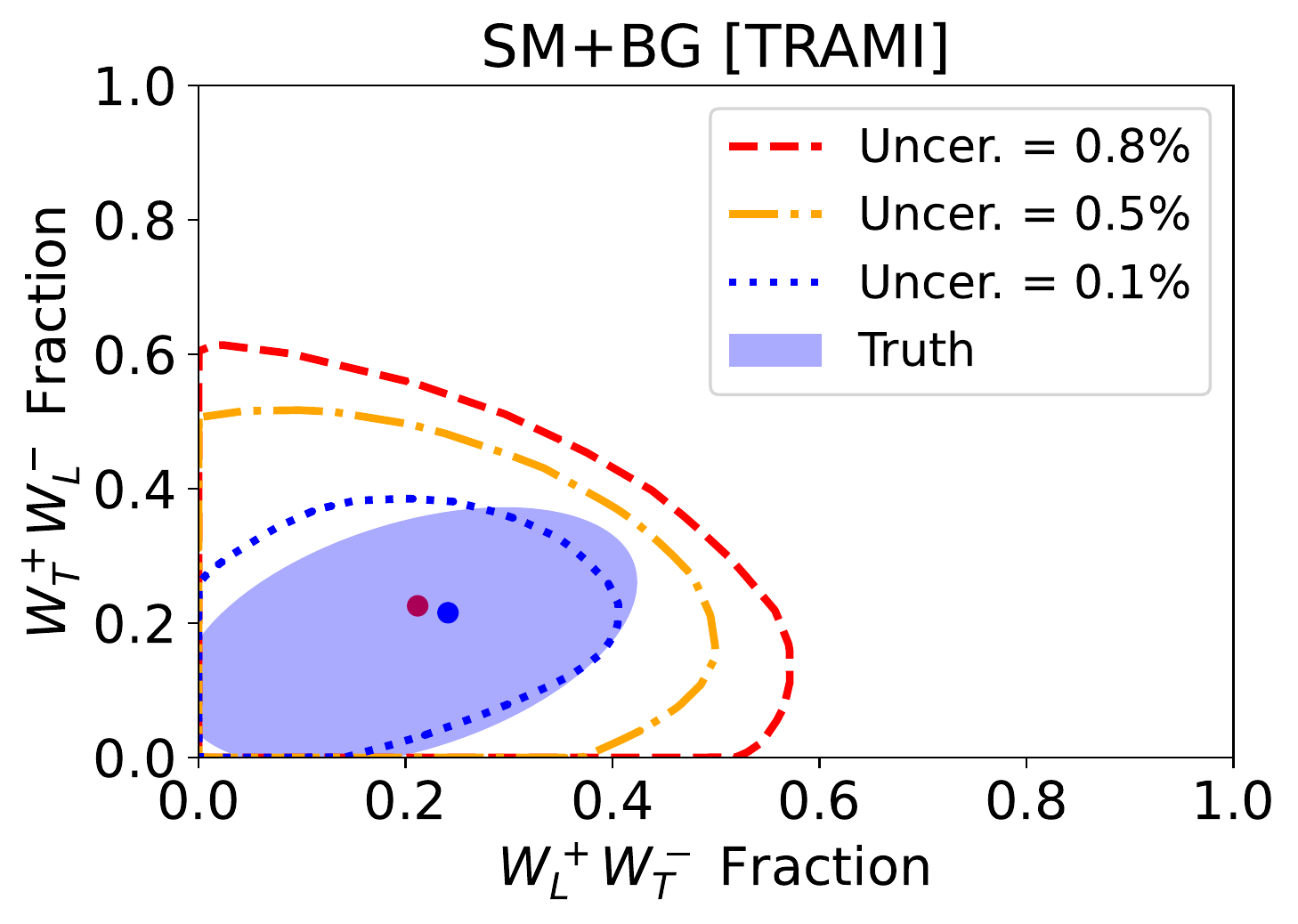}
\includegraphics[width=0.3\textwidth]{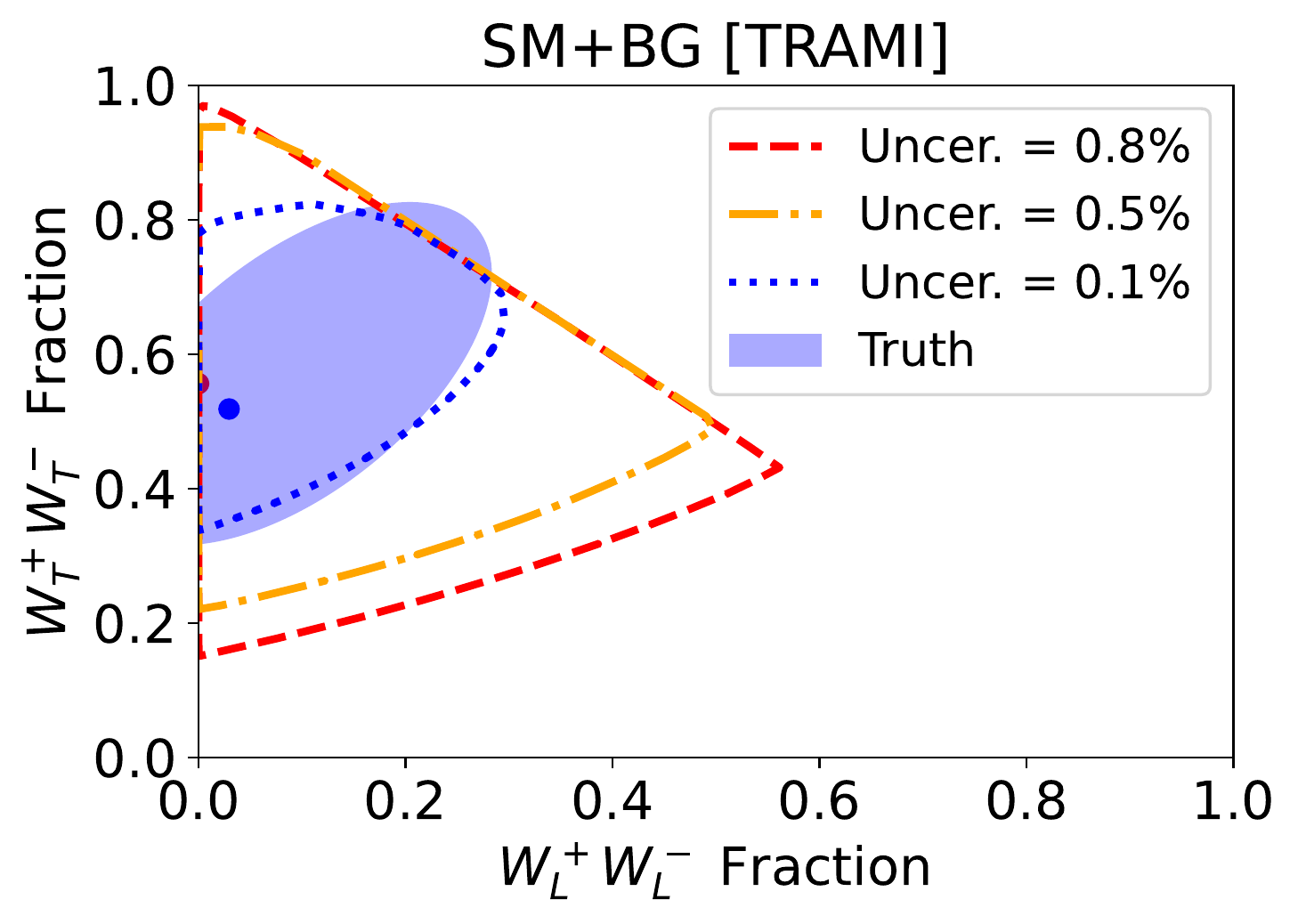}
\caption{The projected lepton angle ($\cos \theta^*_{\ell^{\pm}}$) distributions and fitted $\Delta \chi^2$ contours for the SM $W^+W^-$ scattering at the 13 TeV LHC with integrated luminosity of 3 ab$^{-1}$. The background contributions are subtracted with uncertainties as indicated in the legends. The results of the TRAMI network are shown. \label{fig:smbg13}}
\end{figure}

With background contamination, we adopt the results from the TRAMI network to extract the $W^+W^-$ polarization fractions for the SM production at the 13 TeV LHC. 
The results are shown in Fig.~\ref{fig:smbg13} with varying uncertainties in background subtraction. 
We have assumed uncorrelated systematic uncertainties for the event numbers in lepton angle bins ($10 \times 10$ on the $\cos \theta^*_{\ell^+} - \cos \theta^*_{\ell^-}$ plane). The size of the systematic uncertainty in each bin is indicated in the legend.
The left panel shows the projected lepton angle ($\cos \theta^*_{\ell^{\pm}}$) distributions given by the summed templates with the best-fitted fractions, as well as that obtained at the truth level. 
Since the results with three levels of background uncertainties have similar best fit values, the lepton angle distributions are similar for all three cases. 
However, the total background cross section after preselection is around two orders of magnitude larger than the signal cross section. 
The uncertainties of the fitted fractions are very sensitive to the background uncertainties; ${\it i.e.}$, the size of the $\Delta \chi^2=1$ contour is substantially enlarged for increasing background uncertainty. 
The precision of the extracted fractions is promising only if the background uncertainty in subtraction can be controlled at the 0.1\% level. 
Note that this uncertainty can be much smaller than that of the total cross-section. 
More refined cuts are necessary for large systematic uncertainty of the background. In this case, the template for each polarization should be adjusted accordingly, and the $\chi^2$ fit should be done on the (network predicted) lepton angle distribution after cuts. 
Moreover, with more stringent cuts, a larger number of background events need to be simulated, in order to guarantee relatively small statistical uncertainties in our analysis. 
The main point of the paper is to reproduce the lepton angle distribution, so we decide to leave those more involved analyses for future work.

\section{Conclusion} \label{sec:conclude}

We propose networks composed of a Transformer network and CGAN to predict 
the distributions of the angles between the charged leptons in the gauge boson rest frames and the gauge bosons directions of motion for the dileptonic channel of $W^+W^-$ scattering, so that the polarization fractions of the $W^+W^-$ final state can be obtained from fitting the predicted lepton angle distribution to the given templates. 

There could be unknown new physics contributing to the $W^+W^-$ scattering, which may lead to dramatically different kinematic properties for final states. 
To ensure that the network is able to predict the lepton angle distribution precisely, irrespective of the $W^+W^-$ production mechanism, the loss function of the Transformer network is modified with MI and PCC as defined in Eq.~\ref{eq:lossmi} and Eq.~\ref{eq:losspcc}. 
So that the features produced by the Transformer network contain the lepton angle information as much as possible while decorrelating with other kinematic variables. 
For comparison, three different versions of networks are trained, denoted by TRANS, TRAMI, and TMIPCC.

To illustrate the performances of the networks, we apply them to the events of $W^+W^-$ scattering with dileptonic decay in the SM, in the EFT with non-zero $\bar{c}_H$ as well as in the 2HDM with chosen benchmark points. 
The results are summarized in Tab.~\ref{tab:fracfit}. 

\newcommand{\tabincell}[2]{\begin{tabular}{@{}#1@{}}#2\end{tabular}}
\begin{table}[htb]
\begin{center}
\begin{tabular}{|c|c|c|c|c|c|c|c|c|c|} \hline
 & \multicolumn{2}{c|}{SM} & \multicolumn{2}{c|}{EFT $\bar{c}_H=-1$} & \multicolumn{2}{c|}{2HDM(300, 0.7)} & \multicolumn{2}{c|}{2HDM(300, 0.7)@100 TeV} \\\cline{2-9}
       & Center & \tabincell{c}{Uncertainty\\ 3(30)\ $\text{ab}^{-1}$} & Center & \tabincell{c}{Uncertainty\\ 3(30)\ $\text{ab}^{-1}$} & Center & \tabincell{c}{Uncertainty\\ 3(30)\ $\text{ab}^{-1}$} & Center & \tabincell{c}{Uncertainty\\ 0.5(1)[3]\ $\text{ab}^{-1}$} \\\hline
Truth  & 0.03041 & 0.19(0.06) & 0.11254 & 0.21(0.07) & 0.25959 & 0.17(0.04) & 0.14001 & 0.13(0.09)[0.05] \\
TRANS  & 0.00682 & 0.30(0.09) & 0.09464 & 0.29(0.12) & 0.27940 & 0.31(0.10) & 0.28665 & 0.14(0.10)[0.06] \\
TRAMI  & 0.00040 & 0.23(0.05) & 0.10504 & 0.23(0.08) & 0.22584 & 0.21(0.07) & 0.21123 & 0.12(0.08)[0.04] \\
TMIPCC & 0.01529 & 0.28(0.08) & 0.11879 & 0.30(0.11) & 0.20456 & 0.27(0.08) & 0.21595 & 0.16(0.11)[0.06] \\ \hline
\end{tabular}
\caption{\label{tab:fracfit} Best fit values and 1$\sigma$ uncertainties for the fraction of longitudinal polarized $W^+_L W^-_L$, as shown in Fig.~\ref{fig:smfit}, Fig.~\ref{fig:chm1fit}, Fig.~\ref{fig:2hdmfit} and Fig.~\ref{fig:2hdmfit100}. With different integrated luminosity, the best fit values are kept the same while the uncertainties are different.}
\end{center}
\end{table}

The TRAMI network performs best at 13 TeV for all models, as the features of it have been trained to focus on the lepton angle while not being sensitive to the $W$ boson pair production mechanism. 
The fitting precision of the polarization fraction based on the TRAMI predictions is quite similar to that obtained from using the truth lepton angle, except for the $f_{TT}$ in the 2HDM with $m_{H_2} =300$ GeV and $\sin (\alpha -\beta)=0.7$. 
There is a certain amount of deviation, mainly due to the remaining information of the kinematic variables in the features of the TRAMI network. 
The 1$\sigma$ ranges for the fitted fractions are around 0.2--0.3 for an integrated luminosity of 3 ab$^{-1}$. 
When applying to the events at 100 TeV, the reduced performances of the TRANS network and TRAMI network become visible in the projected one-dimensional lepton angle $\cos \theta^{*}_{\ell^\pm}$ distributions. 
The situation is much improved for the TMIPCC network in which the decorrelation with collision energy is conducted, although there are still mild deviations between the polarization fractions obtained from the TMIPCC network and the truth lepton angle. 
Benefited from the increased production rate at higher collision energy, 
the 1$\sigma$ ranges for the fitted fractions can reach $\sim 0.1$ (0.05) for an integrated luminosity of 1 ab$^{-1}$ (3 ab$^{-1}$).

In practice, the opposite sign dileptonic channel of $W^+W^-$ scattering suffers from backgrounds of dileptonically decaying $t\bar{t}$, $tW$, mixed electroweak-QCD $W^+W^- j j$ as well as $WZjj$ productions.
Considering the uncertainty in background subtraction, the fitting precision of polarization fractions is substantially reduced, mainly due to the relatively small signal to background ratio (after applying the preselections). 


\section*{Acknowledgement}
This work was supported in part by the Fundamental Research Funds for the Central Universities by the National Natural Science Foundation of China (NNSFC) under grant number 11905149.

\phantomsection
\addcontentsline{toc}{section}{References}
\bibliographystyle{jhep}
\bibliography{refer}

\end{document}